\documentclass[graybox, secnum]{svmult}

\usepackage{mathptmx}       
\usepackage{helvet}         
\usepackage{courier}        
\usepackage{type1cm}        
\usepackage{makeidx}         
\usepackage{graphicx}        
\usepackage{multicol}        
\usepackage[bottom]{footmisc}
\usepackage{hyperref}        
\usepackage{soul}            
\hypersetup{colorlinks=true,urlcolor=blue}
\usepackage[square,numbers]{natbib}
\makeindex             

\usepackage{graphicx,psfrag,epsf,color}
\usepackage{amsmath,amssymb,amsfonts}
\usepackage{array}
\usepackage{multirow}
\usepackage{rotating}
\usepackage{slashed}
\usepackage{bm}

\usepackage{enumerate}
\usepackage{tensor}
\usepackage{mathtools}

\usepackage{slashed}
\usepackage[dvipsnames]{xcolor}
\usepackage[capitalize]{cleveref}
\usepackage{dsfont}

\newcounter{MBQ}

\newcounter{RSQ}

\newcounter{PHQ}

\newcounter{PHT}

\newcounter{PHA}

\newcommand{\be}{\begin{equation}}
\newcommand{\ee}{\end{equation}}
\newcommand{\bea}{\begin{eqnarray}}
\newcommand{\eea}{\end{eqnarray}}
\newcommand{\bi}{\begin{itemize}}
\newcommand{\ei}{\end{itemize}}
\newcommand{\ben}{\begin{enumerate}}
\newcommand{\een}{\end{enumerate}}
\newcommand{\bt}{\begin{tabular}}
\newcommand{\et}{\end{tabular}}
\newcommand{\lc}{\left[}
\newcommand{\rc}{\right]}
\newcommand{\lp}{\left(}
\newcommand{\rp}{\right)}

\newcommand{\np}{n_+}
\newcommand{\nm}{n_-}

\def\de{d}
\newcommand{\chris}[2]{\tensor{\Gamma}{^{#1}_{#2}}}

\newcommand{\brac}[1]{\left[#1\right]}
\newcommand{\Riem}[2]{\tensor{R}{^{#1}_{#2}}}

\newcommand{\hinv}{\mathfrak{h}}

\newcommand{\delp}{n_+\partial}
\newcommand{\nip}{n_{i+}}
\newcommand{\nim}{n_{i-}}

\definecolor{navy}{rgb}{0.0,0.0,0.5}

\usepackage{scalerel,stackengine,amsmath}
\newcommand\equalhat{\mathrel{\stackon[2pt]{=}{\stretchto{%
    \scalerel*[\widthof{=}]{\wedge}{\rule{0.9ex}{2.9ex}}}{0.6ex}}}}

\newcommand{\nn}{\nonumber}

\allowdisplaybreaks

\begin{document}
\title*{Soft-Collinear Gravity and Soft Theorems}
\author{Martin Beneke,\thanks{Corresponding author}
Patrick Hager\thanks{Address after October 1st, 2022: 
PRISMA\textsuperscript{+} Cluster of Excellence \& Mainz Institute 
for Theoretical Physics, Johannes Gutenberg University, D--55099 Mainz, 
Germany}  and Robert Szafron}
\institute{M. Beneke, P. Hager \at 
Physik Department T31, 
James-Franck-Stra\ss e~1, 
Technische Universit\"at M\"unchen, 
D--85748 Garching, Germany
\and R. Szafron \at 
Department of Physics, Brookhaven National Laboratory,
Upton, N.Y., 11973, U.S.A.\\[0.3cm]
TUM-HEP-1425/22, 
14 October 2022
}
%
%
\maketitle
\vspace*{-2cm}
\abstract{
This chapter reviews the construction of ``soft-collinear 
gravity'', the effective field theory which describes the 
interaction of collinear and soft gravitons with matter 
(and themselves), to all orders in the soft-collinear power 
expansion, focusing on the essential concepts. Among them are an emergent soft background gauge symmetry, which 
lives on the light-like trajectories of energetic 
particles and allows for a manifestly gauge-invariant 
representation of the interactions in terms of a soft 
covariant derivative and the soft Riemann tensor, and 
a systematic treatment of collinear interactions, which are absent at leading power in gravity. The gravitational 
soft theorems are derived from soft-collinear 
gravity at the Lagrangian level. The symmetries of the 
effective theory provide 
a transparent explanation of why soft graviton emission 
is universal to sub-sub-leading power, but gauge boson 
emission is not and suggest a physical 
interpretation of the form of the universal 
soft factors in terms 
of the charges corresponding to the soft symmetries. The 
power counting of soft-collinear gravity further provides an 
understanding of the structure of loop corrections to the 
soft theorems.
}

\section*{Keywords} 
Gravitation, soft-collinear effective field theory, 
effective Lagrangian, soft and collinear divergences, 
soft theorem, power corrections, 
Einstein-Hilbert theory 

\newpage

\begin{flushright}
\vspace*{0.3cm}
\begin{minipage}{0.6\textwidth}
\it ``My reasons for now attacking this question are: (1) Because 
I can. [...] (2) Because something might go wrong and this would 
be interesting. Unfortunately, nothing does go wrong.''\\[0.2cm]
S. Weinberg, Ref.~\cite{Weinberg:1965nx} 
\end{minipage}
\end{flushright}


\section{Introduction}
\label{sec:introduction}

The gravitational force is widely perceived to be fundamentally 
different from the gauge forces that govern the other microscopic 
interactions of the elementary particles. The gravitational 
interactions are inevitably non-renormalisable, calling for a 
modification at very short distances (or a non-trivial ultraviolet 
fixed point). Their underlying gauge symmetry is related to space-time 
transformations, in contrast to the internal 
SU(3)$\times$SU(2)$\times$U(1) gauge symmetries 
operating on fields in rigid Minkowski space-time.

Yet, from the low-energy perspective and applying the basic 
principles of quantum field theory, the Lagrangian of weak-field 
gravity on Minkowski space follows from the desire to construct 
a consistent theory for a massless spin-2 particle in very 
much the same way as gauge theories do for the case of a massless spin-1 
particle. The universality and space-time symmetry of 
gravitation then arises from the requirement that  
interacting massless fields with spin larger than $\frac{1}{2}$ must couple to a conserved current, which is the 
energy-momentum tensor for spin-2. This motivates a closer inspection 
of the relation between gauge theory and gravitational scattering 
in Minkowski space.

The study of gravitational scattering amplitudes in quantised 
weak-field gravity has attracted much attention after the discovery 
of remarkable relations between graviton and gluon scattering 
amplitudes \cite{Bern:2008qj, Bern:2019prr}, 
which state that 
tree-level amplitudes of the former can be obtained by ``squaring'' 
Yang-Mills tree-level amplitudes and replacing colour factors 
by kinematic ones. The simplest example is the three-point 
amplitude in spinor-helicity notation (reviewed in 
\cite{Elvang:2013cua}):
\begin{eqnarray}
&&\mathcal{A}_{\rm YM}(1^-_a,2^-_b,3^+_c) = -g_s f^{abc} 
\frac{\langle 12\rangle^3}{\langle 23\rangle \langle 31\rangle}
\nonumber\\
&&\hspace*{2.2cm}\big\Downarrow\\ 
&&\mathcal{A}_{\rm EH}(1^{--},2^{--},3^{++}) =\frac{\kappa}{2} 
\frac{\langle 12\rangle^6}{\langle 23\rangle^2 \langle 31\rangle^2}
\nonumber\,.
\end{eqnarray}
Numerous extensions of such ``double copy'' or ``colour-kinematics 
duality'' relations have been found to different gauge/gravity 
theories, to non-trivial classical backgrounds, and to the 
one-loop level.  

\begin{figure}[t]
    \centering
    \includegraphics[width=0.25\textwidth]{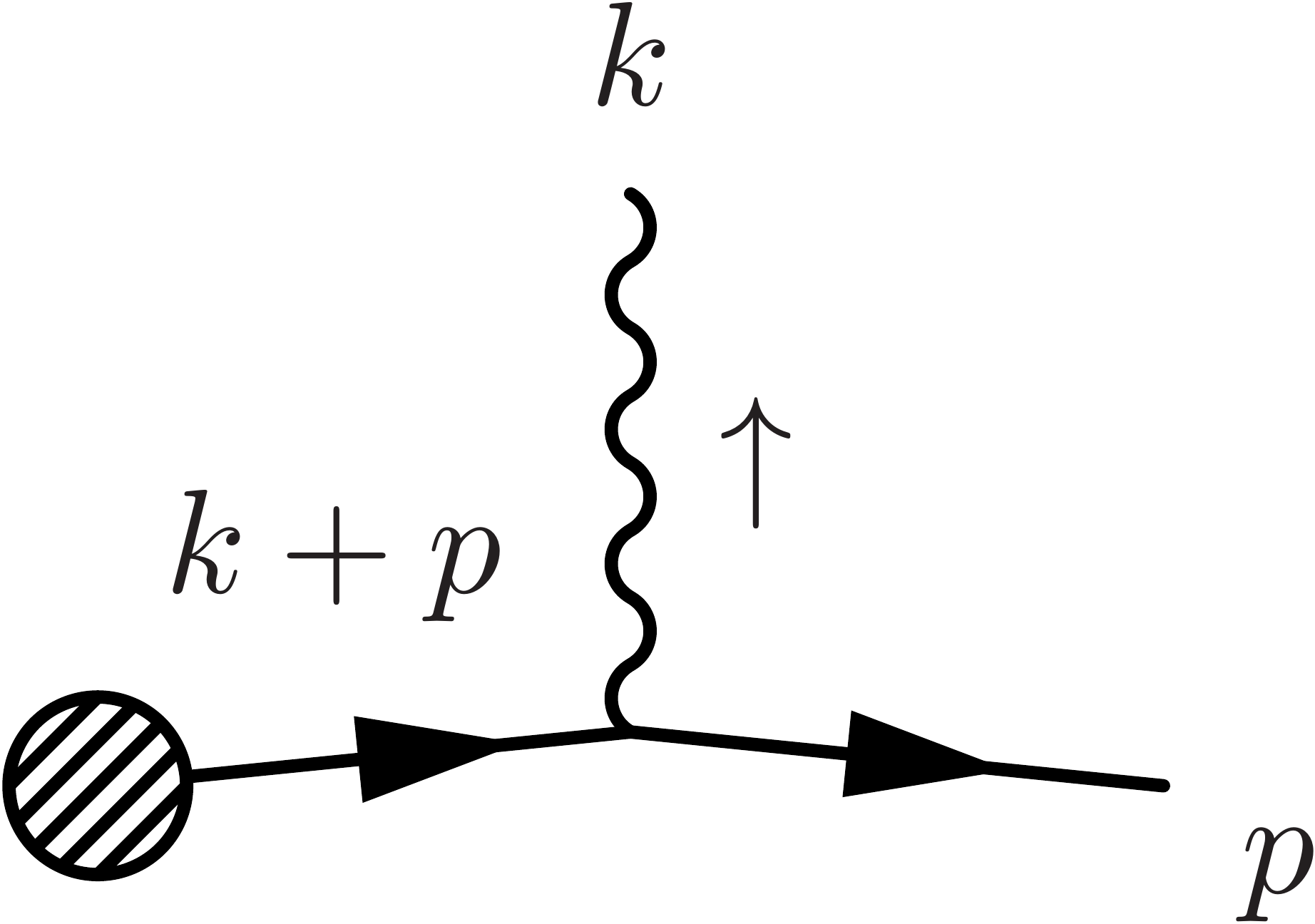}
    \caption{Emission of a boson with momentum $k$ from an 
external energetic particle with momentum~$p$.}
    \label{fig:emission}
\end{figure}

Much can be learnt by looking at the behaviour of quantum 
amplitudes in the infrared (IR). When an energetic, massless 
particle with momentum $p^\mu$ emits another massless 
particle with momentum $k^\mu$, see Figure~\ref{fig:emission}, 
the internal propagator $1/(p+k)^2$ becomes singular in the 
{\em soft} limit $k^\mu\to 0$, and in the {\em collinear} limit 
$p^\mu\parallel k^\mu$. When such configurations are integrated 
over the phase space of the emitted particle, or appear 
inside loops, the result is a logarithmic divergence. It has 
been recognised from the early days of quantum field theory 
that the soft and collinear limits exhibit universal, 
process-independent features 
\cite{Bloch:1937pw,Kinoshita:1962ur,Lee:1964is} 
and that a precise definition of quantum-mechanical observables 
is required to obtain sensible, IR-finite results.
The study of these limits in quantum electrodynamics and 
non-abelian gauge theories accordingly has a long history.

In QCD, which is strongly interacting in the infrared, 
understanding the soft and collinear limit is of paramount 
importance to make predictions for high-energy scattering, 
and culminates in powerful factorisation theorems 
\cite{Collins:1989gx}. In this way, one isolates 
the infrared physics in some well-defined functions, 
while leaving the more process-dependent features for 
perturbative calculations. Given the presence of several 
momentum scales, it is logical to apply effective Lagrangians 
to capture the IR physics, especially as Lagrangians 
are better suited than amplitudes to uncover the gauge-invariance 
and recursive structure of multi-scale problems. 
{\em Soft-collinear effective theory} (SCET)  
\cite{Bauer:2000yr,Bauer:2001yt, 
Beneke:2002ph,Beneke:2002ni}  has therefore 
emerged as an important conceptual and calculational tool 
for factorisation in gauge theories. It is designed to 
precisely reproduce Feynman amplitudes in their soft 
and collinear limits. Moreover, at least in principle, it 
can do so beyond the leading power in the expansion in small 
scale ratios. Although the case of high-energy gravitational 
scattering appears to be of less practical relevance, in view 
of the above mentioned relations between graviton and 
Yang-Mills scattering amplitudes, which is presently not 
understood at Lagrangian level, it is suggestive 
to apply effective Lagrangian techniques to at least the 
soft and collinear limits of graviton amplitudes---which 
is the subject of this chapter. 

These limits already exhibit interesting similarities and 
differences between massless spin-1 and spin-2 particles (gauge 
bosons and gravitons, respectively) coupled to matter.
It has been noted long ago  \cite{Weinberg:1965nx} 
that the ``eikonal'' or 
leading soft limit of gravity is very similar to gauge theory. 
The long-wavelength radiation ``sees'' only the direction 
of motion (classical trajectory) 
and charge of energetic particles. 
Thus,  in the eikonal approximation, the amplitude for radiating a single soft graviton from 
energetic particles with momenta $p_i^\mu$ emerging 
from a hard scattering process, 
\begin{equation}
\label{eq:eikonal-gravity}
\mathcal{A}_{\mathrm{rad}}(p_i;k) = \frac{\kappa}{2}\sum_i \,
\frac{p_i^\mu p_i^\nu\varepsilon_{\mu\nu}(k)}{p_i\cdot k}\,
\mathcal{A}(p_i)\,,
\end{equation}
is obtained from its gauge-theory correspondent, 
\begin{equation}
\label{eq:eikonal-QCD}
\mathcal{A}_{\rm rad}(p_i;k)=-g_s\sum_{i} t^a_i \,
\frac{p_i\cdot\varepsilon^a(k)}{p_i\cdot k}\,\mathcal{A}(p_i)\,,
\end{equation}
by simply replacing the gauge charge (generator) 
by the gravitational charge, momentum, $t_i^a\to p_i^\nu$ 
and adjusting the coupling $g_s$ and polarisation vector 
$\varepsilon_\mu^a(k)$ to the gravitational coupling, 
$\kappa=\sqrt{32 \pi G_N}$, and polarisation tensor, 
$\varepsilon_{\mu\nu}(k)$. Since eikonalised propagators 
$1/(p_i\cdot k)$ are closely related to semi-infinite Wilson line  
operators, we expect soft graviton Wilson lines to 
play a similar role for soft graviton physics  
\cite{Naculich:2011ry,White:2011yy} as they do in gauge 
theories. Eqs.~(\ref{eq:eikonal-gravity}),~(\ref{eq:eikonal-QCD}) 
represent the leading terms in the so-called 
{\em soft theorems}, to which we shall return in a
later section of this chapter.

The collinear limit of graviton amplitudes is, however, very 
different from the one of gauge amplitudes. In fact, in gravity, 
collinear enhancements and singularities are absent altogether.
As a consequence, even if the gravitational coupling was 
not minuscule, i.e. near Planckian scattering energies, energetic 
particles do not produce gravitational jets, which in 
QCD constitute the most visible footprints of the 
non-abelian charges of the quarks and gluons.
The absence of collinear singularities for graviton emission 
was first shown in \cite{Weinberg:1965nx} in the simultaneous 
eikonal limit. Weinberg also noted that it would be rather 
troublesome, if this was not the case, since it would 
prevent the existence of massless particles with gravitational 
charges, that is, any non-vanishing four momentum. 
However, while massless particles with gauge charges do 
not exist in Nature, and hence there is no conflict with the 
existence of collinear singularities in gauge theories, 
there {\em are} massless particles which gravitate, such 
as the photons and the gravitons themselves. 

There is a simple qualitative explanation for the absence of  
collinear graviton singularities based on the classical 
radiation pattern \cite{Beneke:2012xa}. When an energetic particle 
with virtuality much less than its three-momentum squared 
$\mathbf{p}^{2}$ emits a graviton with momentum
$\mathbf{k}$ with small angle $\theta$ between $\mathbf{p}$ and
$\mathbf{k}$, the near mass-shell singularity of the 
emitting particle propagator 
$1/(|\mathbf{p}||\mathbf{k}| (1-\cos\theta))$ 
yields a factor $\theta^{-2}$ for the 
splitting amplitude. Quantising the radiation field in 
the spherical basis with single-particle states 
$|\mathbf{k} j m;\lambda\rangle$, 
where $\lambda$ denotes helicity ($\pm 2$ for gravitons and 
$\pm 1$ for gauge bosons) and $jm$ the angular momentum quantum 
numbers with respect to the quantisation axis $\mathbf{p}$, this implies 
that the emitted graviton must be in a state 
$|\mathbf{k} j \,0;\lambda \rangle$, where $m=0$ due 
to 
angular momentum and helicity conservation. The angular dependence of this 
state is given by the spin-weighted spherical harmonic or Wigner function 
$D_{\pm \lambda,0}^{j}\left(  \mathbf{k}\right)\propto 
\sin^{|\lambda|}\frac{\theta}{2}$, 
which tends to zero as $\theta^{|\lambda|}$ in the 
$\theta\rightarrow 0$ limit.
Thus, the splitting amplitude has no singularity in the 
collinear limit for graviton emission ($\lambda=\pm 2$) in 
contrast to the case of gauge bosons. 

The above argument refers to the physical polarisation states of the 
graviton and thus does not cover the properties of individual 
Feynman amplitudes in general, in particular in covariant gauges, 
which do have collinear divergences. The formal demonstration 
of the absence of collinear divergences without the restriction 
to the eikonal limit adopted in \cite{Weinberg:1965nx} 
has been presented only relatively 
recently \cite{Akhoury:2011kq} with diagrammatic factorisation 
methods. This fact is made evident in the construction 
of the soft-collinear effective Lagrangian for gravity 
(``soft-collinear gravity'') \cite{Beneke:2012xa}: the 
leading effective Lagrangian describing collinear graviton 
self-interactions and their interactions with matter is a free 
theory. This motivates the investigation of collinear 
gravitational physics at {\em sub-leading order} in the 
collinear expansion, where it is non-trivial, and naturally 
leads to the systematic construction of soft-collinear 
effective gravity beyond the leading power in both, the 
collinear and soft limits \cite{Beneke:2021aip}.

The present chapter starts with a review of basic ideas 
and methods for soft-collinear Lagrangians, assuming no 
prior familiarity with the subject. We then provide a  
technically light-weight discussion of soft-collinear 
gravity, focusing on the exposition of the principles 
of the construction, the structure and emergent symmetries 
of the result at 
the expense of many technical details, for which we refer 
to \cite{Beneke:2021aip}. By definition, soft-collinear 
gravity builds an extension of the so-called soft theorems 
to all orders in the loop and soft expansion. It is 
nevertheless of interest to rederive them 
from the effective Lagrangian \cite{Beneke:2021umj}. 
In the last section of this chapter, we briefly cover
how this provides an understanding of why in gravity the 
soft theorem extends to the next-to-next-to-soft order 
(but does not in gauge theory) and how the 
form of the universal terms is related to the (emergent) 
soft gauge symmetries of the effective Lagrangian.
The chapter concludes with a discussion of loop corrections 
to the soft theorem.


\section{Basic ideas and concepts}
\label{sec:basics}

The following section sets up the notation and introduces a number of concepts that arise in effective field theory (EFT) and 
SCET in particular.

\subsection{Perturbative gravity}
\label{sec:PertGrav}

The full theory, from which SCET gravity is constructed, is the Einstein-Hilbert theory with action 
\begin{equation}
\label{eq::GR::EinsteinHilbertAction}
S_{\mathrm{EH}} = -\frac{2}{\kappa^2}\int\de^4x\: \sqrt{-g} \,R\,,
\end{equation}
coupled to matter, here a minimally-coupled scalar field $\varphi$ in the curved space-time with metric tensor $g_{\mu\nu}$.
The matter part is described by the action
\begin{equation}\label{eq::GR::GeometricScalarField}
            S_\varphi = \int\de^4x \sqrt{-g}\: \frac 12 g^{\mu\nu}\partial_\mu\varphi\partial_\nu\varphi\,, 
        \end{equation}
where $g$ denotes the metric determinant.\footnote{In the following, the convention
\begin{equation}
\tensor{R}{^\mu_{\nu\alpha\beta}}= \partial_\alpha \chris{\mu}{\beta\nu}- \partial_\beta\chris{\mu}{\alpha\nu} + \chris{\mu}{\alpha\lambda}\chris{\lambda}{\beta\nu} - \chris{\mu}{\beta\lambda}\chris{\lambda}{\alpha\nu}
\end{equation}
for the Riemann tensor and the metric signature $(+,-,-,-)$ are employed.}
At this point, there is an important notion to clarify: the Einstein-Hilbert action is not renormalisable in the strict sense.
Instead, it
should be treated as the first term in a low-energy EFT of gravity. This idea was pioneered in \cite{Donoghue:1994dn}, and is also well explained in \cite{Donoghue:2017pgk}.
If gravitational loops are included, it is necessary to introduce additional terms to render the theory finite. These higher-order terms correspond to a derivative expansion, and they can be expressed as products of Riemann tensors.
Schematically, the full action then takes the form
\begin{equation}
\label{eq::GR::EinsteinHilbertEFT}
			S_{\mathrm{grav,EFT}} = -\int \de^4x\: \sqrt{-g} \left( \Lambda  + \frac{2}{\kappa^2} R - c_1 R^2 - c_2 R_{\mu\nu}R^{\mu\nu} + \dots\right)\,,
\end{equation}
where $\Lambda$ is the cosmological constant and $R_{\mu\nu} = \Riem{\alpha}{\mu\alpha\nu}$ is the Ricci tensor.
This subtlety is important if one wants to consider higher-loop orders in gravity, though this is not very relevant for the following discussion. 
While the detailed form of the soft-collinear effective theory is determined by the full theory from which it is derived, the \emph{construction} of the effective theory does not depend on the precise loop order that is considered. 
All these higher-order terms respect the same gauge symmetry that is already present in the leading term, the diffeomorphism invariance, and this symmetry forms the guiding principle of the SCET construction.

The diffeomorphisms can be arranged in the form of \emph{local translations}
\begin{equation}
\label{eq::GR::LocalTranslation}
x^\mu \to x^{\prime\,\mu}(x) = x^\mu + \varepsilon^\mu(x)\,,
\end{equation}
where $\varepsilon^\mu(x)$ is some (not necessarily small) vector 
field. Under such a transformation, scalar fields behave as
\begin{equation}
\varphi(x)\to \varphi^\prime(x^\prime) \stackrel{!}{=} \varphi(x)\,.
\end{equation} 
The transformed scalar field $\varphi^\prime(x^\prime)$ can be expressed as
\begin{equation}
    \varphi^\prime(x+\varepsilon(x))
\equiv T_\varepsilon \varphi^\prime(x),
\label{eq:active}
\end{equation}
where the translation operator $T_\varepsilon$ is defined as
\begin{equation}
T_\varepsilon f(x) = f(x) 
+ \varepsilon^\alpha(x)\partial_\alpha f(x) 
+ \frac 12 \varepsilon^\alpha(x) \varepsilon^\beta(x) \partial_\alpha\partial_\beta f(x)+ \mathcal{O}(\varepsilon^3)\,.
\label{eq:Tdef}
\end{equation}
In the following, this active point of view is adopted. That is, 
the local translations correspond to \emph{purely internal} transformations, acting on the field space, instead of actually transforming the coordinates.
In practice, this means that the dynamical fields $\varphi(x)\to\varphi^\prime(x)$ transform, but never the coordinates $x$ themselves.
A scalar field then transforms as
\begin{equation}
\varphi(x) \to \varphi^\prime(x) = \lc U(x)\varphi(x)\rc\,,
\label{eq:scalargaugetrafofull}
\end{equation}
where due to \eqref{eq:active} $U(x)$ is the inverse 
translation, $U(x)=T_\varepsilon^{-1}$.\footnote{The square-bracket notation emphasises that $U(x)$ is a derivative operator that acts only on $\varphi(x)$.}
The metric field transforms as
\begin{equation}\label{eq:basics:Metrictransformation}
    	g_{\mu\nu}(x) \to \lc U(x) \left( \tensor{U}{_{\mu}^\alpha}(x)\tensor{U}{_{\nu}^\beta}(x) g_{\alpha\beta}(x)\right)\rc\,,
\end{equation}
with Jacobi matrices
\begin{equation}
\tensor{U}{^\mu_\alpha}(x) = \frac{\partial x^{\prime\,\mu}}{\partial x^\alpha}(x)\,,\quad \tensor{U}{_\mu^\alpha}(x) = \frac{\partial x^\alpha}{\partial x^{\prime\,\mu}}(x)\,.
\label{eq::GR::JacobianDef}
\end{equation}
It is convenient to adopt this active point of view for two reasons:
first, it emphasises the formal similarity to gauge theories.
Note that the transformation \eqref{eq:scalargaugetrafofull} looks formally the same as the transformation of a matter field $\phi^a(x)$ with respect to a non-abelian gauge symmetry, 
\begin{equation}
    \phi^a(x)\to U^{ab}(x)\phi^b(x)\,,
\end{equation}
which then exposes the similarities and differences between these transformations in gravity and gauge theory. Second, in this active point of view, one never has to worry about transformations of the coordinates, the integral measure $d^4x$ as well as derivatives $\frac{\partial}{\partial x^\mu}$, and only needs to keep track of the transformation of the dynamic fields.
Note that this point of view does not change the \emph{form} of invariant objects. In the passive point of view, $\int d^4x\sqrt{-g}\:\mathcal{L}$ is diffeomorphism-invariant, since the invariant measure $d^4x\sqrt{-g}$ appears in combination with a scalar quantity $\mathcal{L}$.
In the active point of view, the relevant object is the scalar density $\sqrt{-g}\mathcal{L},$ which is gauge-invariant up to total derivatives, rendering the integral manifestly invariant.
This provides a guiding principle for the required manipulations, such as the construction of  gauge-invariant or covariant fields.

To construct perturbative gravity, one assumes small fluctuations of the metric field, and performs a weak-field expansion of
		\begin{equation}
			g_{\mu\nu}(x) = \eta_{\mu\nu} + \kappa h_{\mu\nu}(x)\,,
		\end{equation}
		in $h_{\mu\nu}(x)$ around Minkowski-space with metric $\eta_{\mu\nu}$.
The action then turns into an infinite series in 
$h_{\mu\nu}$, resp. $\kappa$:
		\begin{equation}
		    S = \sum_{k=0}^\infty \kappa^k S^{(k)}\,,
		\end{equation}
where the precise form of $S^{(k)}$ at higher orders depends on which terms one considers to be part of the ``full theory'', i.e. if one only considers Einstein-Hilbert \eqref{eq::GR::EinsteinHilbertAction}, or also takes into account higher-order Riemann terms as in \eqref{eq::GR::EinsteinHilbertEFT}.

In this weak-field expansion, the residual gauge transformations correspond to the translations\footnote{We extract a factor of $\kappa$ 
from $\varepsilon^\mu$, so that $h_{\mu\nu}$ has the linear gauge 
transformation \eqref{eq:Basics:LinearGaugeTrafo} that does 
not contain $\kappa$ explicitly.}
\begin{equation}
            x^\mu \to x^\mu + \kappa\varepsilon^\mu(x)\,.
\end{equation}
In this expansion, the action of $U(x)$ is given by
\begin{eqnarray}
\lc U(x)\varphi(x)\rc &=& \varphi(x) 
    	    - \kappa\varepsilon^\alpha(x)\partial_\alpha\varphi(x) \nonumber \\
&&\hspace*{-1cm}+ \,\frac{\kappa^2}{2} \varepsilon^\alpha(x) \varepsilon^\beta(x) \partial_\alpha\partial_\beta \varphi(x)
    	    + \kappa^2\varepsilon^\alpha(x) \partial_\alpha\varepsilon^\beta(x) \partial_\beta \varphi(x) + \mathcal{O}(\varepsilon^3)\,.\qquad
\label{eq::GR::ScalarTransformExpanded}
\end{eqnarray}
To first order, the transformations \eqref{eq:scalargaugetrafofull}, \eqref{eq:basics:Metrictransformation} reproduce the well-known results
\begin{equation}\label{eq:Basics:LinearGaugeTrafo}
\begin{aligned}
    h_{\mu\nu} &\to h_{\mu\nu} - \partial_\mu \varepsilon_\nu - \partial_\nu \varepsilon_\mu + \mathcal{O}(\varepsilon^2,\varepsilon h)\,,\\
    \varphi
    &\to \varphi - \kappa\varepsilon^\alpha\partial_\alpha \varphi + \mathcal{O}(\varepsilon^2)\,.
\end{aligned}
\end{equation}
An immediate consequence of 
these truncated diffeomorphisms is that objects which are homogeneous in $h$, that is, are monomials in $h$, cannot be gauge-invariant at the same time. 
Gauge-invariant objects, such as the Riemann tensor, are 
represented as a series order-by-order in $h$ or $\kappa$. 
When working with a theory expanded in $h$, sub-leading terms, that is, higher-order terms in $h$, appear in precise combinations to  
yield a gauge-invariant theory. This is a generic feature 
of non-linearly realised symmetries.
    	
\subsection{Basic concepts of SCET}\label{sec:Basics}

SCET is the theory describing the (self-)interactions of soft and collinear particles \cite{Bauer:2000yr,Bauer:2001yt,Beneke:2002ph,Beneke:2002ni}.
It is one of the more complicated effective theories in modern particle physics, and is used to great success in high-energy collider physics.
Despite its technical nature, one can intuitively understand the construction with only a few key concepts.
It is this intuition that is paramount in constructing the gravitational analogue.
This section serves as an introduction into the underlying concepts of SCET. While the section aims to be self-contained, it will focus on exposition rather than derivations. Detailed discussions and computations can be found in \cite{Beneke:2002ph,Beneke:2017mmf,Beneke:2021umj,Beneke:2021aip}.

\begin{figure}
\centering
\includegraphics[width=0.63\textwidth]{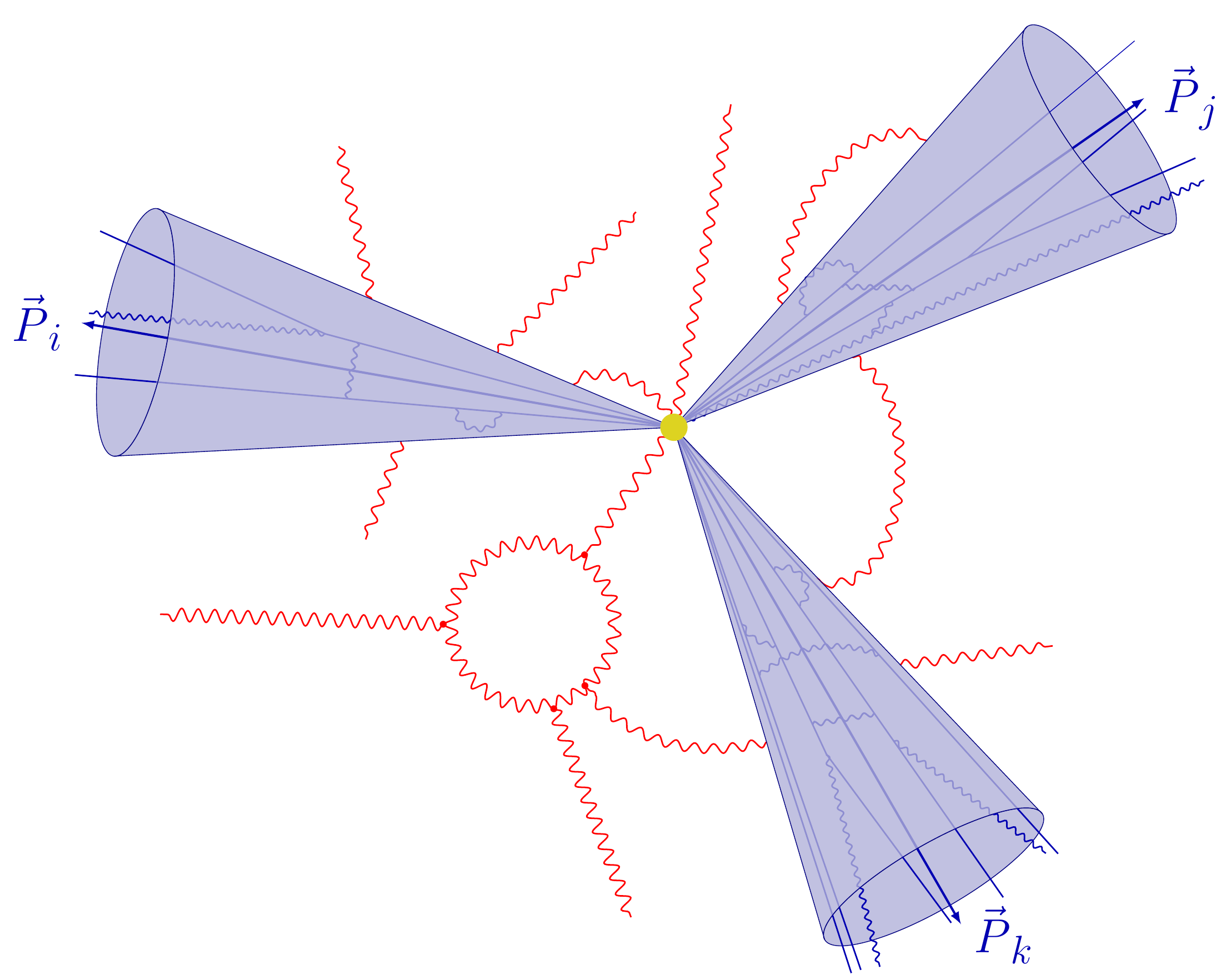}
\caption{The generic kinematic situation of a scattering process 
described by SCET. Blue lines denote collinear particles forming 
jets, emanating from a point-like hard interaction. Red lines 
represent soft modes connecting jets.}
\label{fig:SCETbefore}
\end{figure}

\subsubsection{Kinematics and power-counting}

The general kinematics underlying SCET consist of an energetic scattering, characterised by its large energy scale $Q$ of some hard process, which creates a number of energetic particles, as well as some low-energy, soft radiation. These energetic particles are called ``collinear'' and develop into jets by collinear radiation. The jets are assumed to be well-separated in angle from each other.
The situation is depicted in Figure~\ref{fig:SCETbefore}. 

To be precise, the energetic particles of jet $i$ are taken to be ultrarelativistic (light-like), and characterised by the light-like direction $\nim^\mu$ of the jet with respect to which they have small transverse momentum. Corresponding to each $\nim^\mu$, there is a $\nip^\mu$ such that $\nip\cdot\nim=2$, $n_{i\pm}^2=0$.
These two reference vectors, as well as the two remaining transverse directions, form a basis, with metric tensor
\begin{equation}
    \eta^{\mu\nu} = \frac{1}{2} n^{\mu}_{i+} n_{i-}^{\nu}+\frac{1}{2} n^{\nu}_{i+} n_{i-}^{\mu} + \eta_{\perp i}^{\mu \nu}\,.
\end{equation}
One can decompose a collinear momentum $p^\mu$ as
\begin{equation}
    p^\mu = \nip p \frac{\nim^\mu}{2} + p_{\perp i}^\mu + \nim p \frac{\nip^\mu}{2}\,,
\end{equation}
where the subscript $\perp \hspace{-0.09cm }i$ denotes the transverse components with respect to $\nip^\mu, \nim^\mu$.
Introducing the power-counting parameter $\lambda \sim p_{\perp i}/(\nip p)\ll1$, the components of the collinear momentum $p$ scale as
\begin{equation}\label{eq:basics:collmomscaling}
    (\nip p, p_{\perp i}, \nim p) \sim (1,\lambda,\lambda^2)Q\,,
\end{equation}
and have virtuality $p^2=\lambda^2 Q^2$.
It is a convention to set $Q=1$, which is adopted in the following.
The condition $\lambda\ll 1$ means that the momentum flows in the direction $\nim^\mu$, and only small fractions are deposited in the transverse directions.
For multiple particles in the same collinear sector, it means that these are separated by a small angle, thus they constitute a jet.
Besides the collinear modes, also soft modes can be present, which can interact with collinear modes without changing their collinear nature. This implies isotropic momentum scaling $k^\mu\sim \lambda^2$ and 
virtuality $k^2\sim\lambda^4$, parametrically smaller than collinear 
virtuality. Soft modes can be exchanged between the energetic particles in different directions, as shown in Figure~\ref{fig:SCETbefore}.
This kinematic situation, consisting only of collinear and soft 
modes of different virtuality, is usually denoted as $\text{SCET}_{\rm I}$. In this chapter, we refer exclusively to this situation.

\subsubsection{Field content}

The goal is to construct an EFT that provides a systematic expansion for the full-theory scattering amplitudes in the soft and collinear limits.
The construction differs considerably from traditional EFTs, such as the Fermi theory of weak decays.
In these theories, one is interested in ``light physics'' and integrates out the ``heavy fields'', with masses above some scale $\Lambda$.
This gives rise to an effective Lagrangian wherein only the light fields are dynamical, which now contains (local) higher-dimensional operators, that describe the short-distance physics. The dimension of these operators serves as power-counting parameter for the expansion in~$1/\Lambda$.

In SCET, however, the soft and collinear regions of \emph{all} relevant full-theory particles can contribute to scattering amplitudes.
Instead of integrating out heavy fields, one integrates out certain fluctuations of the fields, or, equivalently, certain \emph{regions} of momenta. Specifically, one integrates out the \emph{hard regions} of momenta, while keeping \emph{collinear} and \emph{soft} ones as the degrees of freedom in the EFT.
In order to achieve this systematically at the Lagrangian level, one needs to split the full-theory fields $\phi_{J}$ into hard, soft and collinear modes $\phi_{J,h}$, $\phi_{J,s}$, $\phi_{J,c_i}$.

To construct a systematic expansion, these modes must be \emph{homogeneous} in $\lambda$, that is, they must scale with a unique power 
of $\lambda$, in order to label them as hard, collinear, and soft.
For this reason, the EFT uses the different fields $\phi_{J,c_i}$, $\phi_{J,s}$, which specifically describe the fluctuations of the original field in the respective kinematic regions.
For SCET, the relevant modes are soft and collinear ones, as depicted in Figure~\ref{fig:SCETbefore}.
In addition, each term in the Lagrangian must be fully expanded 
to also be homogeneous, in the sense that it does not contain any further sub-leading terms.

Since the components $\nip p$ of collinear fluctuations are of the order of the hard scale, one cannot expand in $n_{i+}p/Q$, and the effective Lagrangian cannot be local.
To see this, note that any operator can feature an arbitrary number of large derivatives $\nip \partial$, which all scale as $\nip \partial\sim 1$ and are thus in principle present at any order in $\lambda$.
However, this tower of derivatives can be traded for the non-locality in the $n_{i+}^\mu$ direction.
This is possible simply by rewriting the derivatives as a translation, using
\begin{equation}
    \phi_{J,c_i}(x+tn_{i+}) = \sum_{k=0}^{\infty} \frac{t^k}{k!} (\nip \partial)^k \phi_{J, c_i}(x)\,.
\label{eq:collnonlocal}
\end{equation}
Thus, instead of keeping track of large derivatives $\nip \partial$, one allows for non-localities in the $\nip^\mu$ direction for collinear objects.
When several energetic particles scatter at large angles, the theory is therefore non-local along the directions $\nip^\mu$, i.e. orthogonal to the classical trajectories of the collinear particles.
It is, however, local along these trajectories, resp. the direction $\nim^\mu$ of their momentum. 

In conventional effective theories, the importance of a local operator is tied to its mass dimension.
In SCET, however, the power-counting is not related to the mass dimension. In particular, different components of a field can acquire different scaling in $\lambda$.
To determine the power-counting of these fields, one considers the two-point function
\begin{align}
\langle 0 | T \varphi(x) \varphi(0)  |0 \rangle = \int \underbrace{\frac{d^4 p}{(2 \pi)^4}}_{\lambda^4(\lambda^8)} e^{-ipx} \underbrace{\frac{i}{p^2+i0}}_{\lambda^{-2}(\lambda^{-4})} \sim \lambda^2 (\lambda^4)\,,
\end{align}
where the power-counting of collinear (soft) momenta is given explicitly.
One obtains
\begin{align}
    \varphi_{c_i}(x) \sim \lambda\,, \quad \varphi_s(x) \sim \lambda^2\,,
\end{align}
for the collinear and soft scalar field, respectively.
For the graviton field $h_{\mu\nu}(x)$, one first performs the weak-field expansion and fixes a general de-Donder gauge with parameter $b$.
Then, the two-point function reads
\begin{equation}
  \langle 0 \rvert T h_{\mu\nu}(x) h_{\alpha\beta}(0)\lvert0\rangle = i\kappa^2 \int \frac{d^4p}{(2\pi)^4}\frac{e^{-ipx}}{p^2+i0}\lp P_{\mu\nu,\alpha\beta} + \frac{1-b}{b}S_{\mu\nu,\alpha\beta}\rp\,,
\label{eq:gravitonprop}
\end{equation}
where
\begin{align}
P_{\mu\nu,\alpha\beta} &= \frac 12\lp \eta_{\mu\alpha}\eta_{\nu\beta} + \eta_{\mu\beta}\eta_{\nu\alpha} - \eta_{\mu\nu}\eta_{\alpha\beta}\rp\,,\nn\\
S_{\mu\nu,\alpha\beta} &= \frac{1}{2p^2}\lp \eta_{\mu\alpha}p_\nu p_\beta + \eta_{\mu\beta} p_\nu p_\alpha + p_\mu p_\alpha \eta_{\nu\beta} + p_\mu p_\beta \eta_{\nu\alpha}\rp\,.
\end{align}
Inserting the collinear momentum scaling, one obtains $P_{\mu\nu,\alpha\beta}\sim 1$ if it is non-vanishing, since it does not depend on momenta. The other combination, $S_{\mu\nu,\alpha\beta}$, has non-trivial $\lambda$-scaling.
For example, for the $\perp\!\!+$ and $\perp\perp$ modes,\footnote{For a tensor index $\mu$, $+,-$ means $T_{\pm}\equiv n_{\pm}^\mu T_{\mu}$, while $\perp$ stands for $T_{\mu_\perp}$. In the remainder of Section~\ref{sec:basics}, we drop the collinear direction label $i$ whenever referring to a single collinear sector.} one obtains
\begin{align}
    S_{\perp+,\perp+} \sim \frac{1}{\lambda^2}\,,\quad P_{\perp+,\perp+} = 0\,,\nn\\
    S_{\perp\perp,\perp\perp} \sim 1\,,\quad P_{\perp\perp,\perp\perp} \sim 1\,.
\end{align}
Then, the scaling of the components of the collinear graviton field is easily determined from \eqref{eq:gravitonprop} to be \cite{Beneke:2012xa}
	\begin{align}
           h_{++}&\sim \lambda^{-1}\,, & h_{+\perp}&\sim 1\,, & h_{+-}&\sim \lambda\,,\nn\\
           h_{--}&\sim \lambda^{3}\,, & h_{-\perp}&\sim \lambda^2\,, & h_{\perp\perp}&\sim \lambda\,,\label{eq::CG::GravitonScaling}
		\end{align}
which implies $h^\mu_{\phantom{\mu}\mu}\equiv h \sim\lambda$ for the trace.
Note that the theory contains the $\mathcal{O}(1)$ field component $h_{\mu_\perp+}\sim 1$, and even a power-enhanced component $h_{++}\sim\lambda^{-1}$. This is problematic for the $\lambda$ expansion in the effective theory and must be addressed in the construction of the theory. 
For now, observe that coupling $h_{\mu\nu}$ to a vector $V^\mu$ of the same collinearity yields
        \begin{equation}
            h_{\mu\nu} V^\nu = \frac{1}{2}(h_{\mu+}V_- + h_{\mu-}V_+) + h_{\mu \nu_\perp} V^{\nu_\perp} \sim \lambda V_\mu\,,
        \end{equation}
so index contractions within the same collinear sector are suppressed by a power of $\lambda$. This argument does not hold for couplings between different sectors, where $h_{++}$ could give rise to power-enhancement.
For the soft graviton, it is straightforward to derive the isotropic scaling $s_{\mu\nu}\sim\lambda^2$ from its propagator.

\subsubsection{Gauge symmetry}

Next, it is useful to discuss the gauge symmetry of SCET.
In the full theory, there is only one gauge symmetry, diffeomorphism invariance.
Each full-theory field furnishes some representation of the diffeomorphism group, and comes with its own gauge transformation.
For the scalar field and the graviton, the linear transformation is given in \eqref{eq:Basics:LinearGaugeTrafo}.
Next, one performs the mode split, that is, for each full-theory field $h_{{\rm full},\mu\nu}(x)$ and $\varphi(x)$, one obtains collinear modes $h_{\mu\nu}(x)$, $\varphi_c(x)$ and soft modes $s_{\mu\nu}(x)$, $\varphi_s(x)$.
This has implications for the gauge symmetry in the EFT. Once the split is implemented, e.g. naively as $h_{{\rm full},\mu\nu} = h_{\mu\nu} + s_{\mu\nu}$,
the right-hand side has to transform like the full-theory graviton on the left.
This gives constraints on the allowed gauge symmetry,
since the two fields on the right-hand side are modes with homogeneous power-counting by construction. The soft field can never transform with a gauge parameter that contains collinear fluctuations, since this would turn the soft field into a collinear field. But such gauge parameters are allowed in the transformation of $h_{\mu\nu}$.
The solution is to \emph{extend} the gauge symmetry to two separate, collinear and soft, gauge symmetries such that the collinear fields take the role of \emph{fluctuations} on top of the \emph{soft background} $g_{s\mu\nu} = \eta_{\mu\nu}+\kappa s_{\mu\nu}$.
The technical details will be presented in \cref{sec:softcolgrav}.
The \emph{collinear} gauge transformation then reads
\begin{align}
\kappa h_{\mu\nu} &\to \lc U_c\left( \tensor{U}{_{c\mu}^\alpha}\tensor{U}{_{c\nu}^\beta}(g_{s\alpha\beta} + \kappa h_{\alpha\beta})\right)\rc - g_{s\mu\nu}\,,\nn\\
g_{s\mu\nu} &\to g_{s\mu\nu}\,.\label{eq:basics:ColGaugeTrans}
	\end{align}
The intuition behind these transformations is clear:
the fluctuation $h_{\mu\nu}$ comes with its own gauge symmetry, but 
the background $g_{s\mu\nu}$ is unaffected by this collinear 
transformation, as it must be. The fluctuation $h_{\mu\nu}$ 
transforms in the same way as the $h_{\mu\nu}$ of the weak field 
expansion in the full theory, except that the rigid 
Minkowski background $\eta_{\mu\nu}$ is 
replaced by $g_{s\mu\nu}$ in \eqref{eq:basics:ColGaugeTrans}. 
The soft background field is itself dynamic, with 
$s_{\mu\nu}$ transforming under the soft gauge symmetry $U_s(x)$ 
(to be discussed in \cref{sec:softcolgrav}). 
Let us stress the main message: to consistently implement the split into soft and collinear modes, one treats the collinear fields as fluctuations on top of a soft background.\footnote{
Since the background field method may be more familiar in gauge 
theories, we invite the reader to compare the gauge 
transformations in SCET for QCD \cite{Beneke:2002ni} with those 
for gravity discussed here.}

\subsubsection{Light-front multipole expansion}
\label{sec::multipole}

Due to the power-counting of their momenta (and thus coordinate arguments), products containing both soft and collinear fields are not homogeneous in $\lambda$.
In Fourier space, one finds, for example,
\begin{equation}
\label{eq:ScalarSCETInteractionFT}
    \varphi_c(x)\varphi_s(x) = \int \frac{d^4p}{(2\pi)^4}\frac{d^4 k_s}{(2\pi)^4} e^{-i(p+k_s)\cdot x} \Tilde{\varphi}_c(p)\Tilde{\varphi}_s(k_s)\,.
\end{equation}
Here, the product in the exponent reads
\begin{equation}
    (p+k_s)\cdot x = \frac 12 \underbrace{(\np p + \np k_s)}_{1 + \lambda^2} \nm x + \underbrace{(p_\perp + k_{s\perp})}_{\lambda + \lambda^2}\cdot x_\perp + \frac 12 \underbrace{(\nm p + \nm k_s)}_{\lambda^2 + \lambda^2} \np x\,.
\end{equation}
Only the combination 
\begin{equation}
    n_- p + n_- k_s \sim \lambda^2
\end{equation} 
scales homogeneously as $\mathcal{O}(\lambda^2)$, while the $k_{s\perp}$ and $n_+k_s$ components are suppressed with respect to $p_\perp$ and $n_+p$.
The physical reason for this is that soft fields can only resolve large distances 
\begin{equation}
x_-^\mu \equiv \np x \,\frac{\nm^\mu}{2}\,,
\end{equation} 
whereas collinear fields fluctuate also over smaller distances.
Therefore, one must expand the exponential in \eqref{eq:ScalarSCETInteractionFT} as
\begin{equation}
    e^{-i(p+k_s)\cdot x} = e^{-i(p + \nm k_s \frac{\np}{2})\cdot x} \lp 1 -i\,  k_{s\perp}\cdot x_\perp - \frac{i}{2} \np k_s \nm x + \dots\rp\,,
\end{equation}
keeping only the homogeneous soft momentum $\nm k_s$ in the exponent.
This expansion is equivalent to the {\em light-front 
multipole expansion} \cite{Beneke:2002ph,Beneke:2002ni} 
\begin{equation}
    \varphi_s(x) = \varphi_s(x_-) + (x-x_-)^\mu\lc\partial_\mu \varphi_s\rc (x_-) +  \mathcal{O}(\lambda^2\varphi_s)\,.
\end{equation}
of the soft field on the left-hand side of 
\eqref{eq:ScalarSCETInteractionFT}. This expansion of soft fields 
about the large coordinate $n_+ x$ of collinear fields must be 
applied whenever soft and collinear fields appear in a product. It generates an infinite tower of sub-leading in $\lambda$ soft-collinear interactions, which precisely reproduces the expansion in the small soft momenta $\nip k_s, k_{s\perp}$ of the momentum-space amplitude. One 
may note the similarities to the standard multipole expansion for 
spatially localised systems, which also appears in non-relativistic effective theories.

This has an important implication for the gauge symmetry, since the gauge transformation \eqref{eq:basics:ColGaugeTrans} of $h_{\mu\nu}$ 
contains products of collinear and soft fields. Furthermore, the soft gauge transformation $U_s(x)$ of collinear fields is a product of soft and collinear fields, which still mixes 
different powers in $\lambda$, and therefore must be multipole-expanded. 
For example, the soft gauge transformation of the collinear matter field follows from \eqref{eq:Basics:LinearGaugeTrafo} and reads
\begin{equation}
    \varphi_c(x) \to U_s(x)\varphi_c(x) = U_s(x_-)\varphi_c(x) + x_\perp^\alpha\lc\partial_\alpha U_s\rc(x_-)\varphi_c(x) + \dots\,.
\end{equation}
The additional terms beyond the leading one imply that the 
transformation mixes different orders in $\lambda$ in a way that is incompatible with the multipole expansion of the Lagrangian. To alleviate this, one needs to find a way to obtain collinear fields that have a \emph{homogeneous} gauge transformation, respecting the multipole expansion.
The resolution of this subtlety is technically quite involved, but intuitively very simple:
recall that the multipole expansion is necessary, because soft fields cannot resolve the small-scale fluctuations of collinear modes.
However, in the naive split $g_{\mu\nu}(x) = g_{s\mu\nu}(x) + \kappa h_{\mu\nu}(x)$, one implicitly assumes that this is possible, since the full $s_{\mu\nu}(x)$ appears inside $g_{s\mu\nu}(x)$.
Therefore, the soft gauge symmetry, which is the symmetry of the background metric $g_{s\mu\nu}(x)$, is constructed with respect to the \emph{wrong background field}.
Instead, one needs to identify the appropriate background field $\hat{g}_{s\mu\nu}(x_-)$, whose constituent soft fields can only depend on $x_-^\mu$. In other words, these soft background fields live only on the classical trajectory of the collinear particles.
This proper background field comes with its residual transformation, and it is this residual transformation that is ``homogeneous'' in $\lambda$, respecting the multipole expansion.

A trace of this homogeneous background field can already be seen in the leading-power Lagrangian. By performing the multipole expansion and keeping only the leading terms in \eqref{eq::GR::GeometricScalarField}, one obtains the soft-collinear interaction
\begin{equation}
    \mathcal{L}^{(0)} = -\frac \kappa 8 \nm^\mu \nm^\nu s_{\mu\nu}(x_-) \partial_+\varphi_c \partial_+\varphi_c\,.
\end{equation}
When inserting this expression into a soft-emission diagram, the combination $\varphi_c\partial_+\varphi_c$ generates the eikonal propagator $\frac{i\np p}{2p\cdot k}=\frac{i}{\nm k}$.
Therefore, the leading-power interaction from the homogeneous background field $s_{--}(x_-)$ yields a term proportional to $ \varepsilon_{--} p_+$.
Finding the full expression of the soft background field is more involved, and is explained in \cref{sec:softcolgrav}. Once this proper background field is identified, it will be straightforward to construct the EFT systematically to all orders.

\subsubsection{Basic features of the effective Lagrangian}

The effective theory then takes a simple structure.
The Lagrangian $\mathcal{L}_{\rm SCET}$ splits into a soft-collinear and a purely-soft Lagrangian
\begin{equation}\label{eq:Basics:LagrangianSplit}
    \mathcal{L}_{\rm SCET} = \sum_i\mathcal{L}_{c_i}\lc h_{i\mu\nu}(x),s_{\mu\nu}(x_{i-})\rc + \mathcal{L}_{s}\lc s_{\mu\nu}(x)\rc\,.
\end{equation}
In the collinear part, there is a sum over all collinear sectors 
defined by the directions $\nim^\mu$ of the jets. 
This sum arises because one needs a \emph{hard} scattering to generate particles of different collinear sectors.
Since the Lagrangian does not describe hard scattering, there is no direct interaction vertex between different collinear sectors.
Instead, these processes are allocated to the so-called $N$-jet operators, which generate energetic and soft particles from hard scattering.

The soft-collinear Lagrangian contains also purely-collinear terms.
This purely-collinear Lagrangian, as well as the purely-soft Lagrangian, is then completely equivalent to the original full theory (in weak-field expansion).
This is due to the fact that if only one scale is present, e.g. by only considering purely-collinear or purely-soft modes without any external sources, then there is no Lorentz-invariant notion of soft or collinear.
One could simply perform a Lorentz boost and collinear modes would become soft, and vice-versa.
It is the presence of a source that provides meaning to the notion of soft and collinear in the first place.

Therefore, all the non-trivial physics is contained inside the soft-collinear interaction vertices as well as the ``sources'', which are described by $N$-jet operators.
These vertices stem from the terms in the collinear Lagrangian that are covariant with respect to the non-trivial soft background.

\begin{figure}[t]
    \centering
    \includegraphics[width=1.0\textwidth]{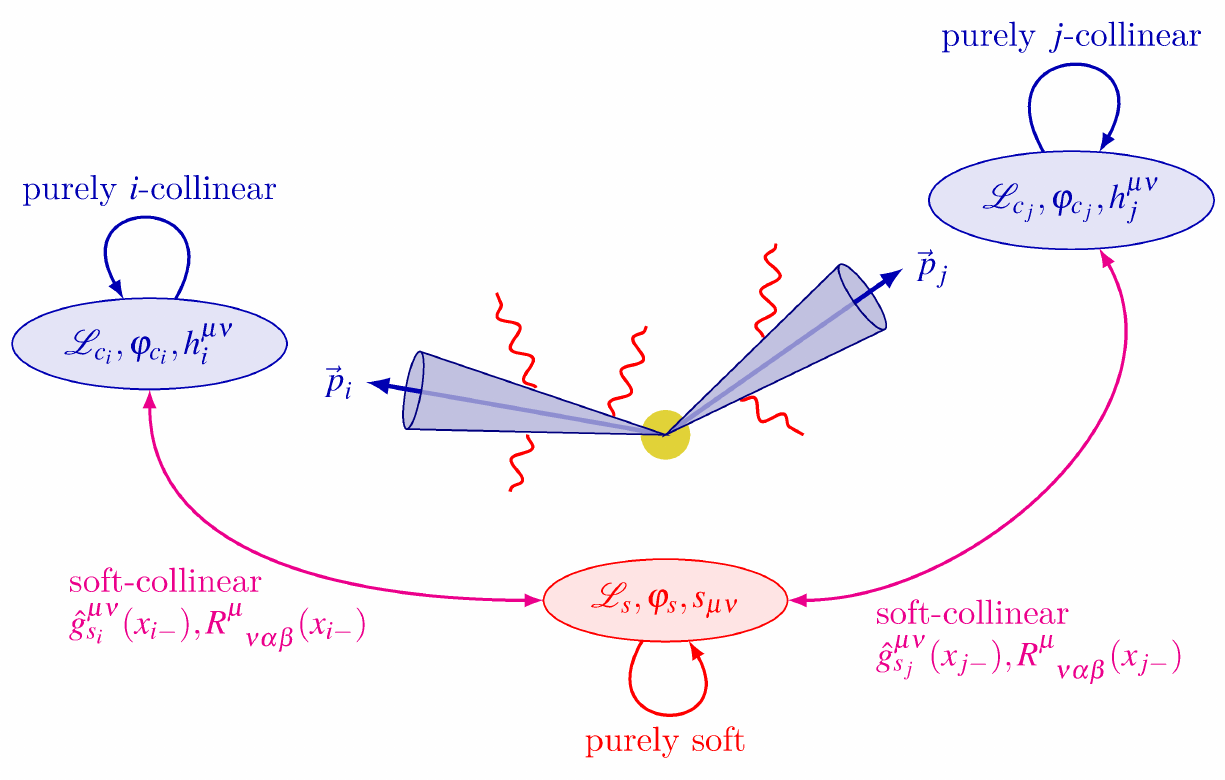}
\caption{An intuitive picture of the form of SCET. The soft 
modes (red) are described by the purely-soft Lagrangian 
$\mathcal{L}_s$, which takes the same form as the full theory.
Each collinear sector (blue) is described by its own 
Lagrangian $\mathcal{L}_{c_i}$, which contains purely-collinear 
and soft-collinear interactions (pink). 
    These soft-collinear terms are covariant with respect to a homogeneous background field $\hat{g}_{s_i}^{\mu\nu}(x_{i-})$, living only on the classical trajectory $x_{i-}^\mu$ of the collinear particles, and describe soft-collinear physics to all orders.}
    \label{fig:SCETpicture}
\end{figure}

From this perspective, the following structure of the theory arises, shown in \cref{fig:SCETpicture}. The collinear sector $i$ is constructed to be covariant with respect to a soft background metric denoted by $\hat{g}_{s_i\mu\nu}(x_{i-})$, which is constructed from the dynamical soft field $\eta_{\mu\nu}+\kappa s_{\mu\nu}(x)$, restricted to the classical light-like trajectories $x_{i-}^\mu$ of the energetic particles. Since due to the multipole expansion, collinear fields interact with soft fields at $x_{i-}^\mu$ only, these are effectively 
$i=1,\ldots, N$ separate soft gauge symmetries $U_s(x_{i-})$. Note that the interactions with the soft field are blind to the non-locality of the collinear sector in the $n_{i+}^\mu$ direction since $(x^\mu+tn_{i+}^\mu)_- = x_-^\mu$. On the other hand, the soft fields have self-interactions, which pervade all of space-time, and are controlled by the soft Lagrangian $\mathcal{L}_s$.

\subsection{Gravity vs QCD: a comparison}

At this point, it is instructive to compare the gravitational situation to the gauge-theory one, where the SCET construction is well understood.
The basic aspects of the construction, as discussed in Section~\ref{sec:Basics}, are the same for both theories, replacing the background field $\hat{g}_{s\mu\nu}(x_-) \to \nm A_s(x_-)$. 

The first main difference lies in the nature of the ``full theory'' itself.
In QCD, the starting point for the EFT construction is Yang-Mills theory, a renormalisable field theory.
Contrast this with gravity: here, the natural starting point, Einstein-Hilbert theory, is not renormalisable.
Instead, one takes the effective action \eqref{eq::GR::EinsteinHilbertEFT}, up to a desired order in the loop-expansion (or curvature, respectively).
In addition, one then performs the weak-field expansion, since the relevant degree of freedom is the fluctuation $h_{\mu\nu}$, the graviton field. 
Therefore, the ``full theory'' underlying SCET gravity is the weak-field expansion of an effective extension of the Einstein-Hilbert action, and thus already defined as an infinite series in $\kappa$ that must be truncated at some order.
This truncated theory is then expanded in $\lambda$. The $\lambda$-expansion shares many features with the $\kappa$-expansion but is not identical to it.

Another major difference is the nature of the underlying gauge symmetry.
In QCD, the gauge symmetry is purely internal, and it has a colour charge with a corresponding generator $t^a\sim \lambda^0$ that is unrelated to the kinematics that define the SCET expansion.
In gravity, however, the charge is related to momenta $P^\mu$ and therefore the gauge symmetry is connected to kinematics.
In particular the collinear momentum components have different $\lambda$-scaling \eqref{eq:basics:collmomscaling}.
This implies that the gauge transformation parameters in gravity also have non-trivial power-counting, and are not homogeneous in $\lambda$.
For example, a collinear matter field transforms under an infinitesimal collinear transformation as
\begin{equation}
    \varphi_c \to \varphi_c - \kappa\varepsilon_c^\alpha\partial_\alpha\varphi_c\,,
\end{equation}
where the second term is suppressed by $\mathcal{O}(\lambda)$ relative to the first due to the scaling of $\partial_\mu$ and the gauge parameter (see \eqref{eq:Col:EpsilonScaling} below).
Therefore, gauge transformations mix different powers of $\lambda$, and no homogeneous scaling can be achieved when manifest gauge-invariance is imposed.
An object can either be homogeneous in $\lambda$, or gauge-invariant, but never both at the same time.

These formal differences aside, one can now take a look at the physical content of both theories.
In QCD, the collinear gluon field $A_c$ contains two physical and two unphysical components. The large component $\np A_c\sim\lambda^0$ as well as the small $\nm A_c\sim\lambda^2$ are unphysical, while the transverse components $A_{c\perp}\sim\lambda$ are physical.
The large component $\np A_c$ is problematic: if this field was an allowed building block, one could add arbitrarily many such fields to any operator while keeping its $\lambda$-counting fixed.
Thus, there would not exist a finite operator basis, and the power-counting would be meaningless, since one would have to perform an infinite number of matching computations already at leading power.
This is alleviated by noting that in light-cone gauge, $\np A_c=0$, thus it is a gauge-artifact.
Therefore, one introduces a Wilson line $W_c$ (definition in \eqref{eq::collWilsonLine} below), which controls these $\np A_c$ to all orders.
These Wilson lines are homogeneous in $\lambda$, but an infinite series in the gauge coupling $g_s$.
In gravity, a similar situation arises, but worse with respect to the power-counting:
the collinear graviton $h_{\mu\nu}$ contains modes $h_{++}\sim\lambda^{-1}$, $h_{+\perp}\sim \lambda^0$, where $h_{++}$ is even power-counting enhanced. Clearly these must be controlled.
Therefore, one employs a similar concept as in QCD, by introducing the analogue of the collinear Wilson line.
In this way, the unphysical $h_{+\mu}$ components can be controlled to all orders.
Just as in QCD, this ``Wilson line'' is an infinite series in the coupling $\kappa$.
Due to the aforementioned inhomogeneity of the gauge symmetry, however, this implies that these gravitational ``Wilson lines'' are also an infinite series in $\lambda$, and no longer homogeneous.
These ``Wilson lines'' can be used to implement a covariant version of light-cone gauge, in the sense that one defines gauge-invariant composite objects that satisfy the light-cone gauge properties.

The suppressed unphysical degrees of freedom, $\nm A_c$ in QCD and $h_{-\perp}, h_{--}$ in gravity, can be eliminated using the equations of motion.
This leaves only the two transverse polarisations $A_{c\perp}\sim\lambda$ and $h_{\perp\perp}\sim\lambda$ as physical degrees of freedom in the effective theory. Since they count with a positive power of $\lambda$, a sensible operator basis exists.

The soft sector is slightly different from the collinear one.
Here, one first needs to identify the appropriate homogeneous background field, and then one organises the sub-leading terms in gauge-covariant objects.
The background field comes with a covariant derivative, which can be eliminated in the sources and the sub-leading Lagrangian terms using the equations of motion.
For the operator basis, only the sub-leading gauge-covariant objects are relevant.
In gauge theory, this gauge-covariant object is the field-strength tensor $F_{s\mu\nu} \sim\lambda^4$, which is the first derivative of the gluon field.
In gravity, the first sensible gauge-covariant object is the Riemann tensor, $\tensor{R}{^\mu_{\nu\alpha\beta}}\sim\lambda^6$, which is the second derivative of the metric fluctuation.
These gauge-covariant objects can appear in the sources and mediate process-dependent soft emissions.
Therefore, already at this stage, one can anticipate that in a soft-emission process, there are universal terms and the non-universality in gravity is more strongly suppressed compared to gauge theory.
The previously discussed features are summarised in \cref{table:QCDvsGrav}.
While there are essential differences between both theories, the construction of effective theories proceeds similarly. This will guide the following sections.

\begin{table}
\begin{center}
\begin{tabular}{ |c|c|c| } 
 \hline
  & QCD & Gravity \\ 
   \hline 
    \textcolor{navy}{gauge symmetry} & $SU(3)$ & $\text{Diff}(M)$\\
  \hline
 \textcolor{navy}{gauge charge} & $t^a \sim \lambda^0$ & $P^\mu \sim (\lambda^0,\lambda,\lambda^2)$\\
 \hline
 \textcolor{navy}{dimensionful coupling} & no & yes\\
 \hline
    \textcolor{navy}{fundamental} & \multirow{2}{*}{$A_\mu \sim p_\mu$} & \multirow{2}{*}{$h_{\mu\nu} \sim \frac{p_\mu p_\nu}{\lambda}$}\\
    \textcolor{navy}{degree of freedom}  & & \\
 \hline
     \textcolor{navy}{unsuppressed} & \multirow{2}{*}{$\np A_c \sim 1$} & \multirow{2}{*}{$h_{++}\sim \lambda^{-1}, h_{+\perp} \sim 1$}\\
    \textcolor{navy}{components}  & & \\
  \hline
     \textcolor{navy}{physical} & \multirow{2}{*}{$A_{c\perp} \sim \lambda$} & \multirow{2}{*}{$h_{\perp\perp}\sim\lambda$}\\
    \textcolor{navy}{degrees of freedom}  & & \\
   \hline
     \textcolor{navy}{redundant} & \multirow{2}{*}{$\nm A_{c} \sim \lambda^2$} & \multirow{2}{*}{$h_{\perp-}\sim\lambda^2, h_{--}\sim\lambda^3$}\\
    \textcolor{navy}{degrees of freedom}  & & \\
 \hline
 \textcolor{navy}{field-strength /}& \multirow{2}{*}{$F_{\mu\nu}\sim\partial A$} & \multirow{2}{*}{$\tensor{R}{^{\mu}_{\nu\alpha\beta}}\sim \partial^2 h$}\\
 \textcolor{navy}{curvature} & & \\
 \hline
\end{tabular}
\end{center}
\caption{A comparison of the main features of gauge-theory (QCD) and gravity from the SCET perspective.}
\label{table:QCDvsGrav}
\end{table}

\section{Collinear gravity}
\label{sec:collinear}

It is convenient to first consider only collinear modes in order to familiarise oneself with the construction. The purely-collinear EFT for gravity is obtained after splitting a generic field $\phi_J = \phi_{J,h} + \sum_i \phi_{J,c_i}$. After integrating out the hard modes, there are no left-over interactions of the collinear fields $\phi_{J,c_i}$ of different sectors, as the sum of the collinear momenta belonging to different directions has hard scaling, and corresponds to off-shell degrees of freedom which have already been eliminated. The Lagrangian is simply the sum  
\begin{equation}
    \mathcal{L}_{\rm SCET} = \sum_i \mathcal{L}_{c_i}\,.
\end{equation} 
For this reason it is possible to focus only on a single collinear direction. To simplify the notation, in the following the subscripts $c, i$ are omitted. 
In addition to the Lagrangian, the theory contains so-called \emph{currents} (aka ``sources'', or ``$N$-jet operators''), which contain products of fields in more than one collinear direction. Their matching coefficients absorb the hard modes that have been integrated out. These are discussed in more detail in \cref{sec:sources}. Here, we focus on the effective Lagrangian.

At this stage, the gauge symmetry consists of $N$-copies of collinear gauge symmetry, such that each collinear sector transforms under its own gauge symmetry and is invariant with respect to all the remaining collinear symmetries.  Collinear gauge transformations must not distort the scaling of the collinear gauge fields. In contrast with SCET QCD, in SCET gravity, the gauge transformation parameter $\varepsilon^\mu$ acquires  $\lambda$-scaling. Enforcing homogeneity of the infinitesimal gauge transformation
		\begin{equation}
			h_{\mu\nu} \to h_{\mu\nu} - \partial_\mu\varepsilon_\nu - \partial_\nu\varepsilon_\mu \sim \mathcal{O}(h_{\mu\nu}) \;,
		\end{equation}
$\varepsilon^\mu$ scales as
		\begin{equation}\label{eq:Col:EpsilonScaling}
			n_+\varepsilon \sim \frac{1}{\lambda}\,,\quad n_-\varepsilon\sim\lambda\,,\quad \varepsilon^{\mu_\perp}\sim 1 \,.
		\end{equation}
Note that $\varepsilon^\mu$ scales as a small translation $\lambda x^\mu$.
Beyond the linear order the transformation of $h_{\mu\nu}$ takes the form
\begin{eqnarray}
h^\prime_{\mu\nu} &=& 
h_{\mu\nu} - \partial_\mu\varepsilon_\nu - \partial_\nu \varepsilon_\mu 
-\kappa\,\Big[\partial_\mu\varepsilon^\alpha h_{\alpha\nu}
-\partial_\nu\varepsilon^\alpha h_{\alpha\mu}
-\varepsilon^\alpha\partial_\alpha h_{\mu\nu}\\
		        &&
		        + \,\partial_\mu\varepsilon^\alpha \partial_\alpha\varepsilon_\nu 
		        +  \partial_\nu\varepsilon^\alpha \partial_\alpha\varepsilon_\mu 
		        +  \partial_\mu\varepsilon_\alpha \partial_\nu\varepsilon^\alpha
		        + \varepsilon^\alpha\partial_\alpha(\partial_\mu\varepsilon_\nu + \partial_\nu\varepsilon_\mu)\Big] + \mathcal{O}(\lambda^3)\,.\nn
\end{eqnarray}

Collinear gauge-invariance 	is not only required from a formal point of view, but also ensures that the EFT power-counting is meaningful, i.e. one cannot generate an infinite number of operators with the help of the unsuppressed collinear fields. For this reason, it is beneficial to introduce the concept of \emph{gauge-invariant building blocks}. An arbitrary current contains fields in multiple collinear directions. Since each sector has its own gauge symmetry, the invariance of the complete operator under each collinear symmetry implies that each sector is separately gauge-invariant. This is automatically achieved if the operator is built from the gauge-invariant building blocks defined as the collinear field dressed with ``Wilson lines'', chosen such that the building block is a gauge-singlet that is always suppressed by at least one power of $\lambda$.
In gauge theory, this idea leads to the concept of the collinear Wilson line~$W_c$
\begin{equation}\label{eq::collWilsonLine}
    W_c(x) = P\exp\lp ig_s\int_{-\infty}^0 ds\: \np A_c(x+sn_+)\rp\,,
\end{equation}
which fulfills the following identity
\begin{align}
    W_c^\dagger n_+D_c W_c = n_+ \partial \;,
\end{align}
such that the collinear gauge-invariant gluon building block is $\mathcal{A}_{\mu} = \frac{1}{g_s} W^\dagger_c \brac{i D_{c\mu} W_c} $.
This leads to the notion of ``covariant light-cone gauge'', since the composite operator $\mathcal{A}_\mu$ built from $A_{c\mu}$ satisfies $\np \mathcal{A}=0$ in \emph{any} gauge.

In analogy to QCD, where the collinear-gauge invariant gluon field satisfies $\mathcal{A}_+=0$, in SCET gravity one constructs the manifestly gauge-invariant graviton building block $\hinv_{\mu\nu}$ \cite{Beneke:2021aip,Donnelly:2015hta,Okui:2017all}, which depends 
on the collinear field $h_{\mu\nu}$, and satisfies $\hinv_{\mu+}=0$, that is, it coincides with the elementary graviton field in light-cone gauge.  
Just like in QCD, where the collinear Wilson line connects the collinear gluon field to the gauge invariant object, in gravity, the collinear graviton field is related to the gauge invariant graviton building block via
\begin{equation}
\label{eq::ginv}
\eta_{\mu\nu} + \kappa\mathfrak{h}_{\mu\nu}(x) = \tensor{W}{^\alpha_\mu}
\tensor{W}{^\beta_\nu}[W_c^{-1}g_{\alpha\beta}(x)]\,,
\end{equation}
where $g_{\alpha\beta}(x) = \eta_{\alpha\beta}+\kappa h_{\alpha\beta}(x)$ and the gravitational collinear ``Wilson line'' is
\begin{equation}\label{eq::CG::WilsonLineDefinition}
 W_c^{-1} = T_{\theta_c}[h] = 1 + \kappa\theta_c^\alpha\partial_\alpha + \frac{\kappa^2}{2} \theta_c^\alpha \theta_c^\beta \partial_\alpha\partial_\beta + \mathcal{O}(\theta_c^3)\,,
\end{equation}
with parameter $\theta_c[h]\equiv\theta_c[h_{\mu\nu}(x)]$ chosen such that, given 
$h_{\mu\nu}$,  the invariant field $\hinv_{\mu\nu}$ satisfies $\hinv_{\mu+}=0$, and $ \tensor{W}{^\alpha_\mu}$ is the Jacobian for the transformation \eqref{eq::CG::WilsonLineDefinition}, defined as in \eqref{eq::GR::JacobianDef}. The above translation corresponds to the transformation to the coordinate system, in which the metric fluctuation satisfies light-cone gauge. Note, however, that we do not actually fix the gauge. Rather, $W_c$ is employed to define the composite operator $\mathfrak{h}_{\mu\nu}(x)$ built from the elementary field $h_{\mu\nu}$ for which no special coordinate system or gauge-fixing is assumed. The ``Wilson line'' transforms as \cite{Donnelly:2015hta}
\begin{equation}
\label{eq::CG::Wilsontransformation}
W_c^{-1} \to W_c^{-1} U^{-1}(x)\,,
\end{equation}
and this transformation precisely cancels the collinear gauge transformation of $h_{\mu\nu}$ in \eqref{eq::ginv}. It is also used to define the collinear gauge-invariant matter field~$\chi_c$,
\begin{equation}
\chi_c = \lc W^{-1}_c\varphi_c\rc = 
\varphi_c+\mathcal{O}(\lambda\varphi_c)\,.
\end{equation}
Note that $W_c^{-1}=1+\mathcal{O}(\lambda)$ in stark contrast to 
\eqref{eq::collWilsonLine}, which is closely related to the absence of 
purely-collinear gravitational interactions at leading power in 
$\lambda$ as will be seen below.

The explicit construction of the collinear ``Wilson line'' requires determining the parameter $\theta_c^\mu[h]$. The closed-form version does not exist in gravity, but instead it can be defined perturbatively in the weak-field expansion
\begin{equation}\label{eq::CG::theta}
    \theta_c^\mu = \theta_c^{\mu(0)} + \theta_c^{\mu(1)} + \dots\,.
\end{equation}
Then, one needs to perform the expansion of~\eqref{eq::ginv},
\begin{equation}\label{eq::CG::hinvexp}
\hinv_{\mu\nu} = h_{\mu\nu} + 
\partial_\mu \theta_{c\nu} +
\partial_\nu \theta_{c\mu} + 
\mathcal{O}(\theta_c^2,\theta_c h)\,,
\end{equation} 
and insert~(\ref{eq::CG::theta}). Requiring $\hinv_{\mu+}=0$ 
determines 	
\begin{align}
\label{eq::CG::colltheta}
\theta^{(0)}_{c\mu} &= -\frac{1}{\delp}\biggl(h_{\mu+} - \frac 12 \frac{\partial_\mu}{\delp} h_{++}\biggr)\,.
\end{align}
The sub-leading (non-linear) terms can be obtained iteratively \cite{Beneke:2021aip}.
The inverse derivative is defined as
\begin{equation}
\frac{1}{i\np\partial+i\epsilon}f(x^\mu) = -i 
\int_{-\infty}^0 \!ds \,f(x^\mu+s \np^\mu)\,.
\label{eq:inverserder}
\end{equation}
Inserting back the leading term in the expansion of $\theta_c$~(\ref{eq::CG::colltheta}) into $\hinv_{\mu\nu}$~(\ref{eq::CG::hinvexp}) leads to the explicit formula for the gauge-invariant graviton building block expressed in terms of the collinear graviton
\begin{equation}
\label{eq::CG::collinvgraviton1}
\hinv_{\mu\nu} = h_{\mu\nu} - 
				\frac{\partial_\mu}{\delp}\left( h_{\nu+} - \frac 12 \frac{\partial_\nu}{\delp} h_{++}\right) -
				\frac{\partial_\nu}{\delp}\left( h_{\mu+} - \frac 12 \frac{\partial_\mu}{\delp}h_{++}\right) +
				\mathcal{O}(\lambda h_{\mu\nu})\,,
\end{equation}
which resembles the analogous expression for the gauge-invariant collinear gluon field,\footnote{Despite the fact that $W_c^{-1}=1+\mathcal{O}(\lambda)$, $\hinv_{\mu\nu}$ differs from $h_{\mu\nu}$ at $\mathcal{O}(\lambda^0 h_{\mu\nu})$ due to the Jacobian terms in \eqref{eq::ginv} multiplying the $\eta_{\alpha\beta}$ 
term in $g_{\alpha\beta}(x)$.}
\begin{equation}
\label{eq::gaugeInvAPerp}
\mathcal{A}_{c\mu_\perp} = \frac{1}{g_s} W^\dagger_c \brac{i D_{c\mu_\perp} W_c} = A_{c\mu_\perp} - \frac{\partial_{\mu_\perp}}{\delp}A_{c+} + \mathcal{O}(g_s A_{c\mu_\perp})\,. 
\end{equation}	
			
Having understood the basic building blocks of the theory, it is straightforward to derive the collinear Lagrangian. The first step involves performing the weak-field expansion of (\ref{eq::GR::GeometricScalarField}). Then one re-expresses the collinear fields in terms of the gauge-invariant building blocks by inserting the gravitational collinear ``Wilson line''~(\ref{eq::CG::WilsonLineDefinition}). 
Since the ``Wilson line'' corresponds to a collinear gauge transformation, one can simply put $h_{\mu\nu}\to\mathfrak{h}_{\mu\nu}$ and $\varphi_c\to\chi_c$.
Then, one can set $\mathfrak{h}_{\mu+}=0$.
Finally, grouping the terms by their order in the $\lambda$-expansion yields
\begin{align}
\label{eq::ManifestlyInvariantCollLag}
\mathcal{L}^{(0)} &= 
\frac{1}{2}\partial_\mu \chi_c \partial^\mu \chi_c\,,\\
\mathcal{L}^{(1)} &= 
- \frac{\kappa}{2} (\mathfrak{h}_{\mu\nu}\partial^\mu \chi_c\partial^\nu \chi_c - \frac{1}{2}\mathfrak{h} \,\partial_\alpha \chi_c\partial^\alpha \chi_c),\\
\mathcal{L}^{(2)} &=  
\frac{\kappa^2}{2}\biggl(\mathfrak{h}^{\mu\alpha}\mathfrak{h}^\nu_\alpha - 
			\frac{1}{2}\mathfrak{h}\mathfrak{h}^{\mu\nu} + 
			\frac{1}{8}(\mathfrak{h}^2 - 2 \mathfrak{h}^{\alpha\beta}\mathfrak{h}_{\alpha\beta})\eta^{\mu\nu}\biggr)\partial_\mu \chi_c\partial_\nu \chi_c\,.
\end{align} 
The leading-power Lagrangian $\mathcal{L}^{(0)}$ describes a free theory of gauge-invariant building blocks. It follows from this simple observation that there cannot be collinear singularities in gravity. Interactions with collinear gravitons start at sub-leading power and all terms are manifestly gauge-invariant. 

The Lagrangian above is expressed entirely in terms of the gauge-invariant building blocks $\chi_c$, $\mathfrak{h}_{\mu\nu}$. This form makes it clear that there is a well-defined expansion in $\lambda$, since 
the metric components unsuppressed in $\lambda$ do not 
appear due to $\mathfrak{h}_{\mu+}=0$. Since the Lagrangian is collinear gauge-invariant, it takes the same form when expressed in terms of the original fields $\varphi_c$, $h_{\mu\nu}$. However, since $h_{++}$ 
and $h_{+\perp}$ do not vanish, this form hides the gauge cancellations 
that occur between the unsuppressed metric components and lacks 
manifest power counting.

The sub-leading components $\mathfrak{h}_{\mu-}$ can be eliminated 
from the Lagrangian by using equations of motion. To their respective 
leading order in $\lambda$, one finds \cite{Beneke:2021aip}
\begin{align}
\label{eq::CG::EOMmetric}
\mathfrak{h}_{\mu_\perp -} &=-2\frac{\partial^{ \alpha_\perp}}{\partial_+}\mathfrak{h}_{ \mu_\perp\alpha_\perp} +\mathcal{O}(\lambda^3)\,,\\
\mathfrak{h}_{--} &= 4\frac{\partial^{ \alpha_\perp}\partial^{ \beta_\perp}}{\partial_+^2}\mathfrak{h}_{\alpha_\perp \beta_\perp}  +\mathcal{O}(\lambda^4)\,,
\end{align}
where the first equation arises from the field equation for
$\mathfrak{h}_{\mu_\perp -}$, and the second from tracing the one 
for $\mathfrak{h}_{\mu_\perp\nu_\perp}$. The Lagrangian is linear in 
$\mathfrak{h}_{--}$, and the field equation for 
$\mathfrak{h}_{--}$ results in the trace constraint 
\begin{align}
\label{eq::CG::EOMh}
    \mathfrak{h} &=\frac{\kappa}{2}\biggl(\mathfrak{h}_{\alpha_\perp \beta_\perp}\mathfrak{h}^{\alpha_\perp  \beta_\perp}-\frac{1}{\partial^2_+}\big[\partial_+\mathfrak{h}_{\alpha_\perp  \beta_\perp}\partial_+\mathfrak{h}^{\alpha_\perp  \beta_\perp}-\partial_+\chi_c \partial_+\chi_c\big]\biggr)+\mathcal{O}(\lambda^3)\,,
\end{align}
which shows that $\mathfrak{h}$ counts as $\mathcal{O}(\lambda^2)$. 
The trace terms in $\mathcal{L}^{(2)}$ can therefore be dropped at this 
order, while the second interaction term in  $\mathcal{L}^{(1)}$ is 
$\mathcal{O}(\lambda^2)$. These simplifications can be made manifest 
by expressing the Lagrangian in terms of 
$\mathfrak{h}_{\mu_\perp\nu_\perp}$ only.

The collinear Einstein-Hilbert Lagrangian for the graviton 
self-interactions, as well as for the scalar field with $\varphi^4$ 
self-interactions are discussed in detail in \cite{Beneke:2021aip}. 
Interestingly, the trilinear graviton self-interaction when expressed 
in terms of $\mathfrak{h}_{\mu_\perp\nu_\perp}$ (equivalently, light-cone 
gauge) has a manifest double-copy structure at the Lagrangian level 
\cite{Beneke:2021ilf}.

\section{Soft-collinear gravity}
\label{sec:softcolgrav}

The full soft-collinear EFT 
can be constructed separately for each collinear sector as seen 
from \eqref{eq:Basics:LagrangianSplit}, therefore the index $i$ will 
be dropped. The first step is to introduce the soft and 
collinear modes. One way to implement the mode split consistently is to treat the collinear modes as small-distance fluctuations on top of the slowly-varying soft background. For the metric field, the decomposition takes the form
\begin{equation}\label{eq::SG::MetricTensorSplit}
    g_{\mu\nu} = \eta_{\mu\nu} + \kappa h_{\mu\nu} + \kappa s_{\mu\nu} \equiv g_{s\mu\nu} + \kappa h_{\mu\nu}\,,
\end{equation}
where by definition all components of the soft field $s_{\mu \nu}$ have isotropic scaling 
\begin{equation}
    s_{\mu\nu}\sim\lambda^2\,.
\end{equation}
It is convenient to work with the soft background in terms of a soft metric tensor $g_{s\mu\nu}$ to leverage the geometric intuition.
Therefore, this first step is fully equivalent to a standard weak-field expansion in $h_{\mu\nu}$ about a non-trivial dynamic background~$g_{s\mu\nu}$.

The gauge symmetry is modified by this decomposition, and changes into the gauge symmetry of the background (the soft gauge symmetry), and the gauge symmetry of the collinear fluctuations, which are covariant with respect to this background.
The soft gauge symmetry is the standard diffeomorphism applied to the slowly varying background field.
The transformations of the soft and collinear fluctuations read
\begin{equation}\label{eq::SG::SoftSymmetry}
	\begin{aligned}
	h_{\mu\nu} &\to \lc U_s \left( \tensor{U}{_{s\mu}^\alpha}\tensor{U}{_{s\nu}^\beta}h_{\alpha\beta}\right)\rc\,,\\
	 \kappa s_{\mu\nu} &\to \lc U_s(\tensor{U}{_{s\mu}^\alpha} \tensor{U}{_{s\nu}^\beta}(\eta_{\alpha\beta} + \kappa s_{\alpha\beta})\rc - \eta_{\mu\nu}\,.
\end{aligned}
\end{equation}
It is understood that the $x$-dependence of $U_s(x)$ is the one of a 
soft field with variations over distances $\mathcal{O}(1/\lambda^2)$.  
The soft fluctuation $s_{\mu\nu}$ has the standard transformation of a 
metric perturbation, whereas the transformation of the collinear 
graviton $h_{\mu\nu}$ is the one of a standard rank-2 tensor, i.e. 
covariant like an ordinary matter field. Note that the soft metric 
tensor $g_{s\mu\nu}$ also transforms like a rank-2 tensor~(\ref{eq:basics:Metrictransformation})
	\begin{equation}
	g_{s\mu\nu} \to \lc U_s \left( \tensor{U}{_{s\mu}^\alpha}\tensor{U}{_{s\nu}^\beta} g_{s\alpha\beta}\right)\rc\,.
	\end{equation}
	The interpretation of these transformations is clear:
	the background geometry is described by $g_{s\mu\nu}$, which transforms like an ordinary metric tensor, and $s_{\mu\nu}$ inherits a non-linear transformation from it.
	From the soft point of view, the collinear gravitons transform just like \emph{ordinary matter fields}, i.e. like a standard rank-2 tensor, and \emph{not} non-linearly like a metric perturbation.
	Therefore, from the soft perspective, any collinear field, regardless if gauge or matter, transforms like a matter field, i.e. according to its tensor representation.
	
The collinear gauge transformations are the transformations of the fields defined on the background metric and were already introduced in \eqref{eq:basics:ColGaugeTrans}. They can be cast in the form
\begin{equation}
\label{eq::SG::CollinearSymmetry}
\begin{aligned}
\kappa h_{\mu\nu} &\to \lc U_c\lp \tensor{U}{_{c\mu}^\alpha}\tensor{U}{_{c\nu}^\beta} \kappa h_{\alpha\beta}\rp\rc 
+ \lc U_c \lp\tensor{U}{_{c\mu}^\alpha}\tensor{U}{_{c\nu}^\beta}g_{s\alpha\beta}\rp\rc	- g_{s\mu\nu}\,,\\
	g_{s\mu\nu} &\to g_{s\mu\nu}\,,
\end{aligned}
\end{equation}
which renders the transformation of $h_{\mu\nu}$ reminiscent of an 
ordinary gauge-field transformation with a covariant and an inhomogeneous term. The fluctuation $h_{\mu\nu}$ has a non-linear transformation, similar to the standard weak-field expansion around flat space~(\ref{eq:Basics:LinearGaugeTrafo}), but with respect to the soft background $g_{s\mu\nu}$, which appears in the inhomogeneous term instead of $\eta_{\mu\nu}$. Observe here that the soft background is necessarily invariant under the collinear transformations, since $U_c(x)$ is a collinear field.

Both gauge symmetries, soft and collinear, have the property that the full metric tensor $g_{\mu\nu}=g_{s\mu\nu}+\kappa h_{\mu\nu}$ transforms like the standard metric tensor.
The transformations of matter fields follow the same logic. They can be found in \cite{Beneke:2021aip}.

However, as already discussed in 
Section~\ref{sec::multipole}, the transformations do not satisfy 
completely homogeneous $\lambda$-scaling, since they contain 
multiplications of collinear and soft fields at point $x$. The following 
construction identifies the background metric and the collinear field 
redefinition that renders the gauge symmetries compatible with 
a manifestly gauge-covariant form of the light-front multipole 
expansion.

\subsection{Riemann normal coordinates}	

To gain some intuition for this procedure in the gravitational context, a simpler example is considered first: the static multipole expansion around the space-time point $x=0$.
The metric tensor in this theory should be multipole-expanded as
\begin{equation}
    g_{\mu\nu}(x) = g_{\mu\nu}(0) + x^\alpha\lc\partial_\alpha g_{\mu\nu}\rc(0) + \frac 12 x^\alpha x^\beta \lc \partial_\alpha \partial_\beta g_{\mu\nu}\rc(0) + \mathcal{O}(x^3)\,.
\label{eq:staticmultipoleexp}
\end{equation}
In gauge theories, the corresponding expansion of $A_\mu(x)$ is conveniently performed using Fock-Schwinger gauge $x^\mu A_\mu(x)=0$. This gauge condition sets $A_\mu(0)=0$, and the higher-order terms in the Taylor-expansion are expressed in terms of the field-strength tensor and its derivatives. Since the resulting expansion is manifestly gauge-covariant, the gauge-fixing can easily be removed at the end. This technique has found numerous applications 
\cite{Novikov:1983gd} to non-perturbative techniques in QCD, in which 
case short-distance fluctuations are treated in the background of the 
soft QCD vacuum. 

In gravity, the analogous gauge corresponds to the familiar Riemann normal coordinates. The ``gauge condition'' reads
    \begin{equation}\label{eq::SG::RNCgaugecondition}
    x^\mu x^\nu \chris{\alpha}{\mu\nu}(x) = 0\,.
    \end{equation}
Compared to Fock-Schwinger gauge $x^\mu A_\mu(x)=0$, one immediately notices that in gravity, the gauge condition starts at second order in $x$, and restricts the derivative of the metric tensor at $x=0$, but not $g_{\mu\nu}(0)$, since $\Gamma \sim \partial g$.
Indeed, the effect of Riemann normal coordinates is to set the first derivative $\lc\partial_\alpha g_{\mu\nu}\rc\!(0) = 0$, and to express the higher derivatives via the Riemann tensor and its derivatives at $x=0$.
The metric tensor near $x=0$ is then expressed as 
\begin{equation}\label{eq::SG::MetricRNC}
        { \tilde g}_{\mu\nu}(x) = g_{\mu\nu}(0) - \frac 13 x^\alpha x^\beta R_{\alpha\mu\beta\nu}(0) + \mathcal{O}(x^3)\,.
\end{equation}

The gauge condition \eqref{eq::SG::RNCgaugecondition} does not fix the
diffeomorphism symmetry completely, but only up to \emph{global} linear transformations $\tensor{A}{_\mu^\alpha}\in \mathrm{GL}(1,3)$,
    \begin{equation}\label{eq::SG::RNCResidualGauge}
        x^\mu \to \tensor{A}{_\alpha^\mu} x^\alpha \,. 
    \end{equation}
Using this residual symmetry, one can transform the leading term 
$g_{\mu\nu}(0)$ to the Minkowski metric $\eta_{\mu\nu}$, and the 
metric tensor in these coordinates reads
\begin{equation}\label{eq::SG::MetricRNCDiagonalised}
        { \tilde g}_{\mu\nu}(x) = \eta_{\mu\nu} - \frac 13 x^\alpha x^\beta R_{\alpha\mu\beta\nu}(0) + \mathcal{O}(x^3)\,,
\end{equation} 
which is the standard form of the metric in Riemann normal coordinates. 
The remaining residual symmetries of the full diffeomorphisms are now 
the ones of the Minkowski metric $\eta_{\mu\nu}$, i.e. global Lorentz 
transformations.

The lesson of this section is the following:
using Riemann normal coordinates, one can express the multipole expansion of the metric tensor in terms of the metric at the origin and Riemann tensor terms.
Then, one can exploit the residual symmetry to diagonalise this metric and change it to the flat-space one at $x=0$.
The result is a covariant multipole expansion that features global Poincar\'e transformations as residual symmetries.
    
\subsection{Fixed-line normal coordinates}	

The above discussion motivates a generalisation of the Riemann normal 
coordinates adapted to the situation where the physical system is 
not localised around a space-time point, but around the light-like trajectory 
$x_-^\mu = \np x \frac{\nm^\mu}{2}$, which allows for the light-front 
multipole expansion of soft fluctuations.
The metric tensor is now expanded as 
	\begin{align}\label{eq:FLMultipole}
    g_{s\mu\nu}(x) &= g_{s\mu\nu}(x_-) + x_\perp^\alpha\lc\partial_\alpha g_{s\mu\nu}\rc(x_-) +
    \frac 12 \nm x \lc\np\partial g_{s\mu\nu}\rc(x_-)\nn\\
    &\quad
    + \frac 12 x_\perp^\alpha x_\perp^\beta \lc\partial_\alpha\partial_\beta g_{s\mu\nu}\rc(x_-)
    + \mathcal{O}(\lambda^3 g_{s\mu\nu})\,,
\end{align}
which differs from the previous one \eqref{eq:staticmultipoleexp} 
in two aspects: first, the expansion is only in the directions orthogonal to the light-like line $x_-^\mu$, and second, the coefficients in the Taylor-expansion are dynamical fields and depend on $x_-^\mu$. In gauge theory, 
the generalisation of Fock-Schwinger gauge is ``fixed-line'' gauge 
\cite{Beneke:2002ph} $(x-x_-)^\mu A_\mu(x_-)=0$. In gravity, this 
suggests the {\em fixed-line normal coordinate} condition \cite{Beneke:2021aip}
\begin{equation}\label{eq:FLNCCondition}
    (x-x_-)^\alpha (x-x_-)^\beta \tensor{\Gamma}{^\mu_{\alpha\beta}}(x) = 0\,,
\end{equation}
which replaces  \eqref{eq::SG::RNCgaugecondition}. Suppose that 
$\tensor{\Gamma}{^\mu_{\alpha\beta}}(x)$ does not satisfy this condition, then one can find new coordinates $\tilde{x}^\mu$, related to $x^\mu$ by
    \begin{align}
\label{eq::SG::FLNCDerivationInversex}
        \tilde{x}^\mu &= x^\mu + \frac 12 (x-x_-)^\alpha (x-x_-)^\beta \chris{\mu}{\alpha\beta}\\
        &\quad+ \frac 16 (x-x_-)^\alpha (x-x_-)^\beta (x-x_-)^\nu \lp \chris{\mu }{\alpha\tau}\chris{\tau}{\beta\nu} + \big[\partial_\nu\chris{\mu}{\alpha\beta}\big]\rp + \mathcal{O}((x-x_-)^4)\,,\nn
    \end{align}
such that $\tensor{\tilde \Gamma}{^\mu_{\alpha\beta}}\!(x)$ fulfills 
\eqref{eq:FLNCCondition}. In the above and for the remainder of 
Section~\ref{sec:softcolgrav}, the following convention is adopted: 
{\em whenever a soft field appears without explicit position argument, the field is evaluated at $x_-^\mu$.}

Once again, the fixed-line normal coordinate gauge condition 
\eqref{eq:FLNCCondition} does not fix the gauge completely.
For example, \eqref{eq:FLNCCondition} leaves the components $\tensor{\tilde{\Gamma}}{^\mu_{--}}\!(\tilde x)$ unconstrained and 
\eqref{eq::SG::FLNCDerivationInversex} does not affect 
$g_{s\mu\nu}(x_-)$, since it is second order in $(x-x_-)^\mu$. The remaining 
diffeomorphisms form a residual symmetry, which  is the analogue of the global symmetry transformations from before \eqref{eq::SG::RNCResidualGauge}, but since the parameters depend on $x_-^\mu$, it is a \emph{gauge} symmetry. As for the Riemann normal coordinates, the residual 
transformations can be used to  further simplify the residual metric 
tensor $g_{s\mu\nu}(x_-)$ by performing a linear coordinate transformation local in $x_-^\mu$ in the components orthogonal 
to $x_-^\mu$. After this step, one can identify the homogeneous 
soft background field. To this end, introduce the vierbein
\begin{equation}
\label{eq::SG::MetricTensorVierbeinFLNC}
    g_{s\mu\nu}(x_-) \equiv \tensor{e}{_\mu^\alpha}(x_-)\tensor{e}{_\nu^\beta}(x_-) \eta_{\alpha\beta}\,,
\end{equation}
and its inverse
\begin{equation}
    \tensor{E}{^\mu_\alpha}(x_-)\tensor{e}{^\alpha_\nu}(x_-) = \delta^\mu_\nu\,.
\label{eq:InverseVierbein}
\end{equation}
These objects can be thought of as matrices that locally diagonalise the metric tensor to express it in terms of the flat-space metric $\eta_{\mu\nu}$. Performing the linear transformation $\tilde{x}^\mu\to 
\check{x}^\mu(\tilde{x})$ defined by 
\begin{equation}
\label{eq::SG::DiagFLNCRelationtonondiag}
n_+\check{x} =n_+\tilde{x}\,,\quad \check{x}^\mu_\perp = \tensor{e}{_{\alpha}^{\mu_\perp}} \tilde{x}^\alpha\,,\quad \nm \check{x} = n_{-\rho}\tensor{e}{_\alpha^{\rho}} \tilde{x}^\alpha\,,
\end{equation}
which leaves $x_-^\mu$ invariant, yields a new coordinate 
system $\check{x}$, the fixed-line normal coordinates (FLNC).
The original coordinate can be expressed in terms of the 
FLNC $\check{x}^\mu$ as 
\begin{equation}
    x^\mu = \check{x}^\mu + \theta^\mu_{\rm FLNC}(\check{x})\,,
\end{equation}
with parameter
\begin{align}
\label{eq::SG::FLNCLLParameter}
    \theta^{\mu}_{\rm FLNC}(x) &= (\tensor{E}{^\mu_\rho} - \delta^\mu_\rho)(x-x_-)^\rho - \frac 12 (x-x_-)^\rho (x-x_-)^\sigma \tensor{E}{^\alpha_\rho}\tensor{E}{^\beta_\sigma}\chris{\mu}{\alpha\beta}\\
    &\hspace*{-1.4cm}+ \,\frac 16 (x-x_-)^\rho (x-x_-)^\sigma (x-x_-)^\lambda \tensor{E}{^\alpha_\rho}\tensor{E}{^\beta_\sigma}\tensor{E}{^\nu_\lambda}(2\chris{\mu}{\alpha\tau}\chris{\tau}{\beta\nu} - [\partial_\nu\chris{\mu}{\alpha\beta}]) + \mathcal{O}(x^4)\,.\nn
\end{align}
Note that every $(x-x_-)^\mu$ generates a contraction with the inverse vierbein as a result of this linear transformation. Returning now to the active point of view, the translation operator $T_{\theta_{\rm FLNC}}(x)$ 
corresponding to the coordinate change to  $\check{x}^\mu$ defines a
 new ``Wilson line'' $R_{\rm FLNC}(x)$ via
\begin{equation}\label{eq::SG::RFLNCDef}
    R^{-1}_{\rm FLNC}(x) = T_{\theta_{\rm FLNC}}(x)\,.
\end{equation}

As an intermediate result, one can compute the metric tensor in fixed-line normal coordinates, denoted by $\check{g}_{s\mu\nu}$, using $R_{\rm FLNC}$ as
\begin{equation}
\label{eq::SG::FLNCMetricDefinition}
\check{g}_{s\mu\nu}(x) \equiv \tensor{R}{^\alpha_\mu}(x)\tensor{R}{^\beta_\nu}(x) \lc R^{-1}_{\rm FLNC}(x) g_{s\alpha\beta}(x)\rc\,,  
\end{equation}
with 
\begin{equation}
\tensor{R}{^\alpha_\mu}(x) = \frac{\partial x^{\alpha}}{\partial \check{x}^\mu}(x)
\label{eq:RJacobian}
\end{equation}
the Jacobian of the transformation. 
As was the case for Riemann normal coordinates, the metric in 
FLNC can now be split into a background metric (simply $\eta_{\mu\nu}$ 
for the former), which forms the leading term, and the multipole series 
of manifestly covariant terms expressed via the Riemann tensor.
Therefore, it is convenient to split the metric tensor as
\begin{equation}\label{eq::SG::BGMetricSplitRiemann}
            \check{g}_{s\mu\nu}(x) \equiv \hat{g}_{s\mu\nu}(x) + \mathfrak{g}_{s\mu\nu}(x)\,,
        \end{equation}
where $\mathfrak{g}_{s\mu\nu}$ are the manifestly gauge-covariant terms, i.e. the Riemann tensor terms, and the background field is $\hat{g}_{s\mu\nu}(x)$.
It is given by
\begin{eqnarray}
\label{eq::SG::ResidualMetric1}
            &&\hat{g}_{s+-}(x) = e_{-+} - (x-x_-)^\alpha \lc\Omega_{-}\rc_{\alpha +}\,,\\[0.15cm]
&&\hat{g}_{s\mu_\perp-}(x) = e_{-\mu_\perp} - (x-x_-)^\alpha  \lc\Omega_{-}\rc_{\alpha\mu_\perp}\,,
\label{eq::SG::ResidualMetric1b}\\
&&\hat{g}_{s--}(x) = \lp\tensor{e}{_-^\alpha} - (x-x_-)^\rho \tensor{\lc\Omega_-\rc}{_{\rho}^\alpha}\rp\label{eq::SG::ResidualMetric2}
        	\lp \tensor{e}{_-^\beta} - (x-x_-)^\sigma \tensor{\lc\Omega_-\rc}{_{\sigma}^\beta}\rp \eta_{\alpha\beta}\,,\quad\\
&&\hat{g}_{s\mu_\perp\nu_\perp}(x) = 
\eta_{\mu_\perp\nu_\perp}\vphantom{\lp\lc\Omega_\mu\rc^{\alpha\beta}\rp}\,,\\
&&\hat{g}_{s+\perp}(x) = \hat{g}_{s++}(x) = 0\,.
        	\label{eq::SG::ResidualMetric3}
\end{eqnarray}
Here, the fields on the right-hand side implicitly depend on $x_-$, i.e. $e_{-+}\equiv e_{-+}(x_-)$ and $\lc\Omega_-\rc_{\alpha\beta}$ is the spin-connection, defined as\footnote{The definitions 
\eqref{eq::SG::MetricTensorVierbeinFLNC}, 
\eqref{eq:InverseVierbein} can be extended from the light-cone 
$x_-^\mu$ to $x^\mu$, in which case 
\begin{equation}
\label{eq:FullSpinConnectionDefinition}
\lc\Omega_\mu\rc^{\alpha\beta}(x) =
\tensor{e}{_\nu^\alpha}\brac{\partial_\mu\tensor{E}{^{\nu \beta}}}(x)
+ \tensor{e}{_\nu^\alpha}\chris{\nu}{\sigma\mu}\tensor{E}{^{\sigma \beta}}(x)\,
\end{equation}
coincides with the usual spin-connection. The background-field 
construction only needs $\lc\Omega_-\rc^{\alpha\beta}(x_-)$, which is 
well-defined on the light-cone, since only the derivative 
$\nm\partial$ along the light-cone appears in 
\eqref{eq::SG::SpinConnectionDefinition}.}
\begin{equation}
\label{eq::SG::SpinConnectionDefinition}
\lc\Omega_-\rc^{\alpha\beta} =
\tensor{e}{_\nu^\alpha}\brac{\partial_-\tensor{E}{^{\nu \beta}}}
            + 
            \tensor{e}{_\nu^\alpha}\chris{\nu}{\sigma-}\tensor{E}{^{\sigma \beta}}\,.
\end{equation}
Note that the components orthogonal to $x_-^\mu$ coincide with the 
flat-space metric, as in these directions the light-front multipole 
expansion is the same as the static one.

The remainder term $\mathfrak{g}_{s\mu\nu}$ in \eqref{eq::SG::BGMetricSplitRiemann} is manifestly gauge-covariant and depends only on the Riemann tensor. Up to $\mathcal{O}(\lambda^4)$, 
it is given by
        \begin{align}
        \mathfrak{g}_{s\mu\nu}(x) &= 
        -\frac{n_{+\mu}n_{+\nu}}{4} x_\perp^\alpha x_\perp^\beta R_{\alpha-\beta-}
        - \frac{n_{+\mu}}{2}\frac{2}{3}x_\perp^\alpha x_\perp^\beta R_{\alpha\nu_\perp\beta-} - \frac{n_{+\nu}}{2}\frac{2}{3}x_\perp^\alpha x_\perp^\beta R_{\alpha\mu_\perp\beta-}\nn \\
        &\quad
        - \lp \frac{n_{+\mu}n_{-\nu}}{4}
        + \frac{n_{+\nu}n_{-\mu}}{4}\rp \frac{2}{3}x_\perp^\alpha x_\perp^\beta R_{\alpha+\beta-}
        - \frac 13 x_\perp^\alpha x_\perp^\beta R_{\alpha\mu_\perp\beta\nu_\perp}
\label{eq::SG::Riemanntermsexpansion}\\
        &\quad- \frac{n_{-\mu}}{2} \frac{1}{3}x_\perp^\alpha x_\perp^\beta R_{\alpha\nu_\perp\beta+}
        - \frac{n_{-\nu}}{2}\frac{1}{3}x_\perp^\alpha x_\perp^\beta R_{\alpha\mu_\perp\beta+}
        - \frac{n_{-\mu}n_{-\nu}}{4} \frac{1}{3}x_\perp^\alpha x_\perp^\beta R_{\alpha+\beta+}\,.\nn
    \end{align}
The explicit expressions of $\hat{g}_{s\mu\nu}$ and $\mathfrak{g}_{s\mu\nu}$ obtained from this construction are important results, while the intermediate $\check{g}_{s\mu\nu}$ is no longer employed in the following.

The result for $\hat{g}_{s\mu\nu}$  contains two gauge fields appearing only in the $n_-^\mu$ components: the vierbein and the spin connection.
They can be expanded in terms of the soft graviton field as
        \begin{align}\label{eq::SG::VierbeinWFExpansion}
            \tensor{e}{_-^\alpha} &= \delta_-^\alpha + \frac{\kappa}{2} s_-^\alpha - \frac{\kappa^2}{8} s_{-\beta}s^{\beta\alpha} + \mathcal{O}(s^3)\,,\\
            \tensor{\lc\Omega_-\rc}{_{\alpha\beta}} &= -\frac \kappa2 \lp \lc \partial_\alpha s_{\beta-}\rc - \lc\partial_\beta s_{\alpha-}\rc\rp + \mathcal{O}(s^2)\,.\label{eq::SG::SpinConnectionWFExpansion}
        \end{align}
Note that the spin connection is defined as the derivative of the 
metric, resp. vierbein. In the EFT, the soft metric and vierbein are  always evaluated at $x_-^\mu$.
Thus, in the EFT, $\partial_\perp s_{\mu\nu}(x_-)=0$ if the 
derivative were taken after setting $x=x_-$. This means that the 
vierbein and the spin-connection above should be interpreted as \emph{two truly independent gauge fields} from the EFT perspective, since the derivative contains additional information that cannot be obtained from the vierbein in the EFT.

Corresponding to these two independent gauge fields, the soft gauge 
transformations induce two {\em emergent} soft gauge symmetries that take 
the form of local translations and local Lorentz transformations on 
the light-cone $x_-^\mu$ of the collinear modes, and global 
transformations on the hyper-planes of fixed $x_-^\mu$. More precisely, 
performing a gauge transformation $U_s(x) = T^{-1}_\epsilon$ with 
local translation parameter $\epsilon_\mu(x)$, $\hat{g}_{s\mu\nu}$ 
does not transform with $T^{-1}_\epsilon$, but instead with 
$T^{-1}_{\epsilon+\omega}$, defined by the parameters 
$\epsilon_\mu(x_-)$ and $ \omega_{\mu\nu}(x_-) = -\frac 12 \lp \lc 
\partial_\mu \epsilon_\nu\rc - \lc\partial_\nu \epsilon_\mu\rc\rp(x_-)$, which again should be regarded as two independent transformations. 
The corresponding infinitesimal coordinate transformation is given by
\begin{align}
x^\mu &\to x^\mu + \kappa\varepsilon^\mu(x_-) + \kappa\tensor{\omega}{^{\mu}_{\nu}}(x_-)(x-x_-)^\nu\,.\label{eq::SG::infinitesimalresidualtransformation}
\end{align} 
The first term stems from the transformation of the vierbein, while the second one, proportional to $(x-x_-)^\mu$, arises from the spin-connection transformation. The two gauge symmetries of the collinear fields in the effective theory 
are {\em emergent}, since they have a different physical meaning 
than the original one.   
The emergent soft gauge symmetries, and consequently the soft fields, 
live only along the classical trajectory of the collinear particles 
$x_-^\mu$ and are ``visible'' only to the collinear fields in 
direction $i$. Other collinear directions see their own soft 
gauge symmetry. The symmetries of the different collinear sectors are 
induced by the gauge symmetry of the full theory, and connected 
through the purely-soft part of the Lagrangian, which continues 
to transform under the full soft symmetry parameter $\epsilon_\mu(x)$.

The effective soft gauge fields $\tensor{e}{_-^\alpha}(x_-)$, 
$\tensor{\lc\Omega_-\rc}{_{\alpha\beta}}(x_-)$ are non-linear functions 
of $s_{\mu\nu}$. 
Since the residual background metric has the property that only $\hat{g}_{s\mu-}$ is non-vanishing, it is possible to define a soft-covariant derivative       
 \begin{align}\label{eq::SG::softcovariantderivative}
            \nm D_s &\equiv \hat{g}_s^{\mu-}\partial_\mu\\
            &\hspace{-0.5cm}=
            \partial_- - \frac \kappa2 s_{-\mu} \partial^\mu + \frac{\kappa^2}{8} s_{+-}s_{--}\partial_+ + \frac{\kappa^2}{16} s_{-\alpha_\perp}s^{\alpha_\perp}_-\partial_+
            + \frac \kappa2 \tensor{\lc\Omega_-\rc}{_{\mu\nu}}J^{\mu\nu} + \mathcal{O}(\lambda^3)\nn\,,
\end{align}
where 
\begin{equation}
J^{\mu\nu} = (x-x_-)^\mu\partial^\nu - (x-x_-)^\nu\partial^\mu
\end{equation}
is the angular momentum (Lorentz generator) operator.
The form of \eqref{eq::SG::softcovariantderivative} is significant and revealing. Usually, a scalar field coupled to gravity does not have a covariant derivative.
However, in SCET gravity, the residual symmetry corresponds to local Lorentz transformations that live only on the classical trajectory of the energetic particles.
A scalar field has a non-trivial transformation under this symmetry, and therefore requires the introduction of a covariant derivative.
Namely, if a collinear scalar field transforms under \eqref{eq::SG::infinitesimalresidualtransformation} infinitesimally as
	\begin{equation}
	    T_{\varepsilon+\omega}^{-1}\varphi_c(x) = 1 - \kappa\varepsilon^\alpha \partial_\alpha\varphi_c(x) - \kappa\omega_{\alpha\beta}\,(x-x_-)^\beta\partial^\alpha\varphi_c(x) + \mathcal{O}(\varepsilon^2)\,,
	\end{equation}
one can check that the derivative \eqref{eq::SG::softcovariantderivative} has the covariant transformation
	\begin{equation}
	    \nm D_s\varphi_c(x) \to T_{\varepsilon+\omega}^{-1}\lc \nm D_s\varphi_c(x)\rc - \kappa\omega_{-\alpha}D_s^\alpha\varphi_c(x) + \mathcal{O}(\varepsilon^2)\,.\label{eq::covariantinftrans}
	\end{equation}
The transverse partial derivatives $\partial_\perp$ and $\np\partial$ already transform as in \eqref{eq::covariantinftrans}, and no modification is necessary.

The soft-covariant derivative \eqref{eq::SG::softcovariantderivative} 
contains the two independent gauge fields. At the linear order in $s_{\mu\nu}$, corresponding to the minimal-coupling terms in the Lagrangian, the first gauge field $e_{_-\mu}$ couples to the momentum $P^\mu \equiv -i\partial^\mu$, while the second gauge field, the spin-connection, couples to the angular momentum of the scalar field.
For representations with non-vanishing spin, the total angular momentum would appear in \eqref{eq::SG::softcovariantderivative}.
In the case of several directions, the object $\nim D_s$ appears inside the $i$-collinear Lagrangian $\mathcal{L}_{c_i}$.
In this sense, one has $i=1,\dots,N$ gauge symmetries that are restricted to their respective collinear trajectories, all ultimately tied to the same background soft field through the purely soft Lagrangian 
$\mathcal{L}_{s}\lc s_{\mu\nu}(x)\rc$ in~\eqref{eq:Basics:LagrangianSplit}.

It is important to stress that, as seen in \eqref{eq::SG::softcovariantderivative} and \eqref{eq::SG::Riemanntermsexpansion}, the fields $\hat{g}_{s\mu\nu}$ and $\mathfrak{g}_{s\mu\nu}$ are expressed in terms of the \emph{original} soft graviton field $s_{\mu\nu}$, where no gauge is fixed.
This is where the residual gauge symmetry stems from.
The FLNC coordinates are merely a tool to construct the covariant light-front multipole expansion of the EFT, and one does not fix this gauge in the final Lagrangian, which is a function of $s_{\mu\nu}$. Similarly to the case of light-cone gauge in the collinear sector, using the analogue of Wilson lines, the FLNC condition can be implemented ``covariantly'' by redefining the collinear fields, without restricting the gauge freedom of the soft field.

\subsection{Soft-collinear Lagrangian}

Equipped with these concepts, it is not difficult to derive the 
effective Lagrangian by purely algebraic manipulations.
As a first step, one redefines the collinear fields using the collinear ``Wilson line'' $W_c$ \eqref{eq::CG::WilsonLineDefinition} and the previously defined $R$-``Wilson line'' \eqref{eq::SG::RFLNCDef}\footnote{In the following, $R_{\rm FLNC}$, its Jacobians $\tensor{R}{_\mu^\alpha}$ and its determinant $\det(\underline{R})$ are understood to be evaluated at $x$ if no argument is given.} to obtain collinear fields that have gauge transformations \emph{compatible} with the multipole expansion.
These new fields are denoted by $\hat{\varphi}_c$ and $\hat{h}_{\mu\nu}$, and are given by\footnote{There is one subtlety: the collinear ``Wilson line'' in the soft-collinear theory is different from the purely-collinear one at the non-linear level, as the soft background field has a non-vanishing $\hat{g}_{s+-}$ component, and therefore appears in the transformation of $\hat{h}_{+-}$. This is explained in \cite{Beneke:2021aip}.}
\begin{eqnarray}
\varphi_c &=& \brac{R_{\rm FLNC} W_c^{-1}\hat{\varphi}_c} = 
\brac{R_{\rm FLNC}\hat{\chi}_c}\,,
\label{eq:mattercollredef}\\[0.1cm]
\kappa h_{\mu\nu} &=& 
\brac{R_{\rm FLNC}\tensor{R}{_\mu^\alpha}\tensor{R}{_\nu^\beta}\left(
	\tensor{W}{^\rho_\alpha} \tensor{W}{^\sigma_\beta}\brac{W^{-1}_c(\kappa \hat{h}_{\rho\sigma} + \hat{g}_{s\rho\sigma}(x))} - \hat{g}_{s\alpha\beta}(x)\right)}
\nonumber\\
&=& \brac{R_{\rm FLNC}\tensor{R}{_\mu^\alpha}\tensor{R}{_\nu^\beta}
\hat{\mathfrak{h}}_{\alpha\beta}}\,.
\label{eq:hcollredef}
\end{eqnarray}
Here the collinear fields on the left-hand side are assumed to be in 
light-cone gauge, but the gauge is not fixed on the right-hand side. 
The right-hand sides also identify the collinear gauge-invariant 
building blocks $\hat{\chi}_c$ and $\hat{\mathfrak{h}}_{\mu\nu}$ 
after the field redefinition, i.e. they are invariant with respect 
to the compatible gauge symmetries.

To illustrate the calculation of the effective Lagrangian, consider 
the collinear matter field in the soft background $g_{s\mu\nu}(x)$ as
defined in \eqref{eq::SG::MetricTensorSplit},\footnote{In \eqref{eq:lphikin} -- \eqref{eq:lepxansion}, 
$g_{s \mu\nu}$ and its determinant
are evaluated at $x$, not $x_-$.} 
\begin{eqnarray}
\label{eq:lphikin}
 \mathcal{L}_{\varphi} &=& \frac 12 \sqrt{-g_s} g_s^{\mu\nu}
\partial_\mu \varphi_c \partial_\nu \varphi_c\,.
\end{eqnarray}
Inserting \eqref{eq:mattercollredef} results in 
\begin{eqnarray}
 \mathcal{L}_{\varphi} &=& 
\frac 12 \sqrt{-g_s}g_s^{\mu\nu}\brac{\partial_\mu 
(R_{\rm FLNC}\hat{\chi}_c)}\brac{\partial_\nu(R_{\rm FLNC} 
\hat{\chi}_c)}\nonumber\\
&=&\frac 12 \det\!\lp \underline{R}\rp  \brac{R^{-1}_{\rm FLNC}
\sqrt{-g_s}}\, 
\tensor{R}{_\mu^\alpha}\tensor{R}{_\nu^\beta} \brac{R^{-1}_{\rm FLNC}
g^{\mu\nu}_s}\,
\partial_\alpha \hat{\chi}_c \partial_\beta \hat{\chi}_c\,.
\end{eqnarray}
The metric $\check{g}^{\alpha\beta}(x)$ from \eqref{eq::SG::FLNCMetricDefinition}
can now be identified, together with the corresponding 
metric determinant. Inserting  \eqref{eq::SG::BGMetricSplitRiemann} 
and dropping the Riemann tensor terms, which are sub-leading 
in the $\lambda$-expansion gives 
\begin{eqnarray}
 \mathcal{L}_{\varphi} &=& 
\frac 12 \sqrt{-\hat{g}_s}\,\hat{g}^{\mu\nu}_s\,
\partial_\mu \hat{\chi}_c \partial_\nu \hat{\chi}_c \nonumber\\
&=& \sqrt{-\hat{g}_s}\left(\frac 12 \np\partial\hat{\chi}_c \nm D_s\hat{\chi}_c + \frac 12 \partial_{\alpha_\perp}\hat{\chi}_c\partial^{\alpha_\perp}\hat{\chi}_c\right).
\end{eqnarray}
The last line follows from \eqref{eq::SG::softcovariantderivative} 
and shows that after the multipole expansion, the leading soft 
interactions are collected into the covariant derivative.

Applying this procedure to the background-field Lagrangian, now including
non-linear terms in the collinear graviton field and keeping the 
Riemann tensor terms in \eqref{eq::SG::BGMetricSplitRiemann}, one finds the soft-collinear Lagrangian in the form
\begin{equation}
\label{eq:lepxansion}
	    \mathcal{L} = \sqrt{-\hat{g}_s} \lp\mathcal{L}^{(0)} + \mathcal{L}^{(1)} + \mathcal{L}^{(2)}+\ldots\rp\,,
	\end{equation}
where
\begin{align}\label{eq::SG::LagrangianCovDevL0}
	    \mathcal{L}^{(0)} &= \frac 12 \np\partial\hat{\chi}_c \nm D_s\hat{\chi}_c + \frac 12 \partial_{\alpha_\perp}\hat{\chi}_c\partial^{\alpha_\perp}\hat{\chi}_c
	    \,,\\
	    \mathcal{L}^{(1)} &= -\frac \kappa2 \hat{\mathfrak{h}}^{\mu\nu}\partial_{\mu}\hat{\chi}_c\partial_{\nu}\hat{\chi}_c
	    + \frac \kappa4 \tensor{\hat{\mathfrak{h}}}{^{\beta_\perp}_{\beta_\perp}}\left(\np\partial\hat{\chi}_c \nm D_s\hat{\chi}_c + \partial_{\alpha_\perp}\hat{\chi}_c\partial^{\alpha_\perp}\hat{\chi}_c\right)
	    \label{eq::SG::LagrangianCovDevL1}
	    \,,\\
	    \mathcal{L}^{(2)} &=
	    - \frac 18 x_\perp^\alpha x_\perp^\beta R_{\alpha-\beta-}(\np\partial \hat{\chi}_c)^2
	    +\frac{\kappa}{2}
	    \hat{\mathfrak{h}}^{\mu\alpha}\tensor{\hat{\mathfrak{h}}}{_{\alpha}^{\nu}}\partial_{\mu}\hat{\chi}_c\partial_{\nu}\hat{\chi}_c
	    -\frac{\kappa^2}{4} \tensor{\hat{\mathfrak{h}}}{^{\alpha_\perp}_{\alpha_\perp}} \hat{\mathfrak{h}}^{\mu\nu}\partial_\mu\hat{\chi}_c\partial_\nu\hat{\chi}_c
	    \nn\\
	    &\quad
	    + \frac{\kappa^2}{16}\left((\tensor{\hat{\mathfrak{h}}}{^{\alpha_\perp}_{\alpha_\perp}})^2 - 2 \hat{\mathfrak{h}}^{\alpha\beta}\hat{\mathfrak{h}}_{\alpha\beta}\right)
	    \left(\np\partial\hat{\chi}_c \nm D_s\hat{\chi}_c + \partial_{\mu_\perp}\hat{\chi}_c\partial^{\mu_\perp}\hat{\chi}_c\right)
	    \,.\label{eq::SG::LagrangianCovDevL2}
	\end{align}
The superscript on $\mathcal{L}$ denotes the lowest order in the 
$\lambda$-expansion at which the respective terms contribute. By 
expanding out the implicit collinear Wilson lines and $n_- D_s$, 
each term generates an infinite tower of higher-order terms in 
$\lambda$. The derivation of the above Lagrangian is presented in 
detail in \cite{Beneke:2021aip}, where one can also find the explicit 
$\lambda$-expansion of the Lagrangian up to $\mathcal{O}(\lambda^2)$ 
in terms of the elementary fields $\hat{\varphi}_c$, 
$\hat{h}_{\mu\nu}$, $s_{\mu\nu}$. In general, there is also the soft 
matter field $\varphi_s$, however, in the absence of scalar 
self-interactions, it does not contribute to the soft-collinear 
Lagrangian at $\mathcal{O}(\lambda^2)$.

The soft graviton field $s_{\mu\nu}$ appears above only inside the 
soft-covariant derivative $\nm D_s$, inside the Riemann tensor and the gauge-invariant building blocks.
This highlights a formal similarity of the gravitational soft-collinear 
interactions to the respective gauge-theory result \cite{Beneke:2002ni}, featuring a leading interaction via a covariant derivative, and sub-leading interactions starting from the quadrupole term (in contrast to the dipole term in gauge theory).
The main difference in gravity is that the covariant derivative contains not one but \emph{two} gauge fields.

Moreover, the Lagrangian above is expressed in terms of the gauge-invariant building blocks  $\hat{\chi}_c$, $\hat{\mathfrak{h}}_{\mu\nu}$.
Similar to the purely-collinear Lagrangian \eqref{eq::ManifestlyInvariantCollLag} in \cref{sec:collinear}, the theory is invariant under the redefined hatted-collinear transformations, and one can decide to employ either the gauge-invariant or non-invariant fields $\hat{\varphi}_c,$ $\hat{h}_{\mu\nu}$. With the latter choice one finds that the collinear 
Wilson line $W_c^{-1}$ drops out of all terms that do not contain the 
soft Riemann tensor. Above, this applies to all but the first term 
in \eqref{eq::SG::LagrangianCovDevL2}, since this term is not minimally-coupled to the homogeneous background field \cite{Beneke:2021aip} and would consequently contain explicit factors of $W_c^{-1}$.

In the purely-collinear sector (setting all soft fields to zero), 
gravity is distinct from gauge theory. 
There is no collinear-covariant derivative $D_c^\mu$ in gravity, and 
correspondingly, leading-power collinear interactions are absent.
In combination with the operator basis discussed below, where the first graviton building block starts at $\mathcal{O}(\lambda)$, this implies that one cannot generate collinear emissions at leading power.
In consequence, there are no jets in gravity and each additional collinear emission costs a power of $\lambda$.
This immediately shows that gravity does not exhibit any collinear divergences, and indeed \emph{nothing does go wrong} in gravity \cite{Weinberg:1965nx}.

Moreover, in the soft-collinear sector the covariant derivative 
can be eliminated iteratively from the sub-leading power 
Lagrangians by making use of the field equation, such that 
it appears only in $\mathcal{L}^{(0)}$. Therefore, gravity features 
an \emph{extended eikonal} interaction compared to gauge theory, 
related to the two independent gauge fields in $n_- D_s$.
This fundamental result follows quite naturally from the SCET 
gravity construction.

\begin{figure}[t]
\centering
\includegraphics[width=0.5\textwidth]{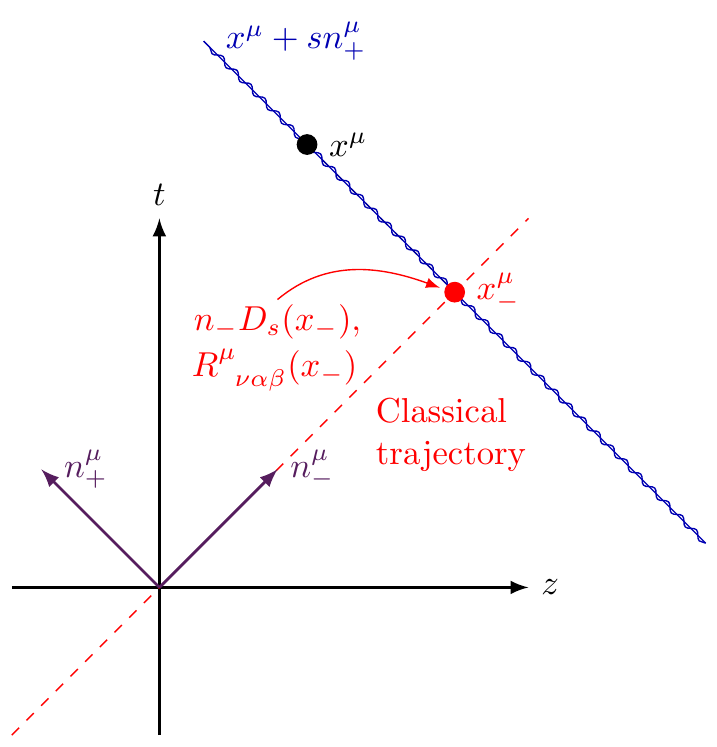}
\caption{See text for explanation.}
\label{fig:lightcone}
\end{figure}

Figure~\ref{fig:lightcone} displays the space-time structure of 
SCET gravity interactions, showing for simplicity only a single 
collinear direction and suppressing the transverse directions. 
The classical trajectory of a massless, energetic (collinear) 
particle passing through $\vec{x}=0$ at $t=0$ is the line 
$x_-^\mu = (n_+ x)\frac{n_-^\mu}{2}$ (dashed red). Collinear 
field operators describing these particles are non-local 
in the opposite light-like direction $n_+^\mu$ (wavy blue)
as discussed around \eqref{eq:collnonlocal} and 
Section~\ref{sec:sources}, but interact with soft fields 
only at $x_-^\mu$ via the extended eikonal interaction 
and the manifestly covariant Riemann tensor terms as a consequence 
of the light-front multipole expansion.

\subsection{Gauge-invariant building blocks}
\label{sec:BBlocks}
In the construction of the Lagrangian, we identified a set of gauge-invariant operators that can be used as ingredients in the hard matching. 
The only collinear gauge-invariant building blocks in the operator basis are (cf. \eqref{eq:mattercollredef}, \eqref{eq:hcollredef})
\begin{equation}\label{eq:gaugeinvariantbuildingblocks}
\begin{aligned}
    \hat{\chi}_c &= \lc W^{-1}_c\hat{\varphi}_c\rc \,,\\ \kappa\hat{\mathfrak{h}}_{\mu\nu} &= \tensor{W}{^\alpha_\mu}\tensor{W}{^\beta_\nu} \lc W^{-1}_c \kappa\hat{h}_{\alpha\beta}\rc + \lp \tensor{W}{^\alpha_\mu}\tensor{W}{^\beta_\nu}\lc W^{-1}_c\hat{g}_{s\alpha\beta}(x)\rc - \hat{g}_{s\mu\nu}(x)\rp\,,
\end{aligned}
\end{equation}
where $\mu\nu \in \{\perp\perp, \perp-, -- \}$ thanks to the 
condition $\hat{\mathfrak{h}}_{\mu+}=0$ defining the collinear Wilson 
line. The sub-leading minus-components $\hat{\mathfrak{h}}_{\perp-}$ and $\hat{\mathfrak{h}}_{--}$ can be eliminated using the collinear equation of motion, which, at leading-order, express these components in terms of derivatives of $\hat{\mathfrak{h}}_{\perp\perp}$, similar to 
\eqref{eq::CG::EOMmetric} in the purely collinear theory. The building blocks $\hat{\chi}_c$, $\hat{\mathfrak{h}}_{\perp\perp}$ both scale as 
$\mathcal{O}(\lambda)$. Additional suppression in 
$\lambda$ is obtained by acting with $\partial_\perp^\mu$ on collinear 
building blocks. Each additional derivative increases the 
$\lambda$-power by one unit. The ordinary derivative is the 
appropriate one, because one operates on gauge-invariant field 
products. 

While the building blocks are inherently inhomogeneous in $\lambda$, one can use the power-counting of the leading term to characterise these operators.
The effect of the infinite tower of terms sub-leading in $\lambda$ is simply to render the leading term gauge-invariant.
If one fixes collinear light-cone gauge, the entire sub-leading tower disappears and one obtains homogeneously-scaling operators.

For the soft sector, one employs gauge-{\em covariant} building blocks, like the soft scalar $\varphi_s$, the derivative $\nm D_s$ or the Riemann tensor $\tensor{R}{^\mu_{\nu\alpha\beta}}.$
However, the soft-covariant derivative can be eliminated from the building blocks by the collinear field equations \cite{Beneke:2021aip}. For the scalar field, at leading power
\begin{align}
    \nm D_s \hat{\chi}_c = - \frac{\partial_\perp^2}{n_+\partial} \hat{\chi}_c\,,
\end{align}
and a similar equation can be derived from the Einstein-Hilbert Lagrangian. 
This shows that $\nm D_s \mathfrak{h}_{\perp\perp} $ is redundant \cite{Beneke:2021aip} and can be traded for (non-local) combinations of the other collinear and soft building blocks, which will contribute to the equation of motion at higher powers. 
This implies that the soft-covariant derivative is not a relevant building block for the $N$-jet operators (sources). When $n_- D_s$ is eliminated systematically, the remaining soft-graviton building block is the Riemann tensor and its derivatives. Therefore, soft gauge-covariance implies 
that soft-graviton building blocks are suppressed by at least 
$\lambda^6$.\footnote{There is an implicit dependence on the soft 
metric field in the definition of the collinear building blocks 
\eqref{eq:gaugeinvariantbuildingblocks} through $W_c^{-1}$ and 
$\hat{g}_{s \mu\nu}$. This dependence is already constrained by 
collinear gauge-invariance and does not give rise to 
soft building blocks. The soft-graviton vertices generated by this 
dependence always contain collinear gravitons as well.}
 
\section{Sources and hard matching}
\label{sec:sources}

The aim of the previous sections was to construct the Lagrangian of an effective theory that reproduces the gravitational scattering amplitudes in the soft and collinear limits, that is when some of the external momenta $k_i$ form small invariants\footnote{Lorentz-invariant products of different momenta.}  $k_i\cdot k_j\ll Q^2$, where $Q$ is the scale of the hard scattering. The soft-collinear effective Lagrangian captures only part of the full scattering amplitude: it describes purely-collinear interactions within a single collinear direction and the soft-collinear interactions. To recover the full scattering amplitude, one also needs to include the effect of the highly virtual propagators and loops, the ``hard'' region. Hard lines and loops always connect different directions and are the source of large-angle scattering of energetic particles in the first place. In the pictorial representation of Figure~\ref{fig:SCETbefore}, the entire hard process is shrunk to the central point. 

In technical terms, in the scattering of multiple collinear sectors, the intermediate highly-virtual states are hard modes due to the presence of the different large momentum components $\nip p_i,n_{j+}p_j,\dots,$ from which one can form hard invariants
\begin{equation}
(\nip p_i)(n_{j+}p_j)(n_{i-} n_{j-}) \sim \lambda^0\,.
\end{equation}
The hard region is integrated out in the EFT by construction and is then encoded in the hard matching coefficients of the operators containing fields of multiple collinear directions. This gives rise to the so-called $N$-jet operators or ``currents'' or ``sources''.\footnote{The reason why the $N$-jet operators do not appear in the Lagrangian as interactions connecting different collinear sectors is that they can appear in a hard process only once, i.e. the hard subgraph must be connected, provided that there are no collinear particles in the same direction in the initial and final state of the scattering. In this case, collinear modes in different directions can never ``meet again'' in a hard process as a consequence of the Coleman-Norton theorem \cite{Coleman:1965xm}.} The full scattering amplitude is then the sum of matrix elements of $N$-jet operators evaluated with the soft-collinear effective Lagrangian. The construction of the effective Lagrangian is a prerequisite to the construction of  $N$-jet operators, because it defines what is the field basis from which the hard sources are constructed and the symmetries they must respect. After having established this in the previous sections, the admissible sources can now be easily specified. Their matching coefficients depend on the particular scattering process, and must of course be calculated in a given context.

The  $N$-jet operators are constructed from the collinear and 
soft building blocks. It corresponds to a 
purely-hard scattering that produces the constituent building-block 
particles of the operator. For example, the tree-level scattering of $N$ widely-separated energetic scalar particles is encoded in a current operator consisting of $N$ copies of the collinear gauge-invariant scalar $\hat{\chi}_{c_i}$, one for each direction, combined with a hard matching 
coefficient. One can then think of the hard matching 
coefficient as the \emph{non-radiative amplitude of the hard process}, while the building blocks correspond to the external legs of the amplitude. 

This notion is made more precise in the following. More details for the gauge theory situation can be found in \cite{Beneke:2017mmf} and for the gravitational case in \cite{Beneke:2021aip}.
The generic $N$-jet operator in gauge-theory is a light-ray operator \cite{Beneke:2017mmf}
\begin{equation}\label{eq:SG:Njetdef}
    \mathcal{J}(0) = \int [dt]_N\: \widetilde{C}(t_{i_1},t_{i_2},\dots) J_s(0) \prod_{i=1}^N J_i(t_{i_1},t_{i_2},\dots)\,,
\end{equation}
$[dt]_N = \prod_{i_k} dt_{i_k}$ and ``0'' refers to the origin $x=0$, where the hard-scattering is supposed to take place.
Here, $\widetilde{C}(t_i)$ is the hard matching coefficient, $J_i$ are composite operators of $i$-collinear fields, and $J_s$ of soft fields.
The variables $t_{i_k}$ correspond to the non-locality of 
collinear operators along the light-cone, for example, 
$J_i(t_{i_1},t_{i_2},\dots)$ might refer to the product 
$\hat{\chi}_{c_i}(t_{i_1}n_{i+})\hat{\chi}_{c_i}(t_{i_2}n_{i+}) 
\ldots$ of fields in the same collinear direction. In momentum space, 
$t_{i_k}\to n_{i+}\cdot p_{i_k}$, which can be expressed in terms of 
the fraction of total momentum in direction $i$ carried by the 
factor $k$ in $J_i$, which provides the relation to the intuitive 
notion of the non-radiative amplitude.

Like the full-theory scattering amplitude, these $N$-jet operators must be gauge invariant. In the effective theory, this means invariance under both the soft and the collinear gauge symmetries.
To ensure collinear gauge-invariance, one uses the gauge-invariant building blocks in the collinear current $J_i$, explicitly given in \eqref{eq:gaugeinvariantbuildingblocks} in the previous section. Soft fields are automatically collinear gauge-invariant.
In addition, the collinear building blocks are soft gauge-covariant, as they are constructed from the hatted fields.
For $J_s$, one should employ soft-covariant building blocks like $\varphi_s$ or the Riemann tensor $\tensor{R}{^\mu_{\nu\alpha\beta}},$ as discussed in \cref{sec:BBlocks}.
The $N$-jet operator as defined in \eqref{eq:SG:Njetdef} is now collinear gauge-invariant and soft gauge-covariant, transforming under $\emph{global}$ soft Poincar\'e transformations $U_s(0)$. The $N$-jet operator as defined in \eqref{eq:SG:Njetdef} is located at $x=0$, however, and hence not translation-invariant.
To render it manifestly gauge-invariant under both soft and collinear gauge transformations, one defines the translation- and therefore gauge-invariant current 
\begin{equation}\label{eq::SG::TranslatedCurrent}
    \mathcal{J} = \int d^4x \,T_x \mathcal{J}(0) T_x^{-1}\,,
\end{equation}
where $T_x = e^{ix\hat{p}}$ is the translation operator.
In practical computations, the integral and translation operators reduces to the momentum conserving $\delta$-function once the amplitude is evaluated.
Therefore, one can choose the simpler $N$-jet operator \eqref{eq:SG:Njetdef} located at $x=0$ and impose momentum and angular-momentum conservation by hand, similar to imposing colour-neutrality in QCD.

As said above, the hard  coefficient is obtained from a process-specific 
matching computation. 
To perform this matching, one follows the standard method of evaluating a suitable on-shell Green function in the EFT and in the full theory and demanding that they are equal up to a desired power in the $\lambda$-expansion. One typically chooses the simplest possible external states that give non-vanishing matrix elements in both theories. Once the matching coefficients are determined, one can use the EFT operators to evaluate arbitrarily complex low-energy matrix elements. 
Note that these sources are the only place in the effective theory where renormalisation takes place. 
The Lagrangian is not renormalised \cite{Beneke:2002ph}, in the 
sense that no further renormalisation is required, when the SCET Lagrangian is expressed in terms of the renormalised full-theory couplings.\footnote{In gravity, one has to keep in mind that the renormalised full theory is constructed up to a certain loop order by introducing additional higher-derivative operators, as explained in \cref{sec:PertGrav}.} The matching coefficients, however, receive loop corrections.

\section{Soft theorem}
\label{sec:softtheorem}

If the radiative amplitude for the emission of (usually) a single 
soft particle from a hard scattering process can be 
expressed in terms of the non-radiative one without detailed 
knowledge of the latter, one refers to the corresponding 
result as a ``soft theorem''. The best known and earliest example 
of a soft theorem is the abelian version of \eqref{eq:eikonal-QCD}, 
which expresses the soft-photon emission amplitude in 
QED in terms of a universal ``eikonal factor''. In QED, the 
absence of photon self-interactions leads to the exponentiation 
of multiple soft photon emissions. Of particular relevance 
to this chapter is the further fact, known as the Low-Burnett-Kroll 
theorem \cite{Low:1958sn,Burnett:1967km}, that the soft theorem 
extends to next-to-leading order in the soft expansion.  

Almost 50 years after Weinberg's first discussion of the 
gravitational soft theorem~\eqref{eq:eikonal-gravity}, Cachazo 
and Strominger \cite{Cachazo:2014fwa} made the remarkable 
observation that its universality extends by {\em two} 
orders in the soft expansion. If $\mathcal{A}(\{p_i\})$ denotes 
the amplitude for a $N$-particle hard-scattering process 
with momenta $p_1,\ldots, p_N$, 
the single graviton emission amplitude at tree-level reads
\begin{eqnarray}
\mathcal{A}_{\mathrm{rad}}(\{p_i\};k) &=& 
\frac{\kappa}{2}\sum_{i=1}^N \bar{u}(p_i) \,\Bigg(\, \frac{\varepsilon_{\mu\nu}(k)p_i^\mu p_i^\nu}{p_i\cdot k} + \frac{\varepsilon_{\mu\nu}(k)p_i^\mu k_\rho J_i^{\nu\rho}}{p_i\cdot k}
\nonumber\\ 
&& +\,\frac{1}{2}\frac{\varepsilon_{\mu\nu}(k) k_\rho k_\sigma J_i^{\rho\mu}J_i^{\sigma\nu}}{p_i\cdot k} +\mathcal{O}(k^2)
\,\Bigg)\,\mathcal{A}(\{p_i\})\,.
\label{eq:SoftTheorem}
\end{eqnarray}
Here 
\begin{equation}\label{eq:J}
J^{\mu\nu}_i = L^{\mu\nu}_i+\Sigma^{\mu\nu}_i= p_{i}^{\mu} \frac{\partial}{\partial p_{i \nu}}-p_{i}^{\nu} \frac{\partial}{\partial p_{i \mu}} + \Sigma^{\mu\nu}_i
\end{equation}
refers to generators of the Lorentz group, in this context 
usually referred to as the ``angular momentum operator'', with 
$L^{\mu\nu}_i$ the orbital angular momentum operator of particle 
$i$ from which the graviton is emitted, and $\Sigma^{\mu\nu}_i$ 
the spin operator in the representation of emitter (``matter'') 
particle $i$.\footnote{All up to a factor of $i$, omitted for 
the convenience of writing \eqref{eq:SoftTheorem} without 
factors of $i$.} $\bar u(p_i)$ denotes the polarisation functions of the 
matter particles, and $\varepsilon_{\mu\nu}(k)$ the graviton 
polarisation tensor. In the following subsections, we shall focus 
on scalar matter, in which case $J^{\mu\nu}_i=L^{\mu\nu}_i$ and 
the $\bar{u}(p_i)$ are trivial. It is worth noting that one 
recovers the Low-Burnett-Kroll theorem for the emission of a 
photon, and its non-abelian generalisation \cite{Casali:2014xpa}, 
from \eqref{eq:SoftTheorem} by substituting $p_i^\mu \to t_i^a$,  $\varepsilon_{\mu\nu}(k)\to \varepsilon_{\nu}(k)$ and  
$\kappa/2\to -g_s$, and dropping the next-to-next-to-soft term 
in the second line of \eqref{eq:SoftTheorem}, which is unique to 
gravitons. The above substitutions are suggestive of 
colour-kinematics duality, but we shall see below that regarding the next-to-soft terms in gauge theory and gravity 
as analogues of each other does not correspond to  
the proper interpretation of their physics origin.  

The discovery of the sub-sub-leading gravitational soft theorem 
\eqref{eq:SoftTheorem} was inspired \cite{Strominger:2013jfa,Strominger:2017zoo} 
by a relation between the leading soft theorem and the Bondi, van der Burg, Metzner 
and Sachs symmetries \cite{Bondi:1962px,Sachs:1962wk}, which consist 
of a subgroup of diffeomorphisms operating on the 
asymptotically flat regions of space-times, the significance of 
which has not yet become completely clear. The actual 
derivation of \eqref{eq:SoftTheorem} in \cite{Cachazo:2014fwa} 
used the spinor-helicity formalism, which is particularly 
elegant for matter particles with spin, as the angular momentum operator in spinor-helicity variables combines the 
orbital and spin parts in a simple way. Subsequently, 
the next-to-next-to-soft theorem was derived in various  
other ways \cite{Broedel:2014fsa,Bern:2014vva,Zlotnikov:2014sva}.

In the spirit of the article, we ask what the effective field 
theory can say about the soft theorem. Obviously, the  
EFT must reproduce it as the special 
case of single-soft emission, but can it provide additional 
insight? In previous sections, the important role of the emergent 
gauge symmetries in the construction of soft-collinear gravity 
has become apparent. The multipole expansion of the Lagrangian 
naturally leads to structures that already resemble the 
angular momentum 
operator, which shows up explicitly in the coupling to one 
of the gauge fields. In the following we therefore focus 
on the fundamental questions: Why are there three and exactly 
three universal terms in the soft theorem? What is the 
origin and interpretation of the angular momentum factors? 
Is the soft theorem corrected by loop effects?

Before approaching these questions from soft-collinear gravity, 
the essence of the derivation of \eqref{eq:SoftTheorem} from the 
explicit expansion of the scattering amplitude will be briefly 
reviewed, following closely \cite{Bern:2014vva}.

\subsection{Soft theorem from graviton amplitudes}

\begin{figure}[t]
\centering
\includegraphics[width=0.38\textwidth]{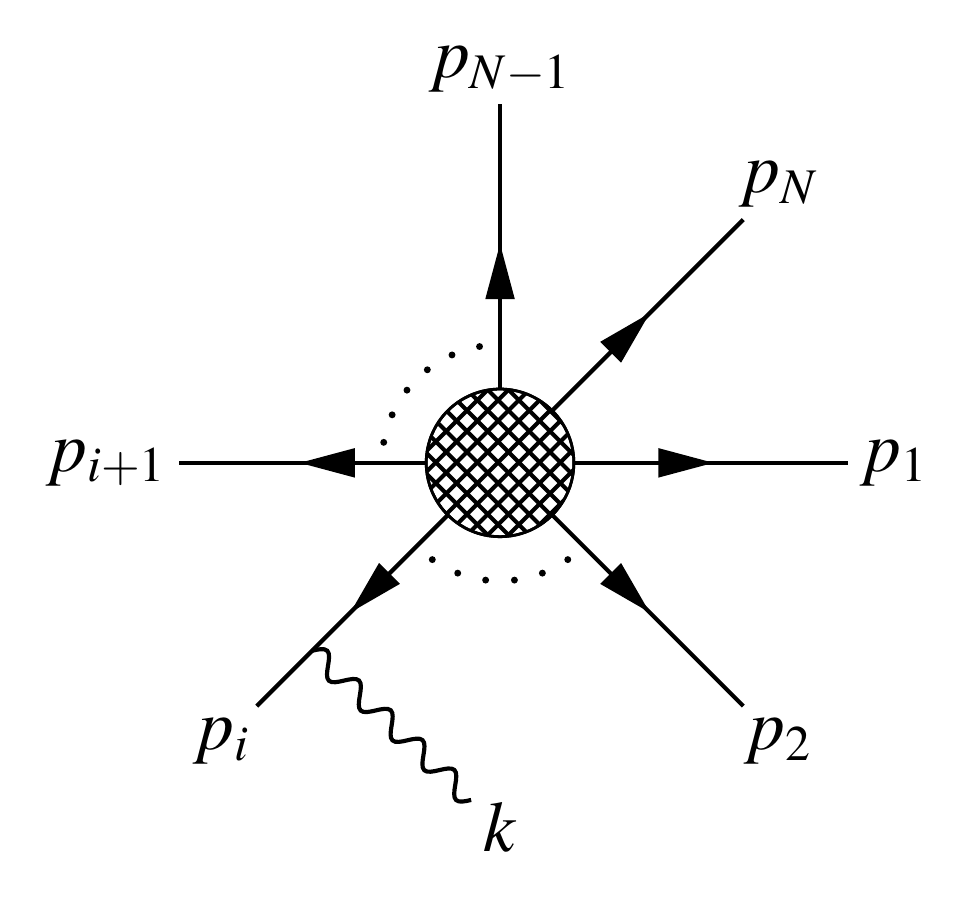}
\caption{Emission from the external leg gives the first term in the 
decomposition of the amplitude in~(\ref{eq:amplitudedecomposition}).}
    \label{fig:emissions}
\end{figure}

Since amplitudes factorise over their poles, the $N+1$-point 
amplitude with emission of a single soft graviton at tree-level 
from a hard $N$-particle scattering process with amplitude 
${\cal A}(\{p_i\})$ can be written as 
\begin{equation}
{\cal A}_{\rm rad}^{\mu \nu } (\{p_i\};k ) = 
\sum_{i=1}^{N} \frac{ p_i^{\mu} p_i^{\nu} }{p_i\cdot k} \,
{\cal A} (p_1, \dots, p_i+k, \dots, p_N) +
\mathcal{B}^{\mu \nu} (\{p_i\};k )\,.
\label{eq:amplitudedecomposition}
\end{equation}
The polarisation tensor  $\varepsilon_{\mu\nu}(k)$ of 
the graviton has been stripped off, as well as the overall 
gravitational coupling factor $\kappa/2$. The first term on 
the right-hand side contains the non-radiative amplitude 
with one momentum shifted by the graviton momentum $k$ 
and fully accounts for the singular $\mathcal{O}(1/k)$ 
term of the amplitude, which arises from the emission off 
an external leg, as shown in Figure~\ref{fig:emissions}. 
The second term starts 
at the next-to-soft order $\mathcal{O}(k^0)$. For simplicity 
of presentation, it is assumed that the $N$ hard particles 
have spin 0.

One now makes use of the invariance of on-shell 
amplitudes under the gauge transformations of weak-field 
gravity, which implies that 
$\varepsilon_{\mu\nu}(k) {\cal A}_{\rm rad}^{\mu \nu } (\{p_i\};k )$
does not change if the graviton polarisation tensor is 
replaced by
\begin{equation}
\varepsilon_{\mu\nu}(k) \to \varepsilon_{\mu\nu}(k) + 
k_\mu \alpha_\nu(\{p_i\};k)\,,
\end{equation}
where $\alpha_\nu(\{p_i\};k)$ is a function of momenta 
satisfying $k^\nu \alpha_\nu(\{p_i\};k)=0$. Since 
 $\alpha_\nu(\{p_i\};k)$ is arbitrary, it follows that
\begin{eqnarray}
0&=& k_{\mu} {\cal A}_{\rm rad}^{\mu \nu } (\{p_i\};k ) 
\nonumber \\
&=& \sum_{i=1}^{N} p_i^{\nu} 
{\cal A} (p_1, \dots, p_i+k, \dots, p_N) +
k_{\mu} \mathcal{B}^{\mu \nu} (\{p_i\};k )\,.
\label{eq:kdotA}
\end{eqnarray}
The amplitudes in this equation are analytic in $k$ for tree-level 
emission. Expanding in $k$ provides relations between 
(derivatives of) $\mathcal{B}^{\mu \nu}$ and ${\cal A}(\{p_i\})$. 
The vanishing of \eqref{eq:kdotA} at $\mathcal{O}(k^0)$ 
is guaranteed by momentum conservation, 
$\sum_{i=1}^N p_i^\nu=0$. The next two orders result in 
\begin{eqnarray}
&& \sum_{i=1}^{N} p_i^{\nu} \frac{\partial }{\partial p_{i \mu}} 
{\cal A}(\{p_i\}) + \mathcal{B}^{\mu \nu} (\{p_i\};0 ) =0\,,
\label{eq:firstder}
\\[0.1cm]
&& \sum_{i=1}^{N} p_i^{\nu} \frac{\partial^2}{\partial p_{i \mu} 
\partial p_{i \rho}} {\cal A}(\{p_i\})
+\left[\frac{\partial \mathcal{B}^{\mu \nu} }{\partial k_{\rho}}+ \frac{\partial \mathcal{B}^{\rho \nu} }{\partial k_{\mu}}\right](\{p_i\};0 )=0.
\quad
\label{eq:secondder}
\end{eqnarray}
The expansion of \eqref{eq:kdotA} does not give the above 
two equations directly, but multiplied by $k_\mu$ and 
$k_\mu k_\rho$, respectively. Removing these vectors is 
justified, because any local gauge-invariant term in 
$\mathcal{B}^{\mu \nu}$ that would be missed in this operation 
can appear only at order $k^2$  \cite{Bern:2014vva}.

Gauge invariance thus allows to express the $k\to 0$ limit 
of $\mathcal{B}^{\mu\nu}$ in terms of the derivative of the 
non-radiative amplitude, as well as a symmetric first-order 
derivative of $\mathcal{B}^{\mu\nu}$ in terms of the second 
derivative. Inserting \eqref{eq:firstder}, \eqref{eq:secondder}
into the expansion of \eqref{eq:amplitudedecomposition} 
in $k$ gives
\begin{eqnarray}
{\cal A}_{\rm rad}^{\mu \nu } (\{p_i\};k ) &=& \sum_{i=1}^{N} 
\frac{p_i^{\nu}}{p_i\cdot k} \left[  p_i^{\mu}  + k_{\rho} J_i^{\mu \rho}\right] {\cal A}(\{p_i\})  \nonumber \\
&&+ \, \frac{1}{2}k_{\rho} k_{\sigma}
\sum_{i=1}^{N} 
\frac{p_i^{\nu}}{p_i\cdot k}\,J_i^{\mu \rho} \frac{\partial}{\partial p_{i\sigma}} {\cal A}(\{p_i\})  \nonumber\\
&&+ \, \frac{1}{2}k_{\rho} \left[ \frac{\partial \mathcal{B}^{\mu \nu}}{\partial k_{\rho}} - \frac{\partial \mathcal{B}^{\rho \nu}}{\partial k_{\mu}} \right](\{p_i\};0 ) +\mathcal{O}(k^2)\,.
\label{eq:protosofttheorem}
\end{eqnarray}
The first line of this equation already shows that the 
next-to-soft term is universal and proves the corresponding term 
in \eqref{eq:SoftTheorem}, while the next-to-next-to-soft 
$\mathcal{O}(k)$ terms in the second and third line 
still contain the anti-symmetric 
first-order derivatives of the non-singular part 
$\mathcal{B}^{\mu\nu}$ of the radiative amplitude.   

It is worth noting that the previous equation can be 
adapted to gauge-boson rather than graviton emission 
by simply replacing $p_i^\nu\to t^a_i$ and removing 
the index $\nu$ on ${\cal A}_{\rm rad}^{\mu \nu }$ and 
$\mathcal{B}^{\mu\nu}$, since the derivation goes through 
for non-abelian gauge invariance under shifts 
\begin{equation}
\varepsilon_{\mu}(k) \to \varepsilon_{\mu}(k) + 
k_\mu \alpha(\{p_i\};k)
\end{equation}
of the emitted soft gauge boson polarisation vector. 
The key point about gravitation is that the radiative amplitude 
{\em also} satisfies $k_{\nu} {\cal A}_{\rm rad}^{\mu \nu } 
(\{p_i\};k )=0$, which can be verified for 
\eqref{eq:protosofttheorem}, but has not yet been used explicitly. 
This fixes the so far undetermined first derivatives in 
terms of the non-radiative amplitude: 
\begin{equation}
\left[ \frac{\partial \mathcal{B}^{\mu \nu}}{\partial k_{\rho}} - \frac{\partial \mathcal{B}^{\rho \nu}}{\partial k_{\mu}} \right](\{p_i\};0 ) = 
\sum_{i=1}^{N} J_i^{\rho \mu}\frac{\partial}{\partial p_{i\nu}}
{\cal A}(\{p_i\}) \,.
\end{equation}
Substituting into \eqref{eq:protosofttheorem} yields the soft theorem
\eqref{eq:SoftTheorem} including three universal terms, that is, 
terms that can be expressed in terms of (derivatives of) the 
non-radiative amplitude. The authors of   
\cite{Bern:2014vva} further checked that there are not enough 
constraints from gauge invariance to determine all second 
derivatives of $\mathcal{B}^{\mu\nu}$, hence the next 
$\mathcal{O}(k^2)$ term in the soft expansion is no longer universal.

The diagrammatic derivation of the soft theorem is remarkably 
simple on the one hand, 
and highlights the crucial role of gauge invariance. 
On the other hand, the number of universal terms appears 
from a rather technical argument and the derivation provides no 
physical 
understanding of why the derivative expansion organises itself 
such that the orbital angular momentum operator arises. 

\subsection{Soft theorem from SCET gravity}

The soft-collinear effective Lagrangian by construction allows 
one to generate from its Feynman rules the soft and collinear 
limits of an amplitude to a desired accuracy in the soft 
and collinear expansion, and the loop expansion. However, it 
organises the derivation of the soft theorem in a different 
form from the above, since the soft factors arise entirely from 
the emission from external legs. For the case of gauge theory this 
was noted first in \cite{Beneke:2017mmf}.\footnote{This statement 
holds in the position-space SCET framework. For a discussion 
of the LBK theorem in the so-called label formalism, see 
\cite{Larkoski:2014bxa}.} 

To begin, it is instructive to formulate Weinberg's leading 
eikonal graviton emission amplitude \eqref{eq:eikonal-gravity} 
in the EFT language by recalling from 
\eqref{eq::SG::LagrangianCovDevL0}, 
\eqref{eq::SG::softcovariantderivative} that an energetic scalar with 
its large momentum $p_i^\mu$ directed along the light-like vector 
$n_{i-}^\mu$ interacts with a soft graviton 
through the effective Lagrangian\footnote{In the following, we drop the hats of the collinear fields.
Furthermore, we consider a \emph{complex} scalar field. This has the advantage that in the convention where all external particles are outgoing, the $N$-jet operator contains one field $\chi_c^\dagger$ for each particle.
Thus, one can always identify $\chi_c^\dagger$ in the Lagrangian as the outgoing field, simplifying the following derivation.}
\begin{equation}
{\cal L}_i^{(0)}=[n_{i+}\partial \chi_{c_i}]^\dagger
\left[-\frac{\kappa}{4} n_{i-}^\mu n_{i-}^\nu 
s_{\mu\nu}(x_{i-})\,i n_{i+}\partial\right]\chi_{c_i}\,. 
\label{eq:softL0grav}
\end{equation}
The structure of \eqref{eq:eikonal-gravity} is already 
manifest in this Lagrangian, which couples the energetic 
particle to the soft gauge field and graviton only 
proportionally to the large momentum 
$p_i^\mu\propto n_{i-}^\mu$.
The content of the leading term in the soft 
theorem can now be stated in operatorial form as 
\begin{equation}
\sum_ii\int d^{4}x\;T\left\{\chi^\dagger_{c_i}(0), \mathcal{L}_i^{(0)}(x)\right\}\Big\rvert_{\rm tree}\,,
\end{equation}
where the sum over $i$ runs over the energetic particles 
created in the hard process. The entire hard, non-radiative 
$N$-particle scattering process is sourced by a gauge-invariant 
product of $N$ $\chi^\dagger_{c_i}$ fields\footnote{Employing the convention that all particles are out-going.} as described in 
Section~\ref{sec:sources}. 
Contracting $\chi^\dagger_{c_i}(0)$ with $\chi_{c_i}(x)$ 
in ${\cal L}_i^{(0)}$ to form the collinear matter 
propagator $i/p^2$, 
and taking the matrix element with $N$ matter particles 
and a soft graviton, immediately yields the 
amplitude~\eqref{eq:eikonal-gravity}.
At this point, it is essential  
that soft gauge bosons and gravitons cannot be emitted 
directly from the hard vertex at this order in the 
soft expansion, since there are no source operators 
containing soft fields that would be invariant under the soft gauge symmetry. 
The entire radiative amplitude originates from the time-ordered product with the universal Lagrangian interaction.
This guarantees the universality of the soft theorem, 
that is, its form is independent of the non-radiative, 
hard process. Briefly,
\begin{displaymath} 
\mbox{Lagrangian insertions$\quad\Rightarrow\quad$ 
universal terms,}
\end{displaymath}
while the source operators in the effective theory have 
process-dependent matching coefficients to the full theory.
Thus, when there is a source operator with $N$ collinear 
fields and a soft graviton, it will contribute a non-universal 
term to the soft expansion of the radiative amplitude.

\begin{figure}[t]
    \centering
   \includegraphics[width=0.38\textwidth]{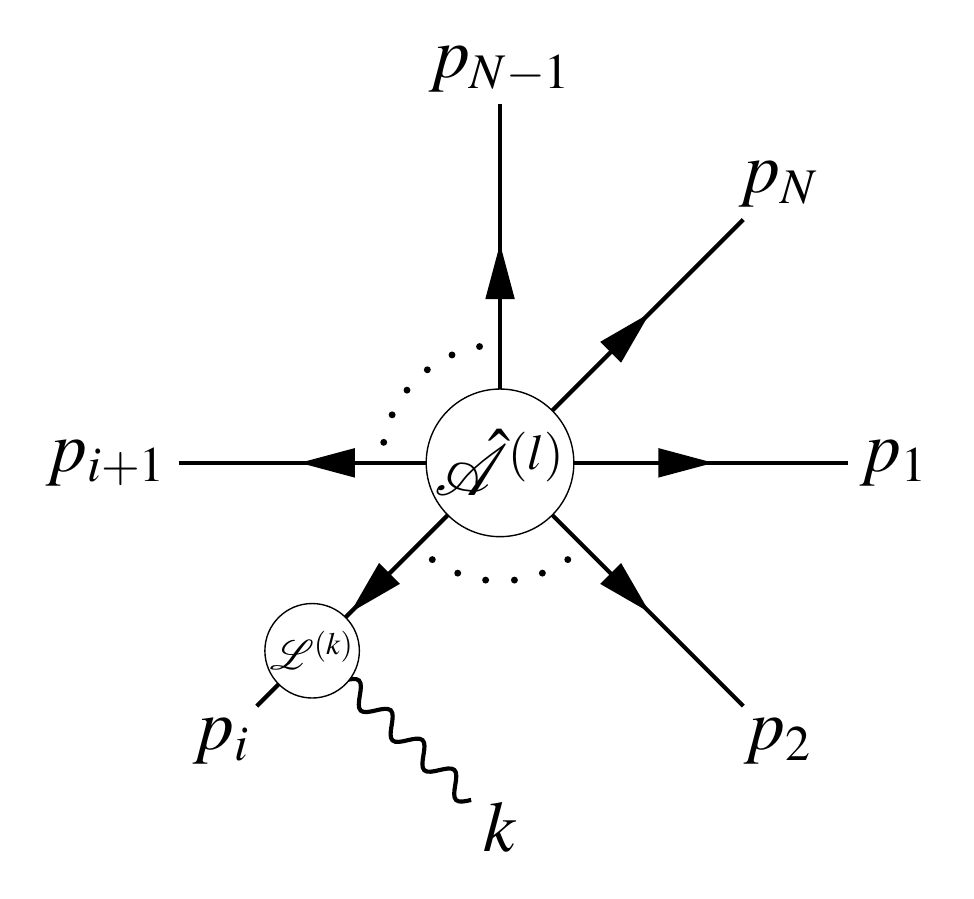}
    \quad
    \includegraphics[width=0.38\textwidth]{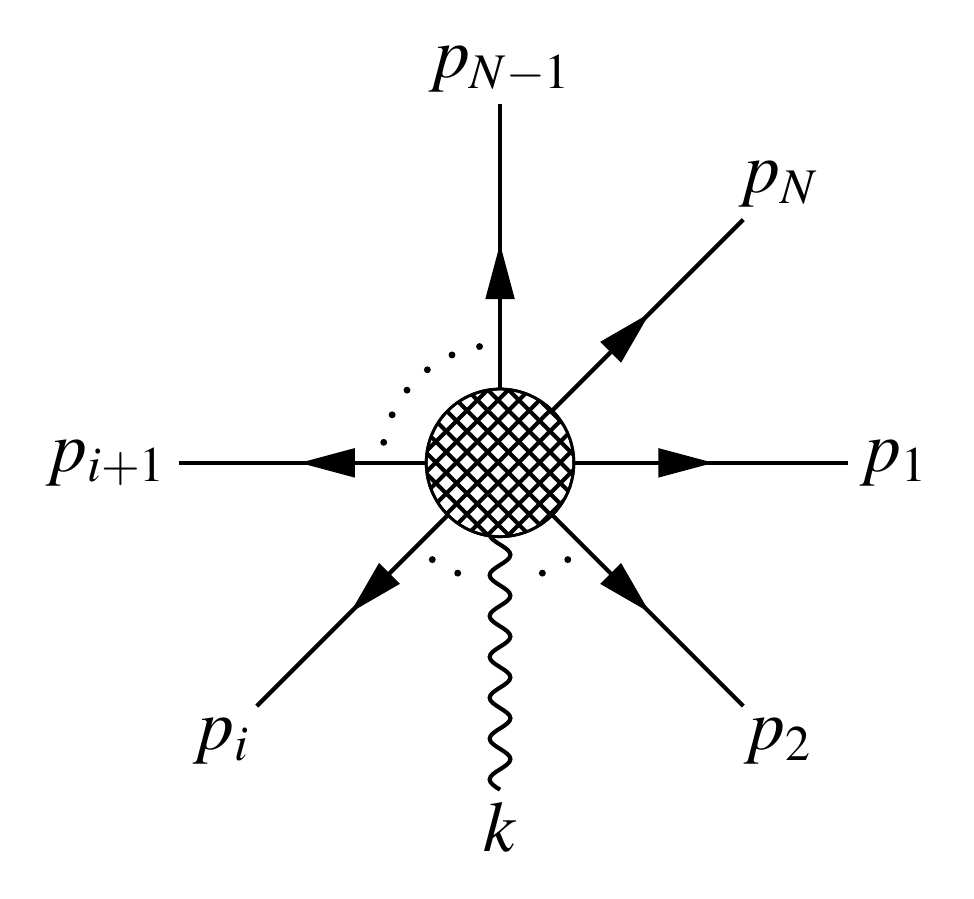}
    \caption{Possible contributions to the radiative amplitude in SCET. The first diagram represents the time-ordered product of the current and the Lagrangian. Both current and Lagrangian can be suppressed by a single power of $\lambda$, or there can be a contribution from the $\lambda^2$-suppressed Lagrangian together with the leading-power current. There are no soft building blocks at order $\lambda^2$ and $\lambda^4$; hence, the second diagram does not contribute to the amplitude in SCET. 
\label{fig:Njet_rad}}
\end{figure}

Unlike for amplitudes, in soft-collinear effective theory 
the separation of the amplitude into emission from the 
external leg and the hard vertex (see Figure~\ref{fig:Njet_rad}) 
is gauge-invariant and corresponds to the separation into 
gauge-invariant Lagrangian insertions 
and gauge-invariant source operators with soft graviton fields. 
The number of universal terms in the soft expansion 
follows from this observation without the need for an 
explicit computation, since, as noted in  
Section~\ref{sec:BBlocks}, the leading gauge-invariant 
soft building block for gravitons is the Riemann tensor, 
which scales as $\mathcal{O}(\lambda^6)$, corresponding 
to a third-order correction in the soft expansion. It is this 
simple consequence of the soft gauge symmetry of the effective 
Lagrangian, which implies that there is 
some form of universal soft theorem including a universal 
next-to-next-to-soft term for gravity, as any soft emission up these orders must arise from universal Lagrangian terms, independent of the source for the energetic particles. The additional 
universal term compared to gauge boson emission follows from 
the observation that the leading soft building block for 
soft gauge fields is the field strength tensor $F^s_{\mu\nu}$,
which  scales as $\mathcal{O}(\lambda^4)$, hence there are 
non-universal contributions already at the 
next-to-next-to-soft order.

This answers the first question. To gain a better understanding of the detailed form of the soft theorem, 
in particular the appearance of the angular momentum operator 
in the sub-leading terms, one must investigate the structure of 
the effective Lagrangian. As before, the specific case of 
scalar matter will be considered. Since the soft theorem 
is a statement about 
tree-level, single graviton emission, two simplifications can 
be made in the general soft-collinear gravity Lagrangian:
\begin{itemize}
\item The collinear graviton field can be set to zero, since 
there are no internal or external collinear graviton lines.
\item Only linear interactions in the soft graviton field 
need to be retained.
\end{itemize}
In the remainder of this subsection, the basic ideas 
will be explained for the next-to-soft term, which 
counts as $\mathcal{O}(\lambda^2)$. For the derivation of the 
sub-sub-leading soft term and more technical detail, we 
refer to \cite{Beneke:2021umj}. This reference also discusses 
the case of non-abelian gauge theory, including the 
case of matter with spin $\frac{1}{2}$ and $1$, which 
demonstrates how the spin term $\Sigma^{\mu\nu}$ in $J^{\mu\nu}$ 
is encoded in the effective Lagrangian.

The soft-collinear effective Lagrangian for the complex gauge-invariant scalar field $\chi_c$ up to $\mathcal{O}(\lambda^2)$, 
with simplifications as stated above already applied, can be 
expressed in terms of the energy-momentum tensor 
(dropping the index $i$ when referring to the Lagrangian, 
since its form is the same for all $i$)
\begin{equation}
    T^{\mu\nu} = \lc \partial^\mu\chi_c\rc^\dagger \partial^\nu \chi_c + \lc \partial^\nu \chi_c\rc^\dagger \partial^\mu\chi_c - \eta^{\mu\nu} \lc\partial_\alpha\chi_c\rc^\dagger\partial^\alpha\chi_c\,
\end{equation}
as
\begin{eqnarray}
\label{eq::GR::StructureLagrangian}
    \mathcal{L} &=& \frac 12 \lc\np\partial\chi_c\rc^\dagger\nm\partial\chi_c + \frac 12 \lc \nm\partial\chi_c\rc^\dagger \np\partial\chi_c
    + \lc \partial_{\mu_\perp}\chi_c^\dagger\rc\partial^{\mu_\perp}\chi_c\\
&&
- \,\frac{\kappa}{4}s_{-\mu}T^{\mu+}
- \frac{\kappa}{4} \lc \partial_{[\mu} s_{\nu]-}\rc\,(x-x_-)^\mu 
T^{\nu+} - \frac 18 x_\perp^\alpha x_\perp^\beta  R^s_{\alpha-\beta-} 
T^{++} + \mathcal{O}(x^3)\,.
\nonumber
\end{eqnarray}
When no space-time argument is specified,  all soft fields 
are evaluated at $x_-^\mu$ {\em after} derivatives are taken.  
In this form, the coupling of $s_{\mu-}$ to the energy-momentum tensor $T^{\mu\nu}$ becomes transparent, as well as the coupling of its derivative $\partial_{[\alpha} s_{\beta]-}$, which is an independent gauge field in the effective theory, to the angular momentum density 
\begin{equation}
    \mathcal{J}^{\alpha\beta\mu} = (x-x_-)^\alpha T^{\beta\mu} - (x-x_-)^\beta T^{\alpha\mu}\,.
\end{equation}
The Lagrangian \eqref{eq::GR::StructureLagrangian} is not yet 
homogeneous in $\lambda$. The $\lambda$ expansion is made 
manifest by decomposing the contraction of the indices $\mu,\nu$ 
in light-cone components.~Then 
\begin{equation}
    \mathcal{L} = \mathcal{L}^{(0)} + \mathcal{L}^{(1)} + \mathcal{L}^{(2)} + \mathcal{O}(\lambda^3)\,,
\end{equation}
where
\begin{eqnarray}
\mathcal{L}^{(0)} &=& \frac 12 \lc\np\partial\chi_c\rc^\dagger \nm\partial\chi_c + \frac 12 \lc\nm\partial\chi_c\rc^\dagger\np\partial\chi_c + \lc\partial_{\mu_\perp}\chi_c\rc^\dagger\partial^{\mu_\perp}\chi_c -\frac \kappa 8 s_{--} T^{++}\,,
\nonumber\\[-0.2cm]
&&\label{eq:GRSCETL0}\\
\mathcal{L}^{(1)} &=& -\frac \kappa 4 s_{-\mu_\perp}\,T^{\mu_\perp +}
- \frac \kappa 8 \lc \partial_{[\mu}s_{-]-}\rc\, x_\perp^\mu T^{++}\,,
\label{eq:GRSCETL1}\\[0.2cm]
    \mathcal{L}^{(2)} &=& - \frac \kappa 8 s_{+-} T^{+-}
    - \frac{\kappa}{4} \lc \partial_{[\mu_\perp}s_{\nu_\perp]-}\rc\, x_\perp^\mu \,T^{\nu_\perp +}
    - \frac{\kappa}{16} \lc \partial_{[+}s_{-]-}\rc\, \nm x \;T^{++}\nn\\
    &&
    -\,\frac  18 x_\perp^\alpha x_\perp^\beta R^s_{\alpha-\beta-} T^{++}\,.
\label{eq:GRSCETL2}
\end{eqnarray}

The source operator $\hat{\mathcal{A}}$ that represents the hard 
scattering process also needs to be expanded in $\lambda$: 
\begin{equation}
\hat{\mathcal{A}}=\hat{\mathcal{A}}^{(0)} + 
\hat{\mathcal{A}}^{(1)}
+ \hat{\mathcal{A}}^{(2)}+ \mathcal{O}(\lambda^3\hat{\mathcal{A}}^{(0)} )\,.
\end{equation}
The $\hat{\mathcal{A}}^{(n)}$ up to this order 
are operators containing 
$i\partial_{\perp i}^{\mu_1}\ldots i\partial_{\perp i}^{\mu_n} 
\chi_{c_i}^\dagger$, which corresponds to the expansion of 
the non-radiative momentum-space hard-scattering amplitude  
in the transverse momentum $p_{i\perp}^\mu$ of particle $i$. 
The operatorial version of the soft theorem then 
amounts to the expansion in $\lambda$ of the right-hand 
side of 
\begin{equation}
\hat{\mathcal{A}}_{\rm rad} = i\int d^{4}x\;T\left\{\hat{\mathcal{A}}, \mathcal{L}\right\}
\label{eq:operatorsoftth}
\end{equation}
in the tree-level approximation and under the assumption that 
the operator is evaluated in the state $\langle \{p_i\};k|
\hat{\mathcal{A}}_{\rm rad}|0\rangle$.

It is always possible to define the $N$ light-cone 
vectors $n_{i-}^\mu$ such that the {\em external} energetic 
matter particle momenta of the radiative amplitude 
satisfy $p_{i\perp}^\mu=0$ for all $i$ with respect to 
their light-cone vectors. This choice simplifies the 
calculations considerably. Many terms vanish, which would otherwise contribute to restoring Lorentz-invariant 
scalar products of the external momenta.  These scalar products do not depend on the specific choice of the reference vectors that break manifest Lorentz invariance. In particular, one finds 
that 
\begin{equation}
T\left\{\hat{\mathcal{A}}^{(n)}, \mathcal{L}^{(0)}\right\} , \;
T\left\{\hat{\mathcal{A}}^{(0)}, \mathcal{L}^{(1)}\right\} 
\end{equation}
evaluate to zero on the right-hand side of 
\eqref{eq:operatorsoftth}, leaving\footnote{The symbol 
$\equalhat$ means "equal up to terms that do not contribute to the specific tree-level matrix elements considered here, i.e.\ with single soft emission, no collinear emissions, and $\perp$-component of the external collinear particle momenta set to zero.''} 
\begin{eqnarray}
\hat{\mathcal{A}}^{(2)}_{\rm rad} & \equalhat &i\int d^{4}x\;T\left\{\hat{\mathcal{A}}^{(1)}, \mathcal{L}^{\left(1\right)}\right\}+i\int d^{4}x\;T\left\{\hat{\mathcal{A}}^{(0)}, \mathcal{L}^{(2)}\right\}
\label{eq:subleadingsoftv1}
\end{eqnarray}
for the next-to-soft term.

Examining \eqref{eq:GRSCETL1}, \eqref{eq:GRSCETL2} one finds 
that the Riemann-tensor term in the last line of 
\eqref{eq:GRSCETL2} does not contribute to the time-ordered 
product with $\hat{\mathcal{A}}^{(0)}$ 
in the frame $p_{i\perp}^\mu=0$, while the remaining terms 
and $ \mathcal{L}^{\left(1\right)}$ can be rewritten after 
integrating by parts into 
\begin{align}
\label{eq::GR::L1orbital}
    \mathcal{L}^{(1)} &\equalhat \frac{\kappa}{2} \lc\partial_\mu s_{\nu-}\rc \chi_c^\dagger \overset{\leftarrow}{L}\!\phantom{\,}_{+\perp}^{\mu\nu} \np \partial\chi_c\,,\\
    \mathcal{L}^{(2)} &\equalhat \frac{\kappa}{2}  \lc\partial_\mu s_{\nu-}\rc \chi_c^\dagger \overset{\leftarrow}{L}\!\phantom{\,}_{+-}^{\mu\nu} \np \partial\chi_c\,,
\end{align}
where 
\begin{eqnarray}\label{eq:Jx}
    L^{\mu\nu}	=x^{[\mu}\partial^{\nu]}
	=\frac{1}{2}x^{[\mu}n_{-}^{\nu]}n_{+}\cdot\partial+\ldots
	=\underbrace{\frac{1}{4}n_{-}\cdot xn_{+}\cdot\partial\;n_{+}^{[\mu}n_{-}^{\nu]}}_{L_{+-}^{\mu\nu}}
	+\underbrace{\frac{1}{2}x_{\perp}^{[\mu}n_{-}^{\nu]}n_{+}\cdot\partial}_{L_{\perp+}^{\mu\nu}} \qquad
\\[-0.5cm]
\nonumber
\end{eqnarray}
is the orbital angular momentum operator in position space and 
the dots after the first equality represent terms that 
vanish in the $p_\perp^\mu=0$ frame. Both time-ordered 
products in \eqref{eq:subleadingsoftv1} can now be 
combined into 
\begin{eqnarray}
    \hat{\mathcal{A}}^{(2)}_{\rm rad} & = &   \int d^{4}x\;T\left\{ \hat{\mathcal{A}},\frac\kappa 2\chi_{c}^\dagger \overset{\leftarrow}{L}\!\phantom{\,}^{\mu \nu} \lc \partial_\mu s_{\nu-}\rc i\np \partial\chi_{c}
    \right\}\,,\label{eq:gr:subleadingOp}
\end{eqnarray}
which is precisely the next-to-soft term in 
\eqref{eq:SoftTheorem}. This answers the second question: 
the angular momentum operator is seen to appear since the 
soft-collinear expansion of the Lagrangian together with 
the light-cone multipole expansion of soft interactions with 
energetic particles naturally provide the right 
structures. At the next-to-soft order, the appearance 
of the angular momentum is already evident
in \eqref{eq::GR::StructureLagrangian} and related to 
the soft gauge symmetry of the effective Lagrangian.

This observation provides a deeper understanding of the 
gravitational soft theorem. Unlike the gauge theory case, 
where the 
next-to-soft term in the LBK theorem can be shown to be 
related to the soft field-strength tensor $F_{\mu\nu}^s$ 
\cite{Beneke:2021umj}, and is therefore gauge-invariant 
without requiring a conserved quantity in the scattering 
process, the next-to-soft term in gravity takes the form of 
an \emph{eikonal term}, just as the leading term. 
The leading term 
\begin{equation}
\varepsilon_{\mu-}\,p^\mu\,\frac{\np p}{p\cdot k}   
\end{equation} 
is related to the coupling of the first 
soft gauge field $s_{-\mu}$ to the energy momentum tensor $T^{\mu+}$ 
in \eqref{eq::GR::StructureLagrangian}. It is gauge-invariant 
due to the conservation of momentum. The first sub-leading 
order in the soft theorem, 
\begin{equation}
k_\rho \varepsilon_{\mu-} J^{\rho\mu}\,\frac{\np p}{p\cdot k}\,,
\end{equation}
is generated by the next term in 
\eqref{eq::GR::StructureLagrangian}. It involves the 
second soft gauge field,  $\partial_{[\mu} s_{\nu]-} 
\equiv \partial_{\mu} s_{\nu-}-\partial_{\nu} s_{\mu-}$ 
(or $\tensor{\lc\Omega_-\rc}{_{\mu\nu}}$, see 
\eqref{eq::SG::SpinConnectionWFExpansion}), 
which couples to the energy-momentum density. The next-to-soft 
term for gravity is only gauge-invariant once angular 
momentum conservation of the scattering process is imposed. Hence it 
is the two-fold emergent soft gauge symmetry that implies two 
eikonal terms in the soft theorem, consistent with the 
covariant derivative 
\begin{equation}
\nm D_s = \partial_- - \frac{\kappa}{2} s_{-\mu} \partial^\mu
- \frac{\kappa}{4} \partial_{[\mu} s_{\nu]-} J^{\mu\nu} + \dots\,.
\end{equation}

Finally, the Riemann-tensor terms, which are present in the 
sub-leading soft-collinear interaction Lagrangians $\mathcal{L}^{(2)},\mathcal{L}^{(3)}$ and $\mathcal{L}^{(4)}$, generate the 
next-to-next-to-soft term \cite{Beneke:2021umj}
\begin{equation}
    \frac{1}{2}\varepsilon_{\mu\nu} k_\rho k_\sigma \,J^{\rho\mu}\frac{J^{\sigma\nu}}{p\cdot k}
\end{equation}
in \eqref{eq:SoftTheorem}. 
Here, one factor of $J^{\mu\nu}$ is related to the charge of one of the 
soft gauge symmetries, while the other arises from the 
kinematic multipole expansion. Originating from the Riemann 
tensor, this term is manifestly gauge-invariant without 
requiring further (non-existent) conserved charges. Thus, 
despite appearance, the {\em next-to-next-to soft term} in gravity 
has the same physical origin as the {\em next-to-soft term} in the 
LBK theorem, while the soft {\em and} next-to-soft term 
should be viewed as the gravitational analogues of the 
familiar eikonal factor in gauge theories.

\subsection{Loop corrections to the soft theorem}

While the soft theorem is agnostic about the details of the 
hard non-radiative amplitude that produces the $N$ energetic 
particles (in widely separated directions), it holds only at 
tree-level in the interactions of the energetic particles 
with soft gravitons. Soft-collinear gravity generates the 
soft and collinear loops in gravitational scattering to any 
order, so it is natural to ask whether it provides insight 
on the loop corrections to the soft theorem. In this subsection, 
we will show that:
\begin{center}
\vskip0.2cm
\begin{minipage}{0.9\textwidth}
\it The leading soft factor is not modified by loop effects. 
The sub-leading factor is only corrected by one-loop, and the sub-sub-leading factor is only modified by one- and two-loop contributions. Higher-order loop corrections do not affect the 
gravitational soft theorem.
\end{minipage}
\vspace*{0.2cm}
\end{center}
This should be compared to non-abelian gauge theory, where 
already the leading term receives loop corrections of any order. 
The above statement was already made  
in \cite{Bern:2014oka}, but the reasoning from EFT provided in  
\cite{Beneke:2021umj} and below is somewhat different, and 
relies only on 1) power counting, 
2) the eikonal identity, 
and 3) the necessity to 
form a soft invariant from a given soft loop integral, as 
otherwise the integral is scaleless and vanishes. It should 
be noted that the radiative amplitude is infrared-divergent at 
loop level, and a regularisation is needed. The above 
statement holds when singularities are regulated 
dimensionally in $d=4-2\epsilon$ dimensions.

In SCET, loop contributions arise from three different loop 
momentum regions, the hard, the collinear and the soft region, 
the latter two corresponding to collinear and soft modes in the 
effective theory. The hard modes are integrated out, thus the contributions of the hard loops are inside the matching coefficients $\widetilde{C}^{X}(t_i)$ and consequently part of the non-radiative amplitudes
$\langle p_1, \ldots, p_N|\mathcal{J}(0)|0\rangle$ defined in 
Section~\ref{sec:sources}. 
Hence, hard loops never affect the soft theorem directly---they modify 
the underlying non-radiative process.

Soft-collinear gravity differs from gauge theory in the purely soft 
and collinear sectors, ultimately due to the dimensionful 
gravitational coupling:
\begin{itemize}
\item[$i)$] There are no collinear singularities. 
In the purely-collinear sector, that is, in the Lagrangian 
terms containing only collinear but no soft fields, there are 
no leading-power interactions. The $\lambda$-expansion 
corresponds to the weak-field expansion, and the first 
collinear interaction appears in $\mathcal{O}(\lambda)$. 
Purely-collinear gravity is an expansion in collinear 
momenta $p_\perp\sim\lambda$.
\item[$ii$)] In the purely-soft sector,  that is, in the Lagrangian terms containing only soft but no collinear fields, there are also no leading-power interactions. Here, the weak-field expansion agrees with the $\lambda^2$-expansion, corresponding to an expansion in soft momenta $k\sim\lambda^2$. Purely-soft interaction vertices thus start at $\mathcal{O}(\lambda^2)$.
\end{itemize}
Hence, whenever a purely-collinear or a purely-soft interaction 
takes place, the contribution is already suppressed by at least 
one order of $\lambda$ or $\lambda^2$, respectively. {\em 
In gravity, only soft-collinear interactions of the eikonal 
type exist at leading power.}

\begin{figure}[t]
\centering
\includegraphics[width=0.34\textwidth]{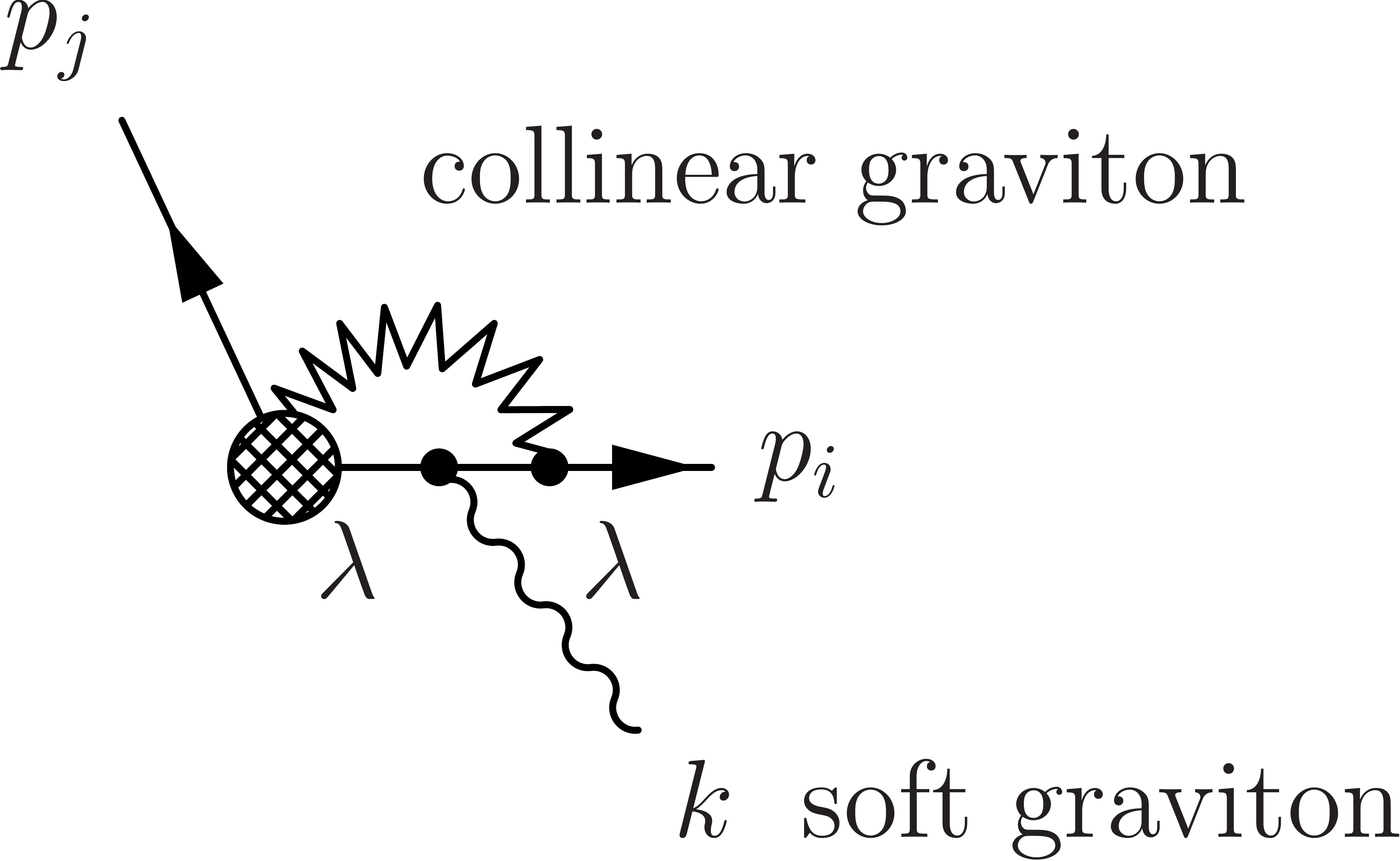}\hskip-0.5cm
\includegraphics[width=0.28\textwidth]{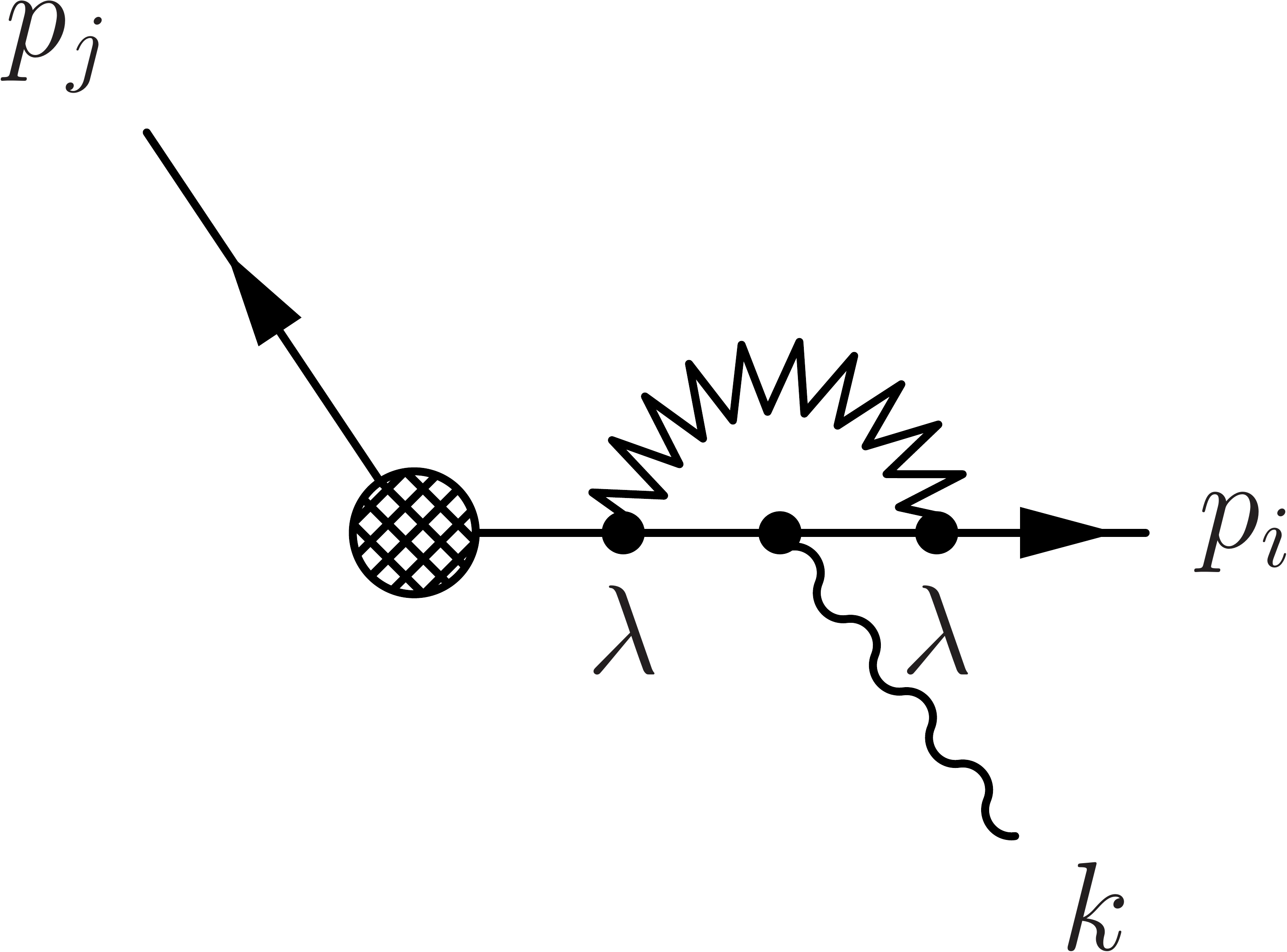}
\includegraphics[width=0.28\textwidth]{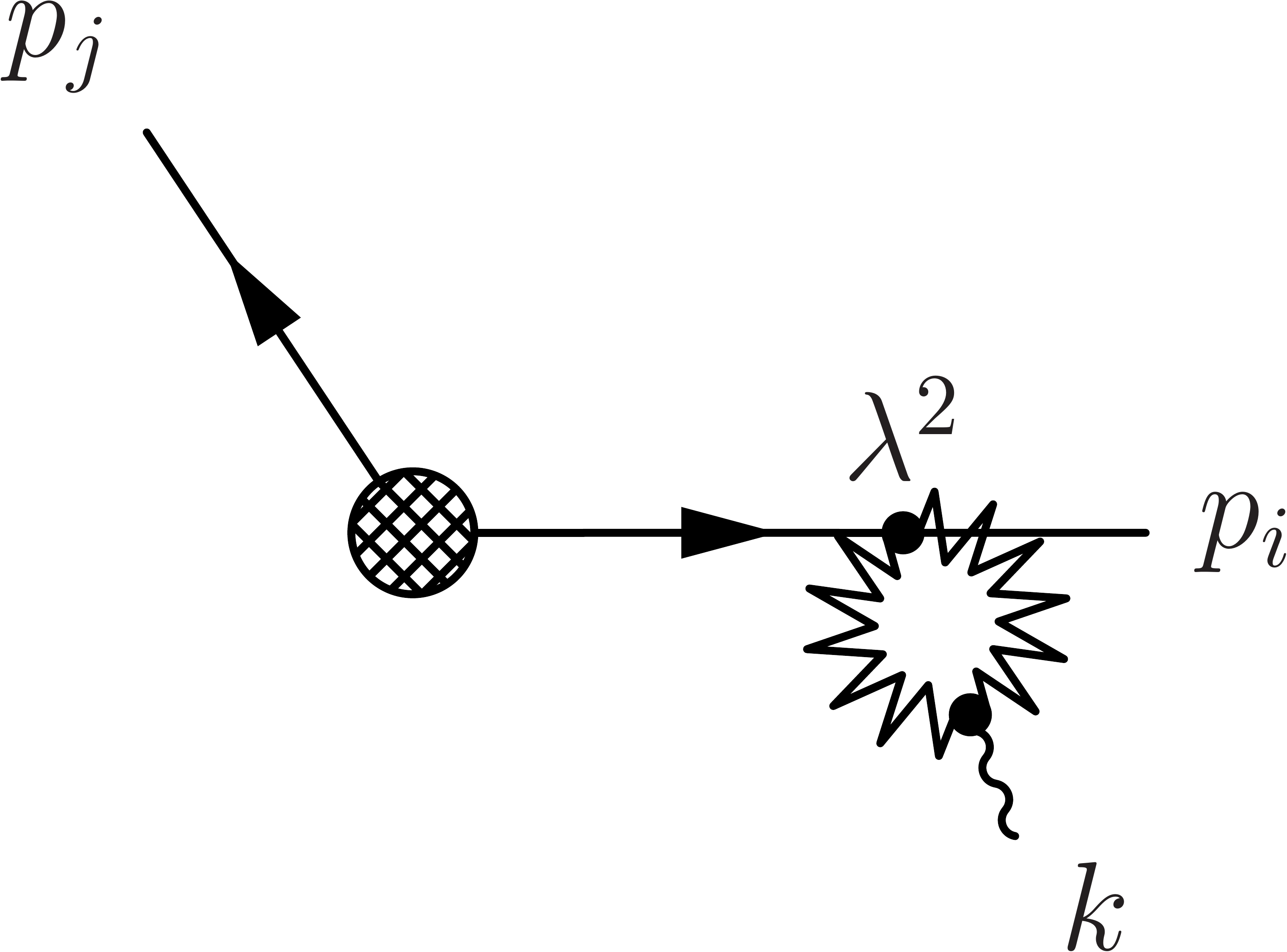}
\caption{
Collinear one-loop corrections. Left: Collinear graviton 
from the source in one of the $N$ given directions. Middle 
and right: Collinear loop corrections to an external leg. 
Two further diagrams similar to the first two but with the soft 
graviton attached to the collinear graviton line are not shown.
}
\label{fig:collinearloops}
\end{figure}

Consider first adding only collinear loops to the single-soft 
emission amplitude. At the one-loop level, there exist two 
possibilities, shown in Figure~\ref{fig:collinearloops}. First, 
a collinear graviton field $\mathfrak{h}_{\perp\perp}$ 
in a given collinear direction $n_{i-}^\mu$ can be added to 
the source, which comes with an extra power of $\lambda$ 
for the hard amplitude. The collinear graviton must be 
attached to the external leg $i$ through a collinear 
interaction vertex, which costs at least another power 
of $\lambda$, resulting in at least $\mathcal{O}(\lambda^2)$ 
suppression of the collinear loop. 
Second, the collinear loop correction can be on the 
external leg $i$ only with no connection to the hard vertex, 
in which case one either needs two purely-collinear 
vertices of at least $\mathcal{O}(\lambda)$ or a four-point 
vertex, which is already $\mathcal{O}(\lambda^2)$. By the 
same argument, adding any further collinear loop incurs 
at least another factor $\mathcal{O}(\lambda^2)$ per loop. 
It follows that the leading term in the soft theorem 
can never be corrected, while a collinear $n$-loop 
correction can contribute only at the $n$th sub-leading order 
to the soft theorem, 
consistent with the claim. Note that the argument holds 
for every collinear direction $i$ separately, since 
the collinear modes from different directions cannot 
interact other than through soft modes, as evident from 
\eqref{eq:Basics:LagrangianSplit},
which implies soft loops. 

\begin{figure}[t]
	\centering
\includegraphics[width=0.17\textwidth]{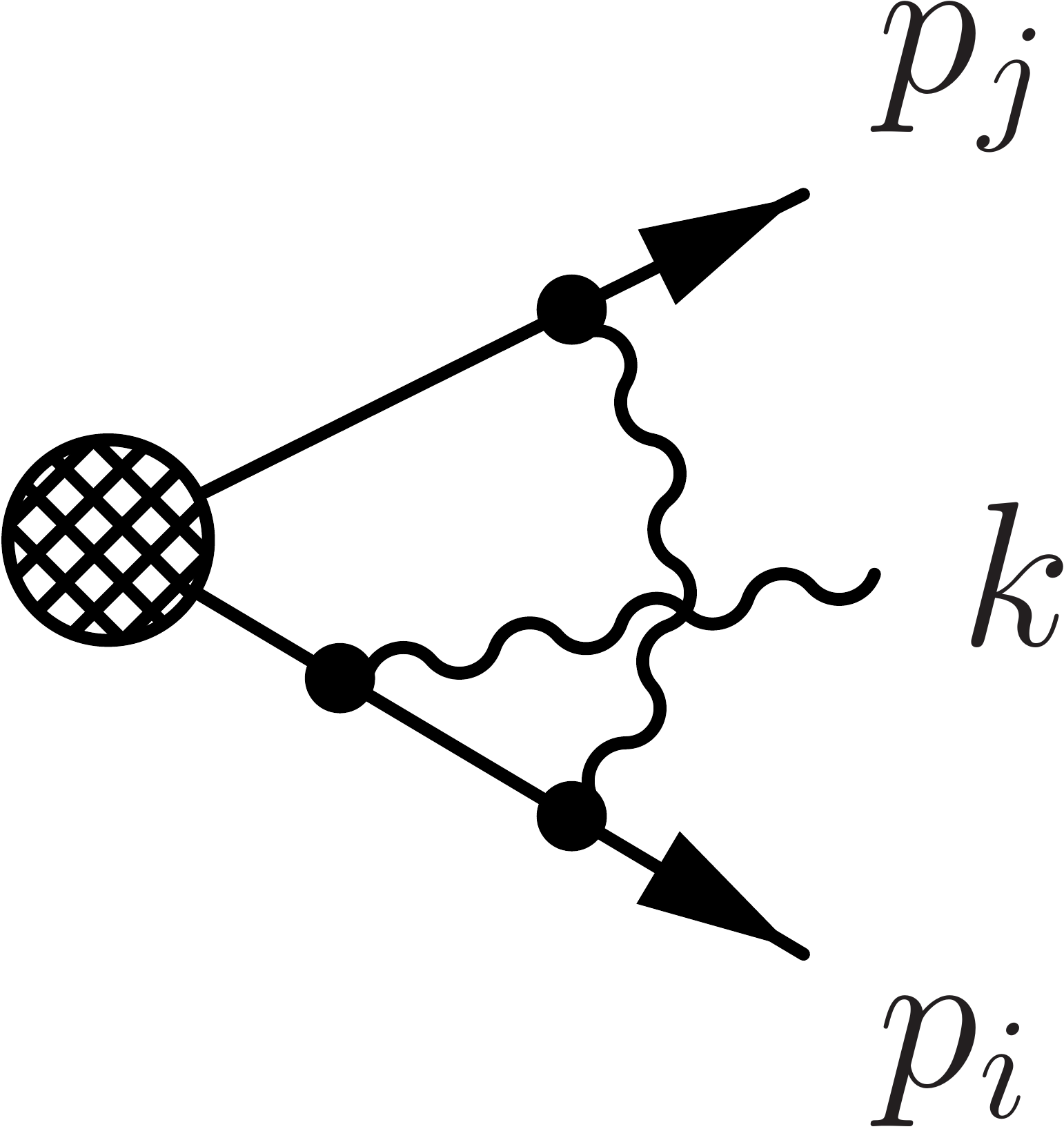}
\qquad
\includegraphics[width=0.17\textwidth]{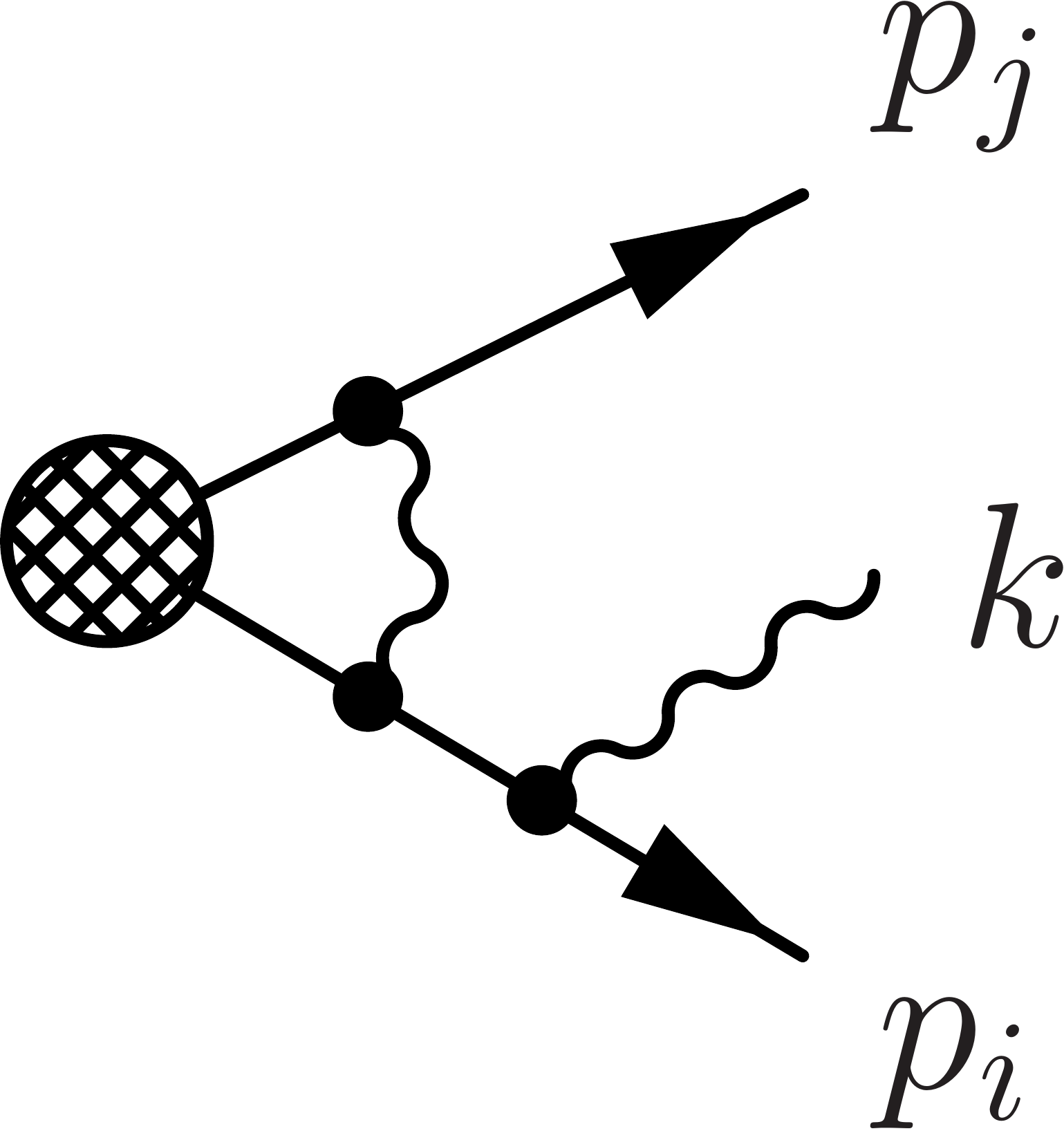}
\qquad
\includegraphics[,width=0.17\textwidth]{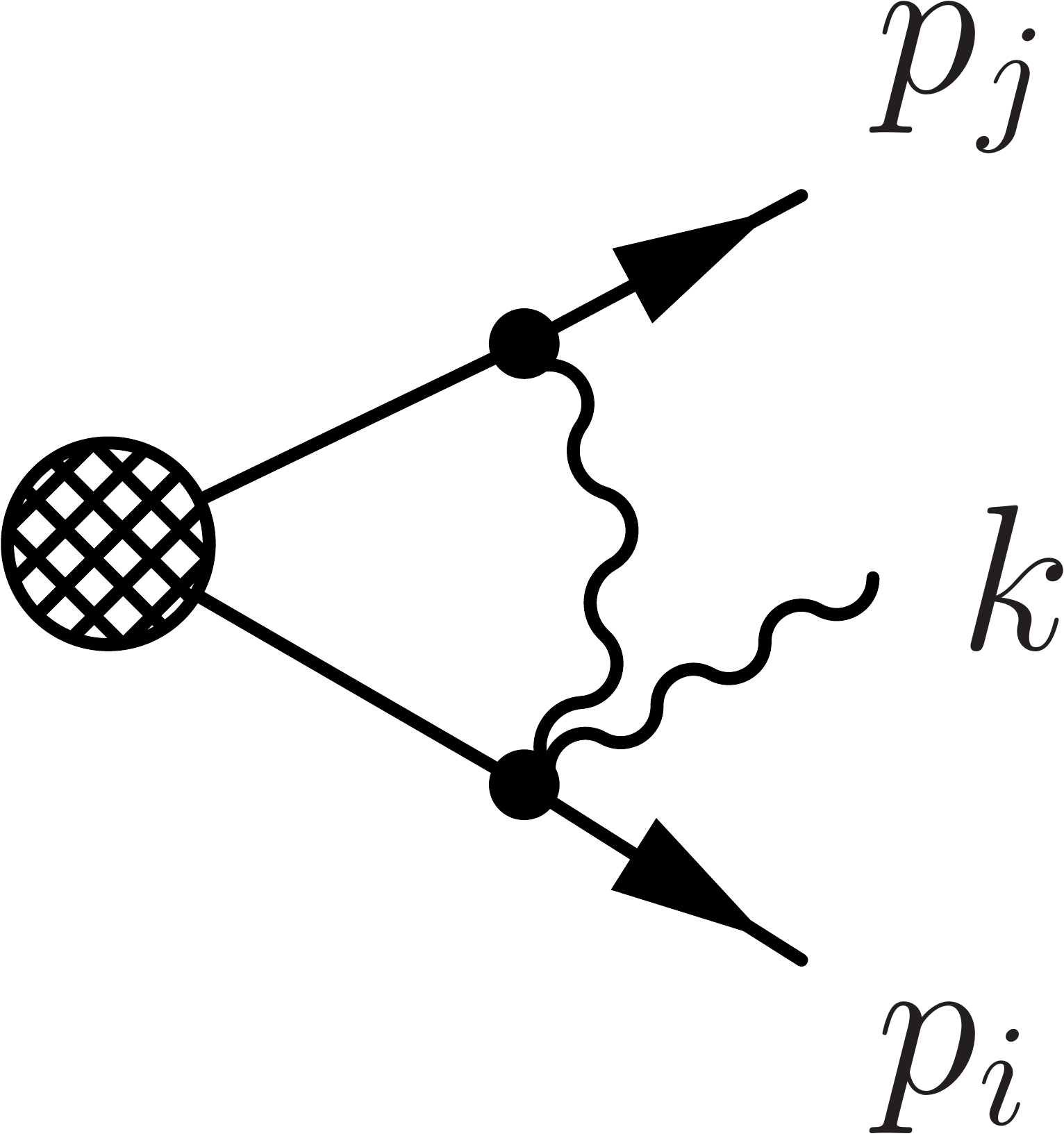}
\qquad
\includegraphics[,width=0.17\textwidth]{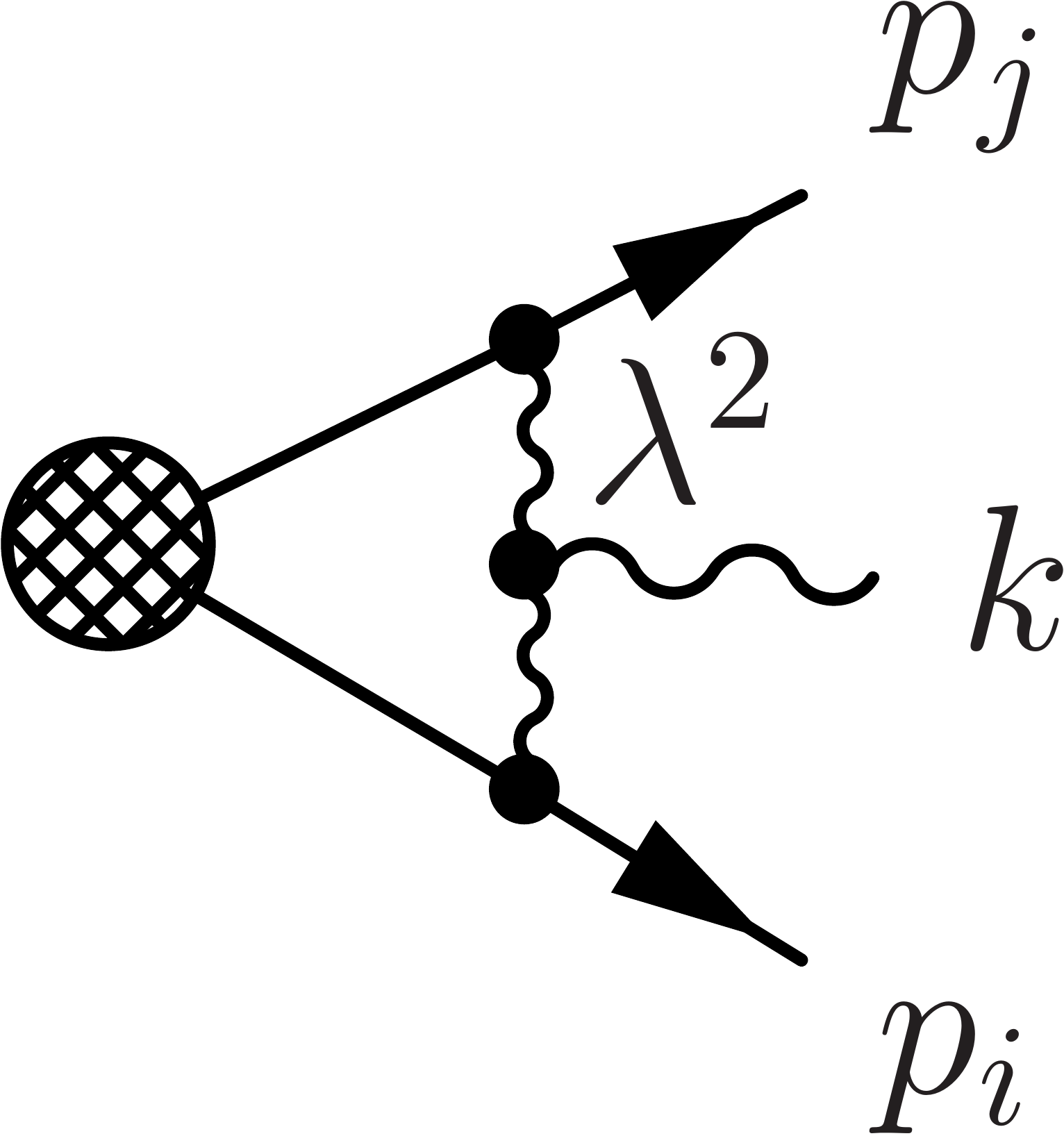}
\caption{Diagram classes modifying the soft emission 
process at the one-loop order. The soft-collinear 
interactions in the first two diagrams are present at 
leading power in $\lambda$. However, due to multipole 
expansion, soft-collinear vertices are only sensitive 
to the $n_{i-}k$ component, and the soft loop vanishes 
unless a soft scale, provided by the injection of the full 
soft momentum $k$, is present. This can only happen by a 
purely-soft interaction vertex. Hence, only the right-most 
diagram is non-vanishing, but is 
at least $\mathcal{O}(\lambda^2)$ suppressed, since 
it contains a purely-soft interaction.}
\label{fig:softloops}
\end{figure}

The case of soft loops is less straightforward.
Since there are no hard vertices with soft fields up to sub-sub-leading 
order,  all soft loops to this order must be built from 
Lagrangian interactions. We first consider the case that 
all loops are soft and none is collinear.

Beginning with one-loop soft corrections, the soft loop can 
connect at most two collinear directions with topologies as 
shown in Figure~\ref{fig:softloops},  
and a similar set of diagrams, where the soft loop  
is attached to a single leg only. These contributions vanish unless the external soft graviton is connected to the loop through a purely-soft interaction, as shown in the right-most diagram in the figure.\footnote{See Appendix~B of 
\cite{Beneke:2019oqx} for a similar discussion for soft interactions in gauge theory at next-to-leading power.}  
The key point is that soft fields are always multipole-expanded 
in interactions with collinear fields, which implies that 
for soft momentum $p_s$, only $(n_{i-}p_s) \,n_{i+}^\mu/2$ enters 
the momentum-conservation delta-function at a soft-collinear 
vertex and hence the collinear propagators in direction $i$. 
The loop depicted in Figure~\ref{fig:softloops}, with the soft graviton emission removed, is given by the (dimensionally-regulated) integral
\begin{eqnarray}
I &\propto& \int \frac{d^d l}{(2\pi)^d} \frac{1}{p_i^2 + n_{i+}p_i n_{i-}l + i0} \frac{1}{p_j^2 - n_{j+}p_j n_{j-}l + i0} \frac{1}{l^2+i0}\,,
\end{eqnarray}
which for on-shell external particles, $p_i^2 = p_j^2 = 0$, 
simplifies to
\begin{equation}
\label{eq::1loopfinal}
I  \propto 
\int \frac{d^d l}{(2\pi)^d} \frac{1}{l^2+i0} \frac{1}{n_{i-}l +i0} \frac{1}{n_{j-}l + i0}\,.
\end{equation}
This integral is evidently scaleless, and vanishes. Note that the 
diagrams of Figure~\ref{fig:softloops} provide 
numerators for this integral, which are polynomial in the 
external and the loop momentum. They have not been written 
explicitly, since whether an integral is scaleless or not 
is independent of such numerators. 

If one now attaches the external soft graviton to one of the 
collinear lines (see first three 
diagrams in Figure~\ref{fig:softloops}), one can always route the 
external soft momentum $k$ such that it appears in the eikonal 
propagators of only one of the legs, say $i$. In this way, the 
loop integral \eqref{eq::1loopfinal} may be modified to include 
eikonal propagators of the form $(n_{i-}(l+k) +i0)^{-1}$. 
Since only the $n_{i-} k \,n_{i+}^\mu/2$ component of the 
soft momentum can ever appear in the denominator, one cannot 
form an invariant scalar product containing $k$ with the required 
soft scaling $\mathcal{O}(\lambda^4)$ (as $n_{i+}^2=0$) and the soft loop 
integral will remain scaleless and vanishing. In order for soft 
loops to yield a non-zero contribution, one needs to bring the 
full external soft momentum $k^\mu$ into the loop integral. This 
requires the external soft graviton to couple to the loop through 
a purely-soft interaction (as in the right-most diagram in the 
figure). Such interaction vertices involve the full momentum 
conservation delta function and lead to propagators $1/(l+k)^2$, 
which allows the soft integral to depend on the soft invariant  
$(n_{i-} k) (n_{j-} k)\,(n_{i+}n_{j+})$ 
and be non-zero. However, by point $ii)$ above, such a 
purely-soft vertex comes at the cost of power-suppression 
by at least $\lambda^2$. Hence, soft one-loop corrections also cannot 
affect the leading term in the soft theorem, however,  
the next-to-soft term can be modified by diagrams with 
the external soft graviton attached to a purely-soft vertex.  

The above argument generalises to the following all-order 
statement: {\it In soft loop-corrections to the soft theorem, contrary to 
the tree-level case, the emitted soft graviton must always 
attach to a purely-soft vertex, and never directly to any of 
the energetic particle lines.} The reason is that soft-collinear 
interactions involve the soft field at the multipole-expanded 
point $x_-^\mu$ to any order in the $\lambda$-expansion. 
Hence, if the emitted graviton couples directly to an energetic 
line, one can always route its momentum such that the entire 
loop integral will depend only on $n_{i-} k \,n_{i+}^\mu/2$ 
of a single collinear direction, $i$, and no soft invariant 
can be formed to provide a scale to the loop diagram.

\begin{figure}[t]
\centering	
\includegraphics[width=0.2\textwidth]{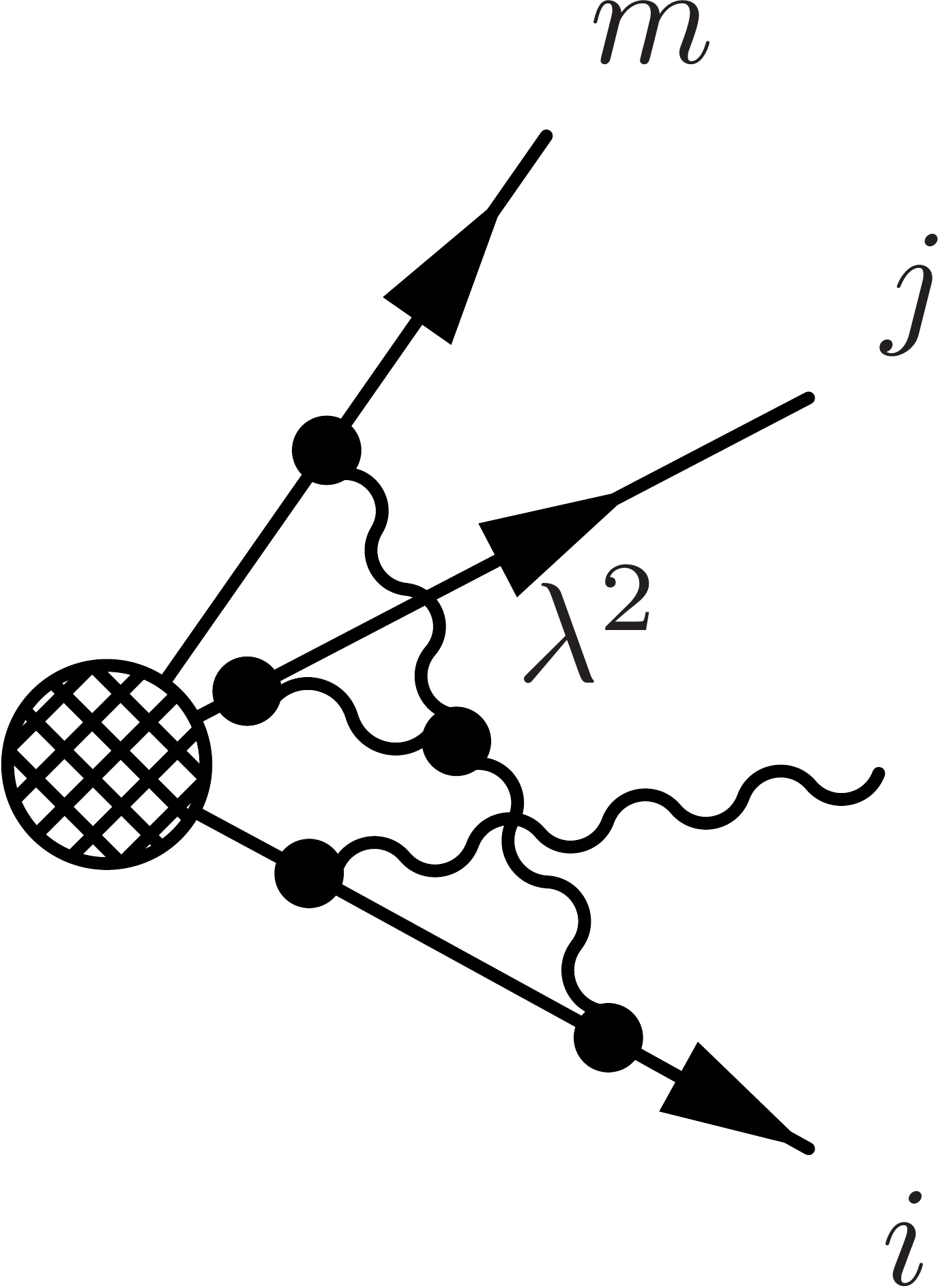}\hskip0.4cm
\includegraphics[width=0.2\textwidth]{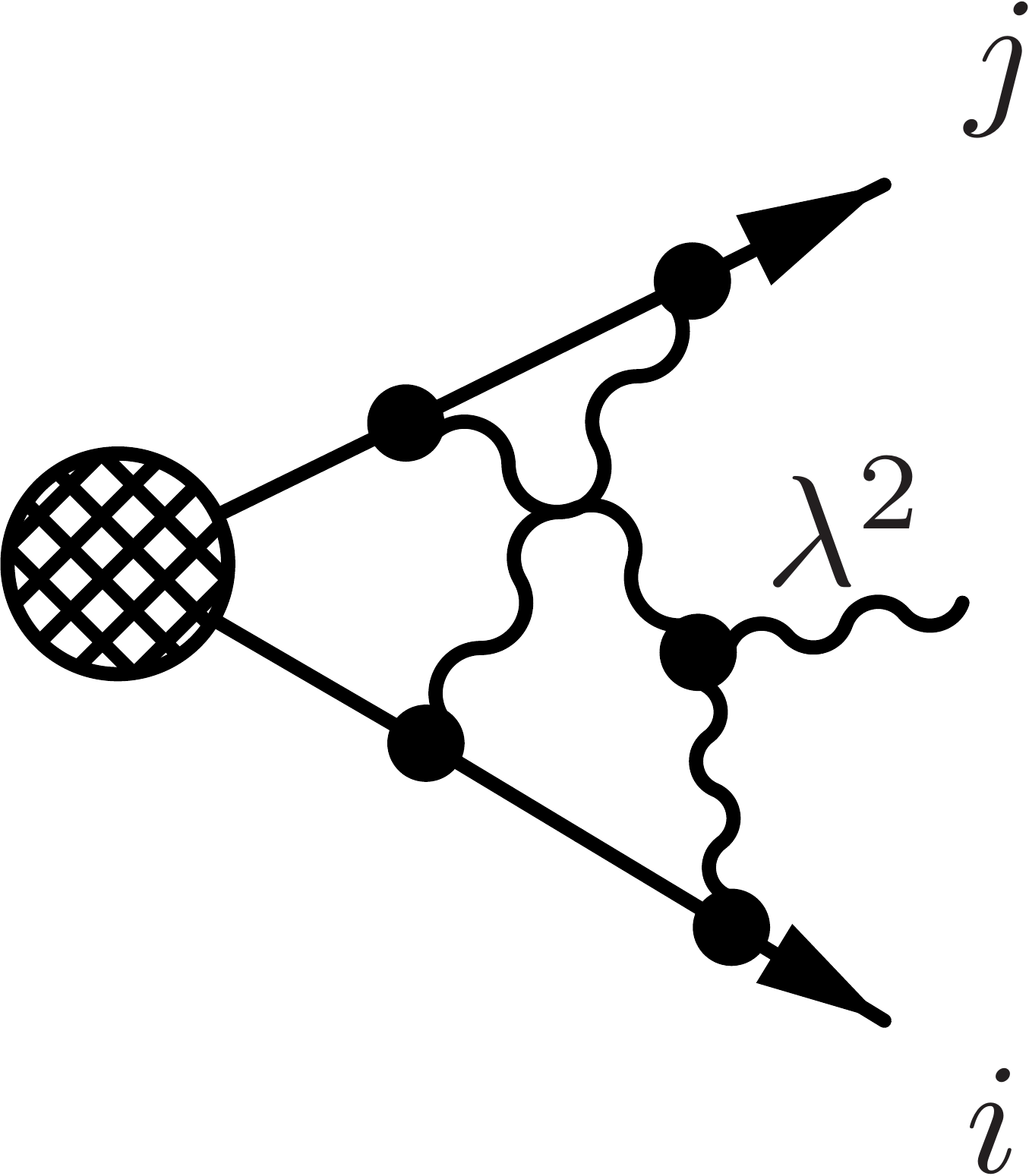}\hskip0.4cm
\includegraphics[width=0.2\textwidth]{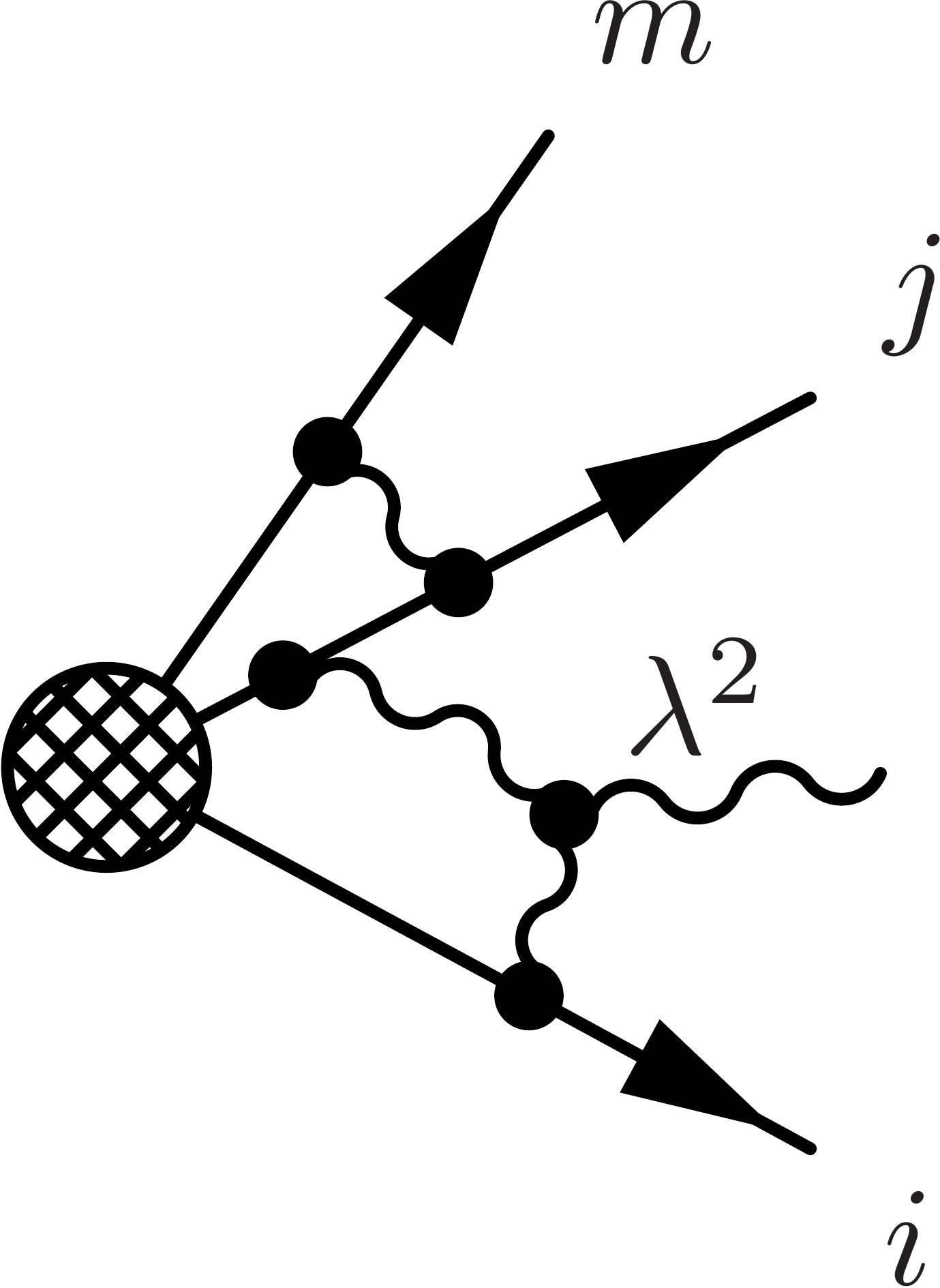}\hskip0.4cm
\includegraphics[width=0.2\textwidth]{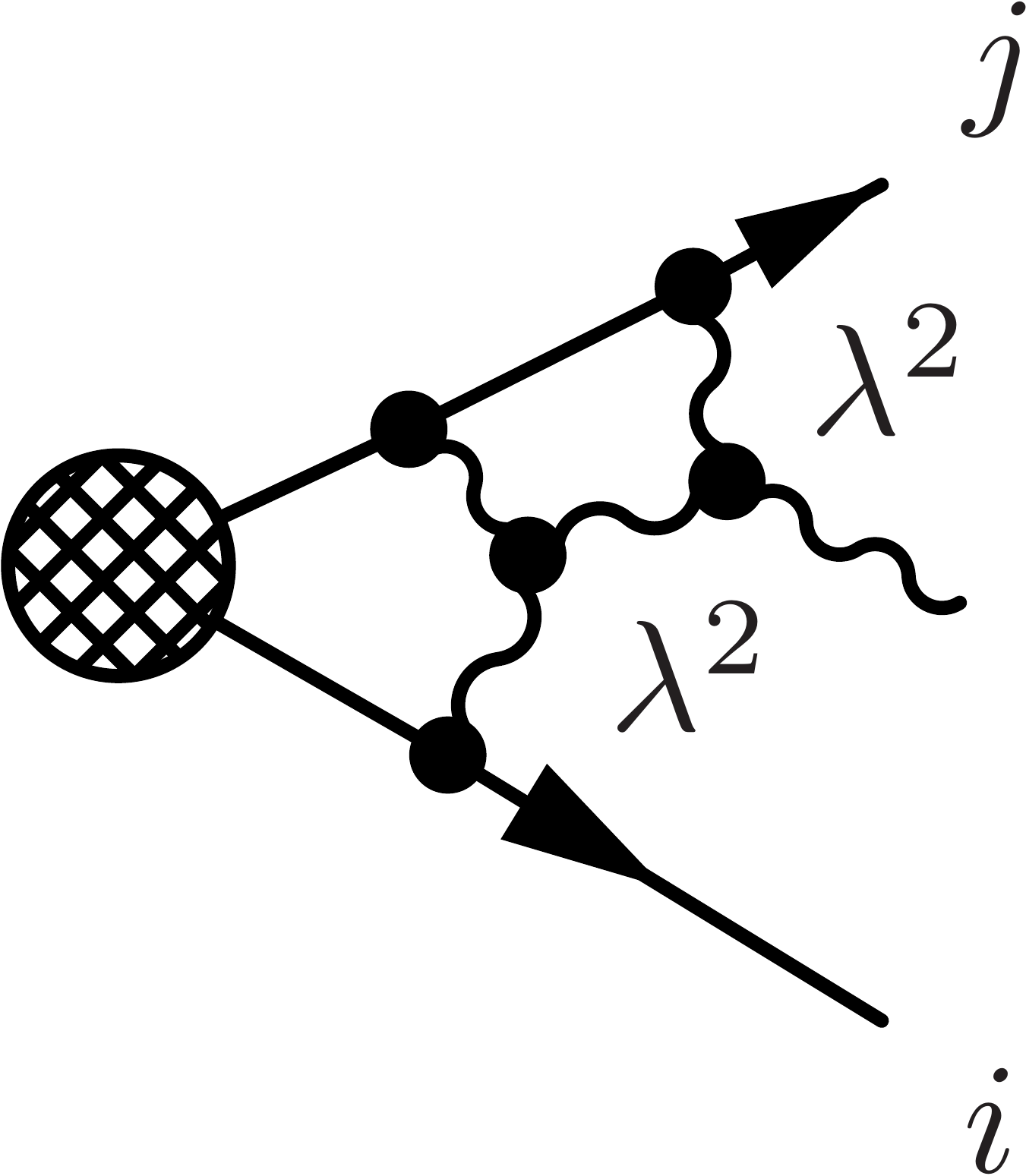}
\caption{Examples of soft two-loop diagrams. Left: Diagram with 
soft emission from a collinear line. Middle: Two diagrams with 
soft-collinear interactions and a purely-soft vertex. 
Right: A non-vanishing 
soft two-loop diagram. In order for the soft scale $k$ to be 
present in both loops, the loops must be connected via 
purely-soft interactions to the emitted graviton, which 
results in a contribution at sub-sub-leading order due to 
$\lambda^4$ suppression. }
\label{fig:softtwoloop}
\end{figure}

Continuing with two soft loops, whenever the diagram contains 
a second purely-soft vertex (as in the right-most diagram of  
Figure~\ref{fig:softtwoloop}), 
there is another 
$\mathcal{O}(\lambda^2)$ suppression factor from this 
vertex and the diagram can at best contribute to the 
sub-sub-leading soft factor, in agreement with the 
assertion. This conclusion can potentially be evaded by 
coupling the soft lines to the energetic lines with 
$\mathcal{O}(\lambda^0)$ or $\mathcal{O}(\lambda^1)$ vertices, 
which adds only linear eikonal-type propagators, possibly raised to 
higher powers due to the multipole expansion. The argument 
in the previous paragraph excludes coupling the emitted 
soft graviton to an energetic line (as in the left-most  
diagram of Figure~\ref{fig:softtwoloop}), but the middle 
two graphs in the figure are still an option. We next show 
that these graphs obtain another $\mathcal{O}(\lambda^2)$ 
suppression from the soft-collinear vertices on the energetic 
lines and they can only contribute to the sub-sub-leading 
term in the soft theorem consistent with the claim.\footnote{This 
and the following arguments fill a loop-hole in the 
discussion of soft loops in \cite{Bern:2014oka}, where it 
is assumed that the dimensionful coupling $\kappa$ that 
comes with a loop must be compensated by a soft momentum 
$k\sim\mathcal{O}(\lambda)$, so a $k$-soft-loop diagram 
must be $\mathcal{O}(\lambda^{2k})$. However, in soft-collinear 
interactions, the factor of $\kappa$ can be compensated by 
powers of the $\mathcal{O}(1)$ collinear momenta as there exists 
a  $\mathcal{O}(\lambda^0)$ vertex. In other words, for 
soft-collinear interactions the loop-expansion in $\kappa$ is 
not equivalent to a $\lambda^2$-expansion.} 

Consider first the case that all soft-graviton couplings 
to the energetic particle lines are through the leading-power 
soft-collinear interaction $\frac{\kappa}{4} s_{--} 
\chi_c^\dagger(n_+\partial^2\chi_c)$ from 
\eqref{eq:GRSCETL0}. This gives 
rise to the eikonal interaction, which has the well-known 
property \cite{Weinberg:1995mt}
that multiple emissions take the form of {\em independent} 
emissions after summing over all possible attachments 
to the energetic lines. For $i=1,\ldots, N$ such lines 
with external momenta $\{p_i\}$, 
the emission of an arbitrary number of soft gravitons
with momenta $\{q_k\}$ is given by the non-radiative 
amplitude multiplied by the eikonal factor:
\begin{equation}
\mathcal{A}(\{p_i\}) \,\sum_{m=0}^\infty \frac{1}{m!}\,
\prod_{k=1}^m\left(\sum_{i=1}^N 
\frac{\kappa}{2} \,\frac{p_i^{\mu_k} p_i^{\nu_k} \epsilon_{\mu_k\nu_k}(k)}
{p_i\cdot q_k+i\epsilon}\right)\,.
\label{eq:eikonalamp}
\end{equation}
To construct a loop diagram with a single emitted soft 
graviton, one must tie together the $q_k$ amongst each other 
with a soft graviton propagator or with the purely-soft  
vertex from which the external graviton is emitted. The 
key observation is that due to the factorised 
form of the denominator of the eikonal amplitude 
\eqref{eq:eikonalamp}, the external momentum $k$ 
does not enter the loop integral of those pairs of 
$q_k$ tied together directly, which therefore can only 
depend on the energetic momenta $p_i$. Now, a 
dimensionally-regulated soft loop integral must be proportional 
to $(S/\mu^2)^\epsilon$, where $S$ is a soft invariant, 
that is, a scalar product of momenta, which scales as 
$\lambda^4$, and has mass dimension two. Since no such 
invariant can be built from the available momenta 
as $p_i\cdot p_j\sim \lambda^0$ 
and $p_i^2=0$, these integrals must be scaleless and 
therefore the entire graph vanishes. This applies to the 
two middle diagrams in Figure~\ref{fig:softtwoloop} when 
all soft-collinear vertices are from $\mathcal{L}^{(0)}$, 
which would otherwise contribute at 
$\mathcal{O}(\lambda^2)$. Note that this vanishing 
also holds if some of the $q_k$ are tied together 
through purely-soft vertices, as long as 
the emitted graviton is not attached to them, since 
in this case $k^\mu$ never enters the corresponding subgraph. 
This results in the following all-order statement: 
{\it If a loop integral involves only the leading-power 
eikonal interactions from $\mathcal{L}^{(0)}$, it vanishes 
unless all soft propagators are connected to the 
external graviton line through purely-soft vertices.} 
As a consequence, any non-vanishing soft $n$-loop graph  
has a suppression of at least $\mathcal{O}(\lambda^{2 n})$ 
from the combination of the required purely-soft vertices.

There is still the possibility that the emitted graviton 
couples to an energetic line through some of the sub-leading 
in $\lambda$ soft-collinear interactions from 
$\mathcal{L}^{(1)}$~(\ref{eq::SG::LagrangianCovDevL1}) and  $\mathcal{L}^{(2)}$~(\ref{eq::SG::LagrangianCovDevL2}). 
The vertices from $\mathcal{L}^{(1)}$ contain 
a single transverse index. In the frame $p_{i\perp}^\mu=0$, 
there is no transverse external vector available, 
hence diagrams with only a single $\mathcal{L}^{(1)}$ insertion 
on an energetic line vanish. It already follows that one cannot 
evade the $\lambda^2$ suppression per loop by the use of 
sub-leading soft-collinear interactions and the two-loop 
diagrams are all at least of order $\mathcal{O}(\lambda^4)$. 
However, these diagrams actually vanish. At the two-loop order, 
the only non-vanishing topologies contain two 
purely-soft vertices connected by a soft propagator as in the 
right-most diagram of Figure~\ref{fig:softtwoloop}, or a purely-soft 
four-graviton vertex, which also scales at least as 
$\mathcal{O}(\lambda^4)$.

\begin{figure}[t]
\centering	
\includegraphics[width=0.5\textwidth]{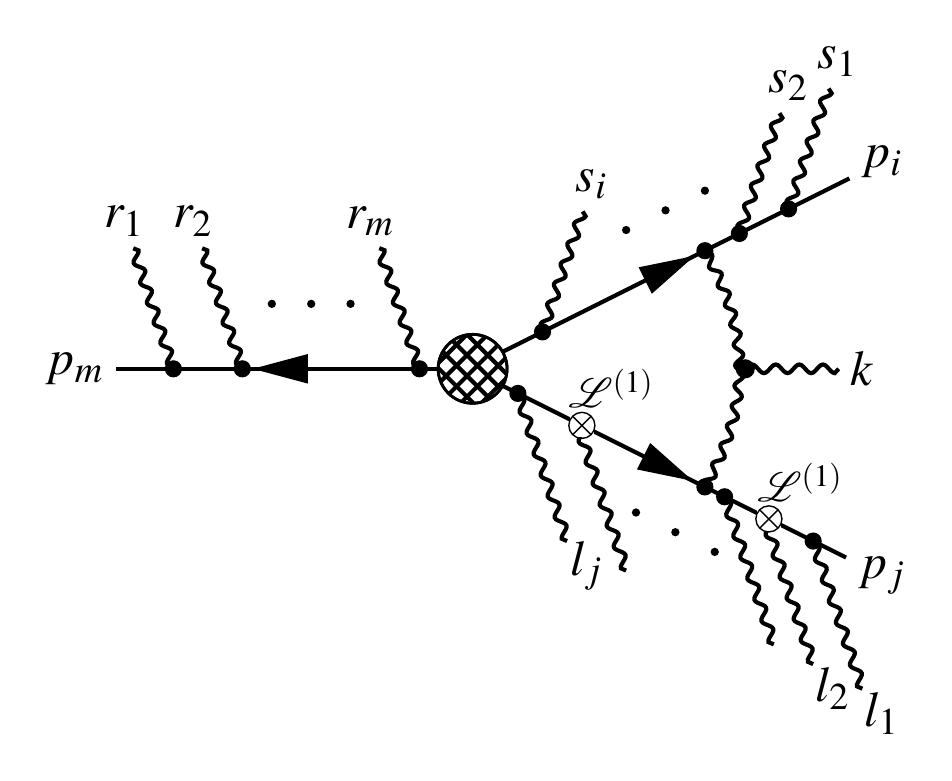}
\caption{$\mathcal{O}(\lambda^4)$ diagram with one purely-soft 
vertex and sub-leading soft-collinear interactions on line 
$j$. Except for the emitted graviton with momentum $k$, all 
other wavy lines have to be tied together amongst each 
other.}
\label{fig:semieikonalsoft}
\end{figure}

To see this, assume that the single purely-soft three-graviton 
vertex connects the emitted graviton to the energetic 
legs $i$ and $j$. Since the sub-leading eikonal vertices that provide 
further $\mathcal{O}(\lambda^2)$ suppression reside on a 
single energetic leg, one can further assume without loss of generality 
that line $i$ has only leading-power eikonal vertices attached to 
it, and then route the external momentum $k^\mu$ through 
this energetic leg. Leg $i$ is therefore an eikonal leg, 
that is, after summing over all permutations of attached 
momenta, the amplitude for this leg takes the form of independent 
emissions. All other external energetic lines also have this 
property, except for the one with sub-leading soft-collinear 
vertices, which may or may not be leg $j$. The 
situation and momentum assignments are illustrated in 
Figure~\ref{fig:semieikonalsoft}. 

At the two-loop order there is only one pair of open lines 
to be tied up with a soft graviton propagator, 
but for later purposes it is convenient to  
proceed with the general situation shown in the figure. 
Because leg $i$ is an eikonal leg, the momenta $k$ and $q$  
appear in the entire diagram only in denominators 
\begin{equation}
\frac{1}{(k+q)^2 q^2}\frac{1}{n_{i-}\!\cdot (k+q)}
\times \prod_k \frac{1}{n_{j-}\!\cdot L_k(q,l_j)}\,,
\end{equation}
where $L_k(q,l_j)$ is the linear sum of a subset (or all) 
of the momenta attached to leg~$j$. Note that this 
expression is general, since it was not assumed that leg $j$ is 
eikonal. Let $l$ be one of the momenta $l_j$. The loop 
integral over $l$ is then of the form 
\begin{equation}
\int\frac{d^dl}{(2\pi)^d}\,
\frac{1}{l^2} \frac{1}{n_{a-}\cdot l} \,
\prod_k \frac{1}{n_{j-}\cdot L_k(q,l_j)}\,,
\label{eq:lloop}
\end{equation}
where $a\not=j$ refers to the eikonal leg to which 
the graviton line $l$ is attached and we used that the emissions 
are independent from this leg. After integrating over $l$, 
the result can only depend on  $(n_{j-}q) \,n_{j+}^\mu/2$ 
and  $(n_{j-}l_j) \,n_{j+}^\mu/2$, from which one cannot 
form a non-vanishing invariant, hence the integral is 
scaleless, and consequently the entire diagram is zero. 
If $a=j$, the expression \eqref{eq:lloop} without the 
factor $1/(n_{a-}\cdot l)$ holds, and the same conclusion is 
reached.

To complete the proof of the main assertion for soft-loop corrections, 
it remains to show that three- and higher-loop diagrams 
are inevitably more suppressed than $\mathcal{O}(\lambda^4)$, so that 
they cannot affect the soft theorem to the sub-sub-leading order. 
The preceding two all-order statements imply that a diagram 
that is  not already excluded by them must 1) have exactly 
one purely-soft vertex of  $\mathcal{O}(\lambda^2)$, to which 
the external soft graviton is attached, 
and 2) $\mathcal{O}(\lambda^2)$ suppression 
from sub-leading soft-collinear 
interactions on a single energetic leg, say $j$, 
plus 3) an arbitrary number of additional 
leading-power eikonal interactions. The property of 
independent soft emission therefore applies to all 
energetic lines except $j$. However, this case is 
covered by the arguments of the previous paragraph, 
which holds to any loop order.

Having established suppression of either only collinear or only soft 
loop corrections in accordance with the claim, we finally turn to 
arbitrary loop diagrams that can have both, which start from two 
loops. It is easy to see that two-loop diagrams must be at least 
$\mathcal{O}(\lambda^4)$, since the external soft graviton must 
still attach to a purely-soft collinear vertex by the same argument 
as before, while the collinear loop still costs a factor of 
$\lambda^2$. The case of higher-loop diagrams is covered by the 
argument of the previous paragraph, which continues to hold  
when the leg with the sub-leading soft-collinear Lagrangian 
insertions is replaced by the leg, to which a collinear loop
is attached. Any mixed collinear-soft loop diagram with more than 
two collinear loops is already beyond the next-to-next-to-soft 
order and does not have to be considered. This concludes the  
characterisation of the structure 
of loop corrections to the soft theorem, as formulated 
at the beginning of this subsection. 

The one- and two-loop corrections to the soft theorem 
are largely universal, since they arise almost exclusively 
from effective Lagrangian terms. The non-universal 
part can be parameterised in terms of the matching 
coefficients $\widetilde{C}^{X}(t_i)$ of SCET source 
operators with up to two additional collinear transverse 
graviton 
fields. An explicit calculation of the loop-corrected 
soft factors has, however, not yet been performed.

\subsubsection*{Acknowledgement}
This work was supported by the Excellence Cluster ORIGINS
funded by the Deutsche Forschungsgemeinschaft 
(DFG, German Research Foundation)
under Grant No. EXC - 2094 - 390783311.
RS is supported by the United States Department of Energy under Grant Contract DE-SC0012704. MB thanks the Albert Einstein 
Center at the University of Bern for hospitality while 
this work was finalised.



\biblstarthook{}

\end{document}